\documentclass[letterpaper]{mn2e}
\usepackage{graphicx,times}
\usepackage{verbatim}

\def \sanch{S\'anchez-Bl\'azquez}

\def \kms {${\rm{km}\,\rm{s}^{-1}}$}

\def \ha   {H\,$\alpha$}

\def \afe  {[$\alpha$/Fe]}

\def \eza   {{\sc EZ-ages}}

\def \apjs{ApJS}

\def \apj{ApJ}
\def \aj{AJ}
\def \mnras{MNRAS}

\def \aanda{A\&A}

\title[Stellar populations of dwarf galaxies in Coma]
{
A spectroscopic survey of dwarf galaxies in the Coma Cluster: Stellar populations, 
environment and downsizing\thanks{Observations reported here were obtained at the MMT Observatory,
a joint facility of the University of Arizona and the Smithsonian Institution. Time was awarded through the National Science Foundation's Public Access Programme.}
}
\author[Russell J. Smith et al. ]
{Russell J. Smith$^{1}$\thanks{Email: russell.smith@durham.ac.uk}, 
John R. Lucey$^{1}$,
Michael J. Hudson$^{2}$, 
Steven P. Allanson$^{2}$,
\newauthor
Terry J. Bridges$^{3}$, 
Ann E. Hornschemeier$^{4}$, 
Ronald O. Marzke$^{5}$, 
Neal A. Miller$^{6}$
\\
$^1$Department of Physics, University of Durham, Durham DH1 3LE\\
$^2$Department of Physics and Astronomy, University of Waterloo, 200 University Avenue West, Waterloo, Ontario N2L 3G1, Canada\\
$^3$Australian Gemini Office, Anglo-Australian Observatory, PO Box 296, Epping NSW 1710, Australia\\
$^4$NASA Goddard Space Flight Centre, Code 662.0, Greenbelt, MD 20771, USA\\
$^5$Department of Physics and Astronomy, San Francisco State University, San Francisco, CA 94132, USA\\
$^6$Department of Physics and Astronomy, Johns Hopkins University, 3400 North Charles Street, Baltimore, MD 21218, USA\\
}

\date{MNRAS accepted} 
\pagerange{\pageref{firstpage}$-$\pageref{lastpage}} \pubyear{2008}

\begin{document}

\label{firstpage}

\maketitle

\begin{abstract}
We investigate the stellar populations in a sample of 89 faint red galaxies in the Coma cluster, 
using high signal-to-noise spectroscopy from the 6.5m MMT. 
Our sample is drawn from two one-degree fields, one centred on the cluster core and the other located a degree 
to the south west of the cluster centre. The target galaxies are mostly 2--4\,magnitudes fainter than $M^*$; 
galaxies with these luminosities have been previously studied only using small samples, or at low
signal-to-noise ratio ($S/N$). 
For a comparison sample we use published high-$S/N$ data for red-sequence galaxies in the Shapley Supercluster. 
We use state-of-the-art stellar population models (by R. Schiavon) to interpret the absorption line indices and infer
the single-burst-equivalent age and metallicity (Fe/H) for each galaxy, as well as the abundances of the
light elements Mg, Ca, C and N. 
The ages of the Coma dwarfs span a wide range from $<2$\,Gyr to $>10$\,Gyr, with a strong environmental 
dependence. The oldest galaxies are found only in the core, while most of the galaxies in the outer south-west field 
have ages $\sim$3\,Gyr.
The galaxies have a metallicity range $-1.0\la$\,[Fe/H]\,$\la$\,0.0, 
and follow the same age--metallicity--mass plane as high-mass galaxies, but with 
increased intrinsic scatter. The Mg/Fe abundance ratios are on average slightly super-solar, and span a range 
$-0.1\la\,$[Mg/Fe]$\,\la+0.4$. The highest Mg enhancements are found only in the cluster core, while solar ratios
predominate in the outskirts. 
We show that parametrized models with more complex star-formation histories perform no better 
than single-burst models in reproducing the observed line indices. 
Assuming a star-formation history dominated by a single burst, the number of dwarf galaxies on the red
sequence in the Coma core has doubled since $z\approx0.7$. Assuming instead an abruptly-truncated constant star-formation 
rate, the equivalent redshift is $z\approx0.4$. These estimates bracket the red-sequence growth timescales 
found by direct studies of distant clusters.
In the south-west field, the red sequence was established only at $z\approx0.2$ for a burst-dominated star-formation
history ($z\approx0.1$ for the truncated case). Our observations confirm
previous indications of very recently quenched star formation in this part of the cluster. 
Our results strongly support the scenario in which much of the cluster passive dwarf population (in this luminosity range) was generated by 
environment-driven transformation of infalling late-type galaxies.
\end{abstract}

\section{Introduction}

Today's passive galaxies are relics of complex star-formation and chemical enrichment processes operating over the 
entire history of the Universe. 
Careful comparison of observed galaxy spectra against sophisticated stellar population models can 
reveal the characteristic stellar ages of galaxies, as a function of their mass, morphology, 
environment or other properties. Measurements of elemental abundances can further constrain the formation 
histories of galaxies, being sensitive to the duration of star-formation events. These so-called 
``archaeological'' constraints provide a means to study galaxy evolution which is complementary
to the ``look-back'' approach of quantifying galaxy populations at significant redshifts, where
only more superficial observations are possible. 

In practice, much of the information in spectra of unresolved galaxies is concentrated in a small number of 
absorption features, which have been used to establish systems of line-strength indices for stellar 
population analysis
(e.g. Burstein et al. 1984; Rose 1985). Worthey et al. (1994) demonstrated the power of combinations of these
indices to break the degeneracy between age and metallicity in the unresolved optical light from
galaxies. Further progress has been made by including the effects of non-solar element abundance patterns
(e.g. Trager et al. 2000a; Thomas, Maraston \& Bender 2003; Thomas, Maraston \& Korn 2004; Schiavon 2007).

These analysis methods were initially applied to samples comprised mainly of giant elliptical and S0 galaxies
(e.g. Trager et al. 2000b; Kuntschner et al. 2001; Thomas et al. 2005). Subsequent work extended the 
methods to galaxies of lower mass and luminosity (e.g. Caldwell et al. 2003; Nelan et al. 2005; 
Smith, Lucey \& Hudson 2007), and incorporated variable $\alpha$-element abundances in a systematic way, 
using the models of Thomas et al. (2003, 2004; collectively TMBK hereafter). 
The enlarged  baseline helped to reveal
the overall scaling relations of age, metallicity and \afe, as a function of velocity dispersion, interpreted as a 
proxy for mass. On average, less massive galaxies are younger, less metal-rich and less $\alpha$-enhanced (or alternatively less
Fe-deficient), than the giant ellipticals. The \afe\ trend is widely interpreted as evidence for a longer timescale of 
star formation in lower-mass systems, allowing more pollution by Fe from Type Ia supernovae (Thomas et al. 2005). 

Given these results, it is natural to extend the stellar population analysis methods to even lower luminosity, to include 
the dwarf galaxy population\footnote{We will use the term ``dwarf galaxy'' loosely for objects fainter than 
$\sim{}M^\star+2$, where $M^\star$ is the characteristic break in the luminosity function. This corresponds to $M_r\ga-19$.  We will not make a 
strict distinction based on profile shape, in contrast to the traditional definition of dwarf ellipticals (e.g. Ferguson \& Binggeli 1994).
The profile shapes are briefly discussed in Section~\ref{sec:sigphotom}.
}. 
Dwarfs are the focus of a number of key questions about galaxy evolution. Their demographics are highly dependent on environment, 
with {\it passive} dwarfs being extremely numerous in clusters (e.g. Binggeli, Sandage \& Tammann 1985) and present in groups (Ferguson \& Sandage 1991) and as 
satellites of massive galaxies (Binggeli, Tarenghi \& Sandage 1990), 
but essentially absent from the field galaxy population (Haines et al. 2006). Tully et al. (2002) note that the cluster dwarf
luminosity function has similar slope to the mass function of dark matter halos, suggesting that the dwarfs in clusters cooled 
prior to reionization, and avoided the photo-ionization suppression suffered by field dwarfs. If so, the cluster dwarfs (or at least
those which survived without subsequent disruption or merging) would be relics from the earliest phases of galaxy formation. 

Alternatively, some or all of the cluster dwarfs could have been formed by transformation of late-type galaxies
through interaction with the cluster potential or the intra-cluster medium, or via encounters with other cluster members. 
Numerous physical mechanisms
for such transformation have been discussed, including ram-pressure stripping of cold or hot gas, or both, and star-formation triggered by tidal
interactions. (A thorough review is given by Boselli \& Gavazzi 2006). These processes become more efficient at later epochs ($z\la1$), 
as the deep potential wells of rich clusters develop. The environmental ``quenching'' of star formation to form passive dwarf galaxies 
may account for the strong evolution in blue galaxy fraction (Butcher \& Oemler 1984) and red-sequence luminosity function 
(e.g. De Lucia et al. 2007; Stott et al. 2007) out to modest redshifts $z\sim0.5$. 
 
Observationally, the study of integrated stellar populations in dwarfs has been less extensive than for giants, owing to the difficulty 
of obtaining high signal-to-noise spectra for galaxies which are faint and {\it also} generally of low surface brightness. 
A number of studies have targeted small samples, usually 10--20 galaxies, in Virgo or at comparable distances (Geha, Guhathakurta \& van der Marel 2003; 
van Zee, Barton, \& Skillman 2004; Michielsen et al. 2008; Sansom \& Northeast 2008). Other authors have observed larger samples of dwarfs
in more distant clusters, by exploiting the multiplex gain of wide-field multi-fibre spectrographs 
(e.g. Poggianti et al. 2001; Chilingarian et al. 2008). In general, these studies find a 
wide range of ages for passive dwarfs,
in contrast to the results for giant ellipticals where most galaxies are old. The reported $\alpha$-element abundance ratios are generally 
solar or modestly $\alpha$-enhanced, consistent with extended star-formation histories. The young ages and low $\alpha$/Fe found in at 
least some cluster dwarfs favours
the picture where many passive dwarfs are the quenched remnants of infalling disk galaxies. There may, however, be a greater proportion of old dwarfs
near the centres of clusters (Michielsen et al. 2008), which could indicate a mixture of origins. This latter scenario finds support in the multiple 
sub-populations of Virgo dwarfs identified morphologically by Lisker et al. (2007).

At $\sim$100\,Mpc distance, Coma is among the nearest very rich galaxy clusters, with two central D galaxies NGC 4874 and NGC 4889. 
X-ray observations revealed a subcluster projected $\sim$1.5\,Mpc to the south west of the core, centred on NGC 4839, which appears to be 
merging with the main cluster (Briel, Henry \& B\"ohringer 1992). In the south-west region, between the dominant cluster and
the merging group, Caldwell et al. (1993) reported an excess of early-type galaxies with recent or ongoing star formation, compared to the generally
old spectra in the cluster core. The Caldwell et al. result has been influential in subsequent studies of Coma galaxy populations. In particular, 
the spectroscopic survey of Mobasher et al. 2001 (see also Poggianti et al. 2001, Carter et al. 2002) similarly focused on the core and south-west regions. 
Working from this data, Poggianti et al. (2004) identified faint ``post-starburst'' galaxies, apparently coincident with substructures in the X-ray gas, 
suggesting the bursts were triggered by interactions with the intra-cluster medium. 

Here, we present and analyse new spectroscopic data from the 6.5m MMT for a sample of red dwarf galaxies in the core and south-west 
regions of the Coma cluster. The key improvements over the work of Poggianti et al. (2001) are: (i) a five-fold increase in signal-to-noise ratio, 
resulting in spectra of excellent quality for precise age and metallicity estimates in the dwarf galaxy regime, and (ii) use of  
state-of-the-art stellar population models (Schiavon 2007) to determine detailed elemental abundance patterns. 
The scope of this paper is to describe the observations and derivation of the stellar population parameters, 
and to analyse the distributions of age and metallicity, with reference to models and observations of clusters at higher redshift. 
Some initial results on cluster-centric dependence of the stellar populations have been 
published by Smith et al. (2008a), based on the data reported here. A companion paper (Smith et al. 2008b)
analyses the light element abundance ratios in greater detail. Further observations are already underway to extend our work 
to additional outer fields in Coma, which will help to 
establish whether the south-west part of the cluster harbours stellar populations which are 
distinct from those in other parts of the cluster periphery. 

The outline of the paper is as follows: 
Section~\ref{sec:data} describes the sample construction, observations and measurements of line-strength indices. 
In Section~\ref{sec:poppars} we use the absorption line data to estimate the age, metallicity and abundance
ratios for each galaxy, by comparison with the Schiavon stellar population models. 
Section~\ref{sec:results} presents the ages and metallicities and their correlations with galaxy mass, and considers the
robustness of the results when different stellar population models are used. 
In Section~\ref{sec:discuss}, we compare the results with previous spectroscopic work on cluster dwarfs and with look-back studies
of distant clusters, and consider the results in the context of formation models for the dwarf population. 
Our principal conclusions are reviewed in Section~\ref{sec:concs}. 

For conversion to physical units, we assume cosmological parameters from Hinshaw et al. (2008), i.e. 
$(\Omega_M,\Omega_\Lambda,h)=(0.28,0.72,0.70)$. 

\section{Data}\label{sec:data}

\subsection{MMT Spectroscopy}

Spectra were obtained at the 6.5m MMT using the Hectospec fibre-fed spectrograph
(Fabricant et al. 2005), in parallel with a redshift survey of faint candidate cluster members (Marzke et al., in preparation). 
The observations were performed in queue-mode over the period 2007 February--April, mainly within five days of new moon. 
Hectospec deploys 300 fibres over a 1\,deg diameter field of view (corresponding to 1.75\,Mpc at Coma); 
the fibre diameter is 1.5\,arcsec (0.7\,kpc at Coma). 
Two fields were observed, one centred on the cluster core, and an outer field centred $\sim$0.9\,deg to the south west. 
The choice of the south-west field was motivated by available membership information and other supporting data, and 
ultimately influenced by the work of Caldwell et al. (1993). 

\begin{figure}
\vskip 3mm
\includegraphics [angle=270,width=85mm]{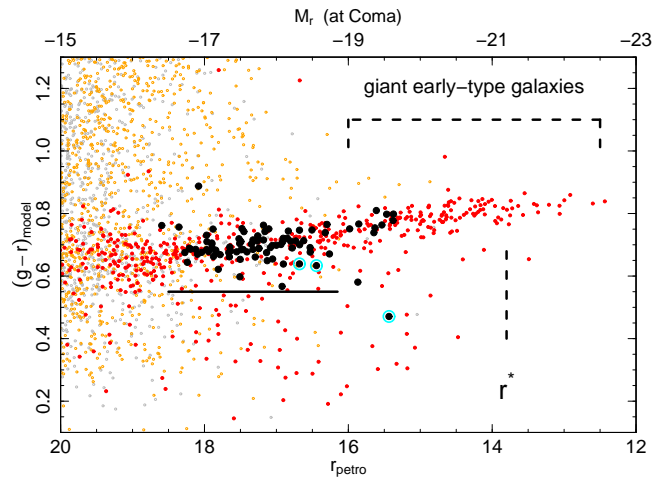}
\caption{The $g-r$ colour magnitude diagram for SDSS galaxies within the two Hectospec fields. The sample of galaxies
analysed in this paper is indicated by large black points. Small points are confirmed members (red), 
confirmed non-members (yellow) and galaxies of unknown membership status (grey). 
Target galaxies with emission lines are highlighted with cyan circles. 
The horizontal line indicates the colour selection limit for the primary sample, which
was not applied to the comparison set of brighter objects. 
The unusually ``red'' target galaxy is GMP3206, which seems to have been 
broken into two sources by SDSS, yielding unreliable photometry. 
For reference, we note the position of the luminosity function break, $r^*$, and the magnitude range for ``giant'' 
early-type galaxies, having velocity dispersions $\sigma>75$\,\kms.
}\label{fig:samplecmr}
\end{figure}

To study the stellar populations of dwarf galaxies, we observed 79 known cluster members with luminosities $2-4$\,mag fainter than $M^*$, 
plus ten brighter galaxies ($r=15.5-16.0$) for overlap with previous studies (e.g. \sanch\ et al. 2006a). 
The faint target galaxies were selected from SDSS imaging data according to colour ($g-r>0.55$) and Petrosian magnitude 
($16.25 < r < 19.00$) (Figure~\ref{fig:samplecmr}). 
Additionally, we imposed cuts on the SDSS 3-arcsec fibre magnitude: $r_{\rm fib}>18.50$ to avoid 
scattered light from bright targets interfering with the parallel faint redshift survey, and $r_{\rm fib}<19.75$ to remove 
low surface-brightness targets which would not yield sufficient $S/N$ in the Hectospec spectra. 
The faint fibre-magnitude cut is equivalent to Petrosian magnitudes $r\sim18.5$, so it is this criterion, not the total magnitude cut,
which effectively sets the faint magnitude limit. Finally, we required the targets to have a measured redshift compatible
with membership of Coma. Since the selection was made prior to availability of redshifts from SDSS for this region, the outer
targets mostly have membership confirmation from Mobasher et al. (2001). 

The 270\,line\,mm$^{-1}$ grating was used, resulting in a wide wavelength coverage (3700--9000\,\AA) at 
a  spectral resolution of 4.5\,\AA, FWHM. 
The redshift survey strategy required reconfiguration of the spectrograph fibres after each hour of integration. For the 
stellar populations study, we repeatedly allocated some of the fibres to the same targets in different configurations, with
generally more repeat allocations for fainter galaxies. 
(Note that a given galaxy is not in general observed through the {\it same} fibre in its 
repeated observations.) The total integration time for the faint galaxies ranges from 1.7 to 20.3 hours,
with a median of 6.9 hours. The signal-to-noise ratio  at $\sim$5000\,\AA\ is 28--63\,\AA$^{-1}$, with median 43\,\AA$^{-1}$. 
The ten brighter galaxies were observed for 0.7--2.0 hours. 
The data were reduced using {\sc hsred}, an automated {\sc idl} package based on the SDSS pipeline, 
provided by Richard Cool\footnote{http://mizar.as.arizona.edu/rcool/hsred}. 
Relative flux calibration was imposed using F stars with  photometry from Sloan Digital Sky Survey 
(SDSS, Adelman-McCarthy et al. 2007), observed simultaneously with the galaxies in each configuration.

Our treatment of the one-dimensional spectra follows methods outlined by Smith et al. (2007) and previously
applied to AAOmega spectra for galaxies in the Shapley Supercluster. 
In combining the spectra from multiple configurations, we use a low-order correction to match to a common continuum, 
and reject a very small fraction of pixels which deviate from the other spectra by three times the standard error.
This produces very clean final data for the fainter galaxies, which are based on many stacked spectra. (By contrast the 
sample of brighter galaxies, with only one or two observations each, suffer more from bad pixels, cosmic ray hits etc.) 
In addition to the default spectra 
which include all the observations, we also constructed two independent spectra for each galaxy, using  
random subsets of half the exposures obtained for it. These spectra are propagated through identical reduction and measurement
pipelines. to assess systematic errors due to different seeing, fibre placement errors, flux calibration uncertainties etc, 
which vary from visit to visit. 

\begin{figure*}
\includegraphics [angle=270,width=180mm]{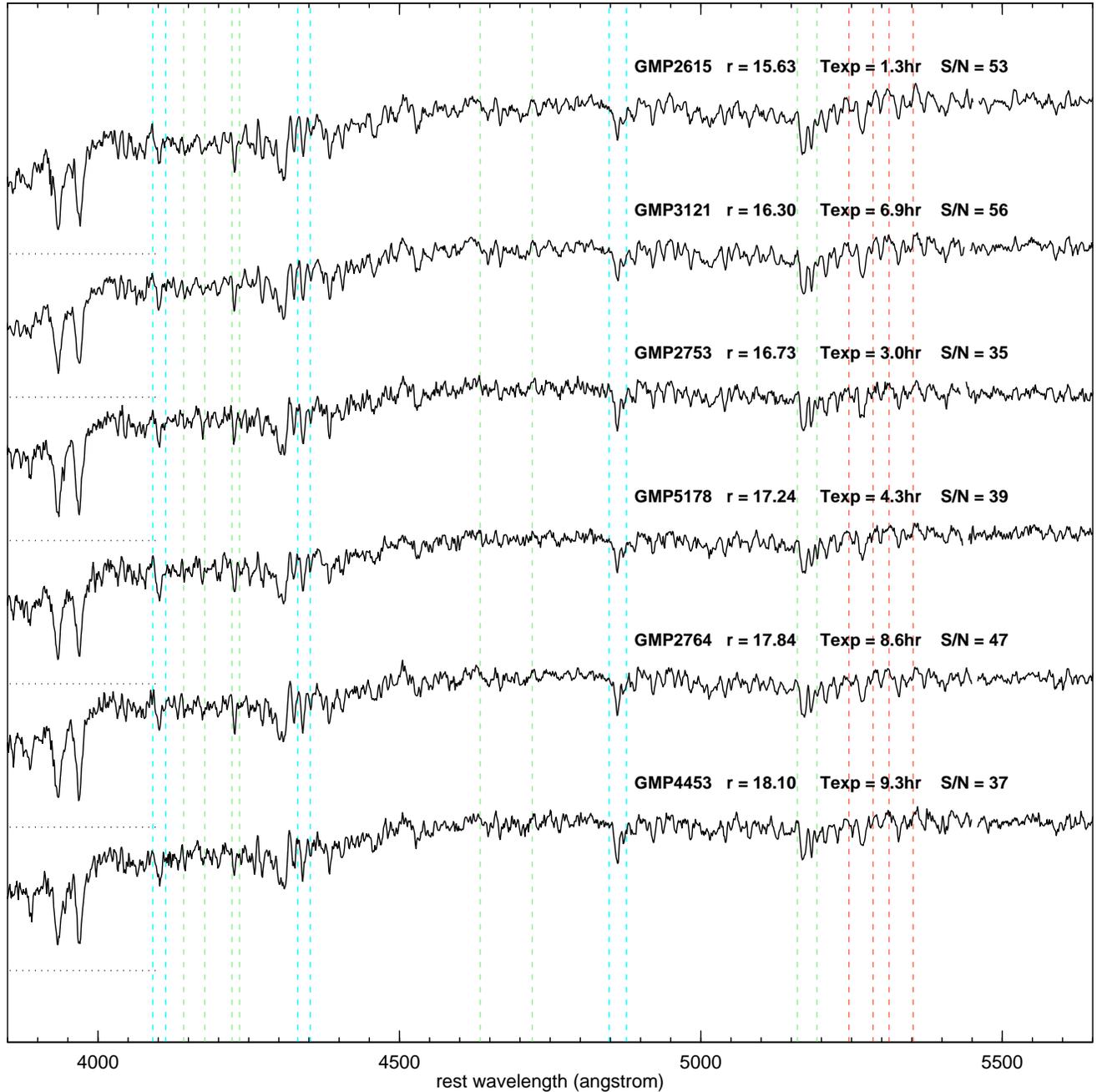}
\caption{Rest-frame spectra of representative sample galaxies, identified by 
Godwin, Metcalfe \& Peach (1983) catalogue number. The SDSS r-band Petrosian magnitude is noted, together with
the Hectospec total integration time, and the average signal-to-noise ratio per angstrom, over the range 4400--5400\,\AA\ (rest frame).
Vertical dashed lines show the central pass-bands of the indices used for measuring stellar population parameters. 
The Fe/H indicators (Fe5270, Fe5335) are shown in red, 
the age-sensitive Balmer indices (HdF, HgF, Hbeta) in blue and the indices used to derive light-element abundances 
(CN2, Ca4227, Fe4668, Mgb5177) in green. A horizontal dotted line segment at the blue end shows the zero level 
for each spectrum.}
\label{fig:repspec}
\end{figure*}

\begin{figure*}
\includegraphics [angle=270,width=180mm]{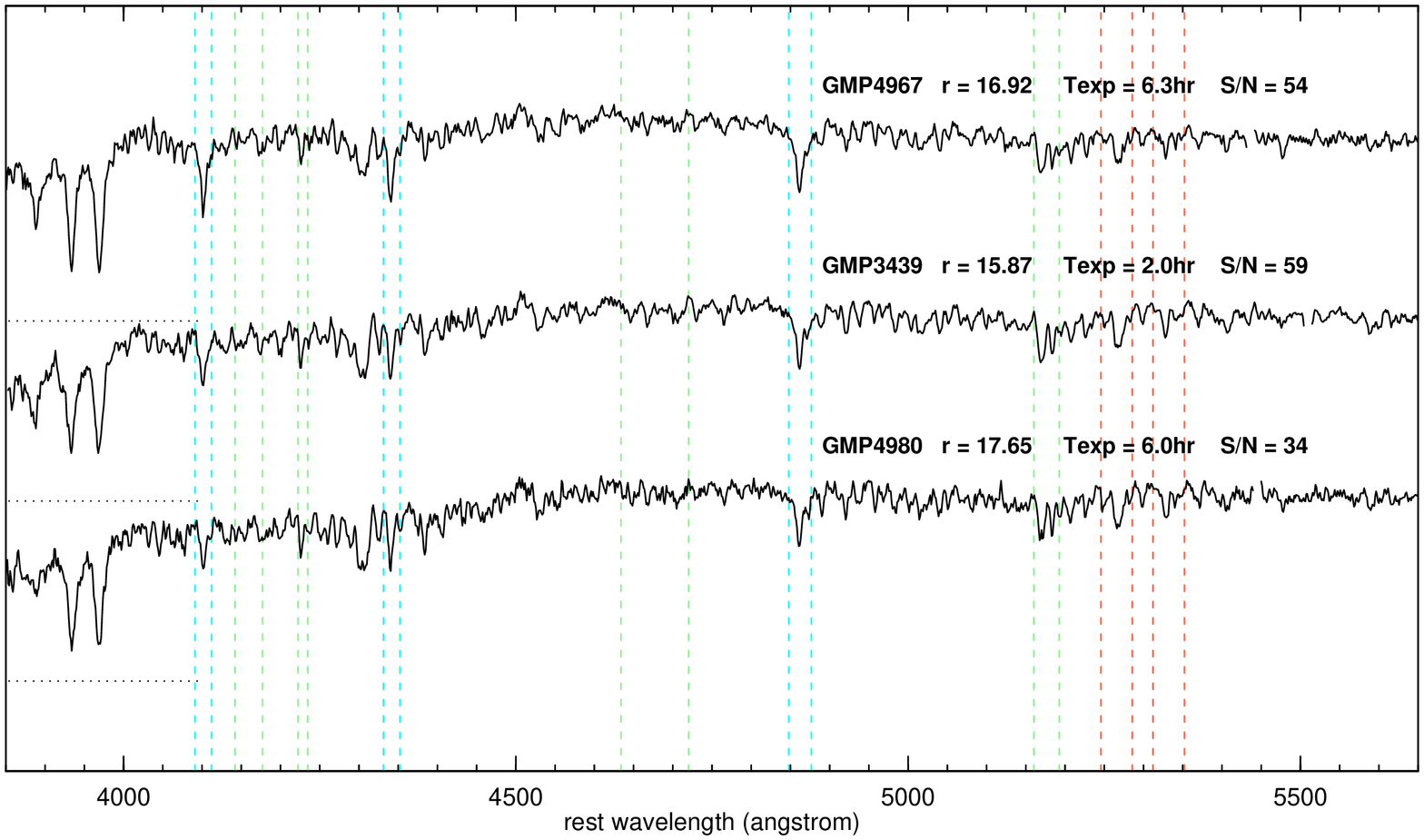}
\caption{Rest-frame spectra of Balmer-strong galaxies. Specifically, these are the three galaxies with largest
Hbeta index. Annotations as in Figure~\ref{fig:repspec}.}
\label{fig:balspec}
\end{figure*}

Figure~\ref{fig:repspec} shows a selection of representative galaxy spectra, while Figure~\ref{fig:balspec} highlights
three galaxies with stronger Balmer absorption lines. 

\subsection{Velocity dispersions, photometry and morphology}\label{sec:sigphotom}

The spectral resolution of our data, with $\sigma_{\rm inst}=115$\,\kms, is too low to measure reliable 
velocity dispersions for dwarf galaxies. 
We have instead compiled velocity dispersion data from the literature, yielding values for 47 of the sample galaxies. 
The data sources are 
J\o rgensen, Franx \& Kj\ae rgaard (1995), Hudson et al. (2001), Moore et al. (2002), Smith et al. (2004), Matkovi\'c \& Guzm\'an (2005); 
Adelman-McCarthy et al. (2007) and Cody et al. (2008).  The availability of velocity dispersion measurements is 
strongly biased to the cluster core, especially at low luminosities.
Corrections were applied to remove systematic offsets in $\sigma$ measured from different data sources. 
The offsets were small in all cases except for Cody et al., which were corrected upwards by $\sim$0.15\,dex. 
Most of the velocity dispersions for the dwarf galaxies are in the range 20--80\,\kms, with median 45\,\kms.

To compute luminosities, we adopt the SDSS $r$-band Petrosian magnitudes in the AB system, and correct by 35.08\,mag for a luminosity 
distance of 104\,Mpc 
(assuming zero peculiar velocity and cosmological parameters from Hinshaw et al. 2008), and a mean galactic extinction of $A_R=0.02$\,mag 
(Schlegel, Finkbeiner \& Davis 1998). 
For reference to solar values we adopt $M_\odot,r=4.64$ in the AB system (Blanton \& Roweis 2007).

We have performed Sersic profile fits to the SDSS $r$-band images for our sample of 89 galaxies, using the public code 
{\sc galfit} (Peng et al. 2002). The Sersic exponent $n_{\rm ser}$ has median 2.0 and interquartile range 1.5--2.6. The
median effective radius is 3.0\,arcsec, with interquartile range 2.4--3.9\,arcsec. 
We will avoid using the term ``dwarf elliptical'' or dE to refer to galaxies in our sample, because they were not selected according to 
morphological or structural criteria, but it is of interest to know how the observed galaxies relate to this class. 
In particular, the dE nomenclature often refers not simply to luminosity, but rather to membership of a structural sequence in 
some ways distinct from the sequence of giant early-type galaxies (Ferguson \& Binggeli 1994). Specifically, dEs have luminosity profiles closer 
to exponential than to the de Vaucouleurs $r^{1/4}$ law,
and have decreasing effective surface brightness towards fainter magnitudes, opposite to the trend for ``giants''. Graham \& Guzm\'an (2003)
show the latter trend is driven partly by the systematic decline in Sersic profile index from $\sim$4 in giant ellipticals to $\la1$ in faint dEs. 
In the luminosity range of our 
sample, $-19.0\la{}M_r\la{}-16.5$, the two sequences overlap, and indeed the red sequence luminosity function in low-redshift clusters
shows a dip or levelling-off in this regime, with the dwarf population starting to dominate from $M_r\ga-18$. (Popesso et al. 2006).
The question, then, is whether the sample is dominated by (bright) dEs or by normal (but faint) early-types. 
Based on our Sersic fits to the SDSS images, the Hectospec sample galaxies with $-19.0<M_r<-16.5$ 
have effective surface brightness $20.5<\langle\mu_r\rangle_{\rm e}<23.0$, and Sersic indices $1<n<3$.  
This range is an excellent match to the dwarf sequence at the corresponding luminosities 
(e.g. comparing to Figure 9, panels a, d and g, of Graham \& Guzm\'an, with appropriate colour corrections). 
By contrast, faint ``compact'' ellipticals, which lie on the extrapolation of the giant sequence, at the same luminosity would 
have $\langle\mu_r\rangle_{\rm e}\la19.5$, and $n\sim4$. In summary, most of our sample galaxies occupy the bright end of the dE galaxy sequence, 
rather than the faint end of the giant sequence.

\subsection{Comparison galaxy sample}

Our sample in Coma is limited to rather a narrow luminosity range in the dwarf galaxy regime. 
As a reference sample of more massive galaxies, we use data from Smith et al. (2007) for red-sequence galaxies 
in the Shapley supercluster. The advantages of this sample are: very high $S/N$; a broad luminosity coverage overlapping
with the bright end of the Hectospec range; similarly wide range in measured indices; H$\alpha$ data for emission selection; 
generally very similar data reduction processes\footnote{These criteria are not simultaneously met by existing studies of giant 
galaxies in Coma itself, e.g. Moore et al. (2002), \sanch\ et al. (2006a). An alternative would have been to define a giant-galaxy 
comparison sample from SDSS spectra in Coma, matched to the Hectospec area. Such a comparison set would be smaller than the
Shapley sample, and have much lower average S/N.}.
The main drawback in using a comparison sample from outside of Coma is that the Smith et al. data sample a larger physical aperture 
(2\,arcsec fibre diameter corresponding to 1.9\,kpc at Shapley, compared to 0.7\,kpc for Hectospec at Coma). 
The physical effective radius of the Shapley galaxies is typically larger as well (2--4\,kpc in Shapley, 1--2\,kpc in Coma), 
so that the fraction of galaxy light sampled is similar in each case, and in this sense the aperture effects are smaller than 
had we used giant galaxies in Coma itself.  
However, for galaxies of similar luminosity (e.g. at $M_r\approx-19$ where the samples overlap), the factor of $2-3$ in aperture size, coupled with 
typical metallicity gradients for giant early-type galaxies, would generate an offset in derived metallicity of 
$\sim$0.1\,dex (e.g. Rawle et al. 2008), with lower measured metallicity in the Shapley objects.
We will not apply corrections for this aperture effect in the tabulated data, or in figures showing only index data.  
A simple 0.1\,dex shift in Fe/H {\it is} applied to the Shapley data in figures showing the recovered Fe/H, and explicitly noted in the captions. 
The aperture correction for $\sigma$ is small ($\la0.02$\,dex) and neglected here. No aperture 
correction is necessary for age or abundance ratios, since there is little evidence for strong gradients in these parameters, on average.

A further complication is that at the distance of Shapley, either the Fe5270 or the Fe5335 index is contaminated for many galaxies
by sky-subtraction residuals from the 5577\,\AA\ line. These indices will be used as our primary tracers of the Fe abundance, so in 
Section~\ref{sec:poppars} we further restrict the comparison sample by requiring a redshift such that Fe5270 is uncontaminated. 
This selection, together with a cut on  $S/N>25$\,\AA$^{-1}$ (to match the Hectospec data quality), 
selects 75 galaxies with $cz<14120$\,\kms\ in Shapley, with median $S/N=74$\,\AA$^{-1}$.

The Shapley photometry is taken from the NOAO Fundamental Plane Survey (Smith et al. 2004). We correct the total $R$-band Vega-based 
magnitudes to $r$-band AB magnitudes by adding 0.11\,mag (Blanton \& Roweis 2007), 
and correct for an average galactic extinction of $A_R=0.14$\,mag (Schlegel et al. 1998). 
We adopt a luminosity distance 214\,Mpc, 
corresponding to the median redshift of the full Shapley sample (not the redshift-limited subset with measured Fe5270). 

In this paper, we will resist fitting relationships simultaneously to the Hectospec and AAOmega samples.
Instead, we will generally make fits to the Hectospec data alone, 
and comment on how these results compare to extrapolation of the trends obtained for more luminous galaxies.

\begin{figure*}
\includegraphics [angle=270,width=180mm]{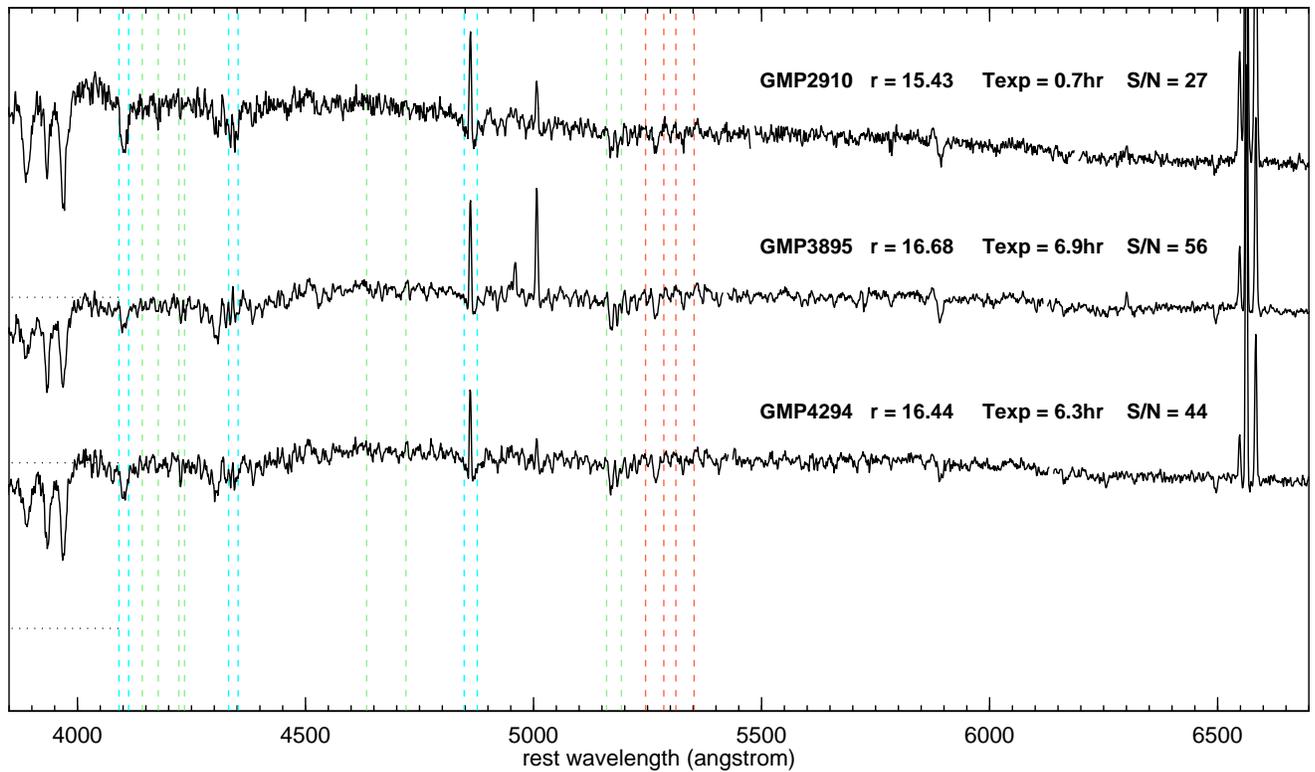}
\caption{Rest-frame spectra of the three galaxies with emission lines. The annotations are as in Figures~\ref{fig:repspec} and \ref{fig:balspec}, but 
the wavelength axis has been expanded to include the \ha\ region.}\label{fig:emlspec}
\end{figure*}

\subsection{Emission lines and absorption indices}

Nebular emission can cause contamination of the Hbeta and higher-order Balmer absorption features, leading
to overestimates of the characteristic stellar age. 
As discussed in detail by Smith et al. (2007), the most reliable method for identifying potentially-affected
galaxies is by their H\,$\alpha$ emission, which is 4--5 times stronger than the emission at H\,$\beta$, and
generally dominates over the stellar H\,$\alpha$ absorption line. 
Emission lines for the Coma sample were measured as in Smith et al. (2007) for the Shapley sample, after dividing out the 
best-fitting stellar continuum. We estimate that \ha\ emission with equivalent width greater than 0.5\,\AA\ can be 
readily identified by this method, corresponding to a maximal H\,$\beta$ contamination of $\sim$0.1\,\AA. 
In this worst case, the effect on the Hbeta-derived ages is $\la15$\,\%. 
In the Coma sample, only three galaxies exhibit significant \ha\ emission: GMP2910, GMP3895 and GMP4294.  
In all of these cases, the emission is strong enough, with EW(\ha)\,$\ga$20\,\AA, that the effect at H\,$\beta$
is obvious. 
The three galaxies with emission are excluded from the subsequent stellar population analysis.

Absorption indices were measured, using the {\sc indexf} 
programme\footnote{http://www.ucm.es/info/Astrof/software/indexf/indexf.html} (see Cenarro et al. 2001), from the
fully-combined spectra, with errors based on the associated error spectra. The indices were corrected to zero velocity dispersion 
and to the Lick resolution as in Smith et al. (2007). Briefly, this method uses SSP models to establish a
linear relation between indices measured at the observed resolution (i.e. instrumental and velocity broadening) 
and at the Lick resolution (without velocity broadening). 
The correction can then be applied without convolving the spectrum, preserving the original
noise properties. The velocity broadening corrections (handled simultaneously with the resolution correction) are 
minimal, since the characteristic velocity dispersions are $\la{}50$\,\kms, compared to the Lick resolution of $\sim$230\,\kms.
Appendix~\ref{sec:appixcomp} presents comparisons between indices measured from the half-exposure spectra for each galaxy.
From the repeatability of these measurements, we confirm the formal error estimates to within $\sim$10\% for most indices. 
An external comparison with the high-S/N dataset of \sanch\ et al. (2006a) similarly shows agreement within the expected
uncertainties in most cases. 

\begin{figure*}
\includegraphics [angle=0,width=175mm]{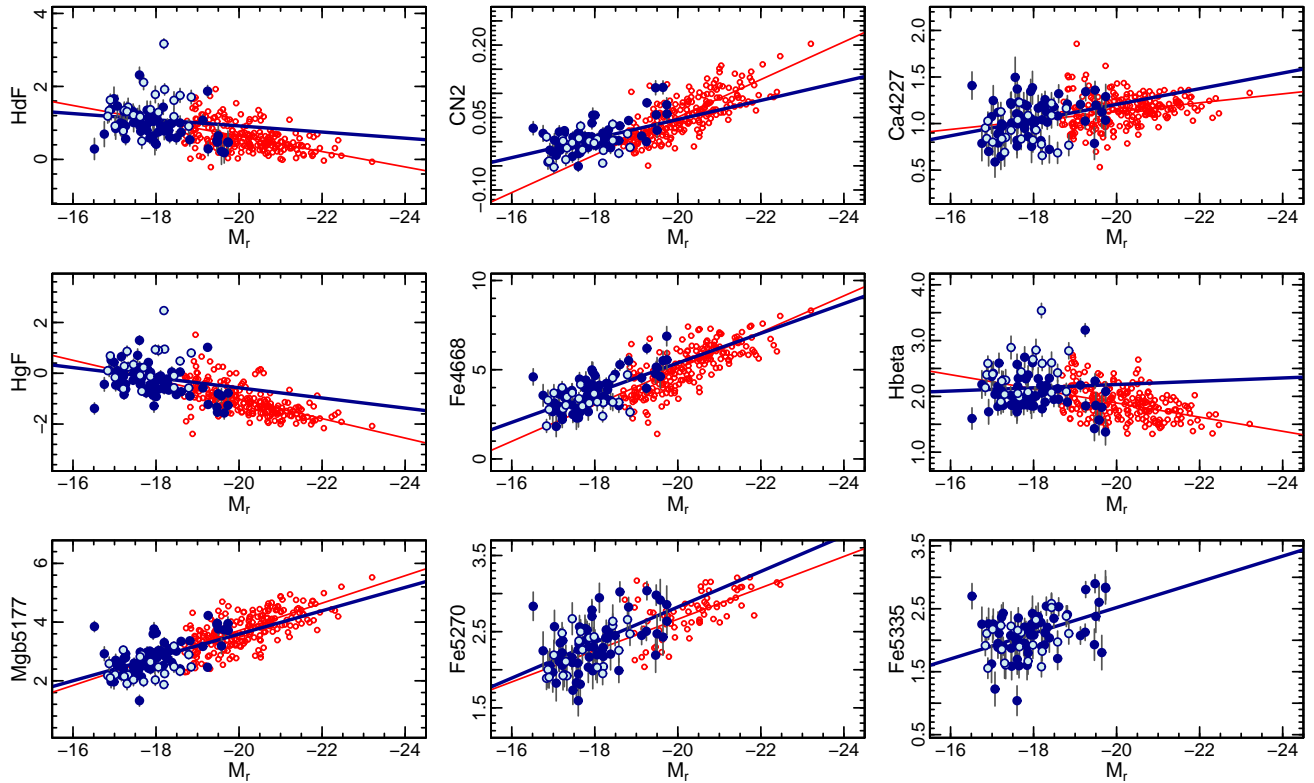}
\caption{Absorption line strengths versus luminosity for the most important indices, i.e. those used in our 
stellar population model inversions (CN2, Ca4227, Fe4668, Hbeta, Mgb5177, Fe5370, Fe5335), 
plus the higher-order Balmer indices (HdF, HgF). The panels are presented in order of wavelength. 
The dark blue symbols indicate Coma dwarf galaxies within 0.7\,Mpc from the cluster centre, while
the light blue points are Coma dwarfs beyond this radius. The blue line is a fit to all of the Coma sample. 
For comparison, the red points (without errorbars, for clarity) show the comparison sample of more luminous galaxies in 
Shapley; the red line is a fit to this sample. There are no Fe5335 measurements in the Shapley sample, due to sky-line contamination. 
No aperture corrections have been applied. The three galaxies with emission lines have been removed from all panels. 
}\label{fig:ixlum}
\end{figure*}

\begin{figure*}
\includegraphics [angle=0,width=175mm]{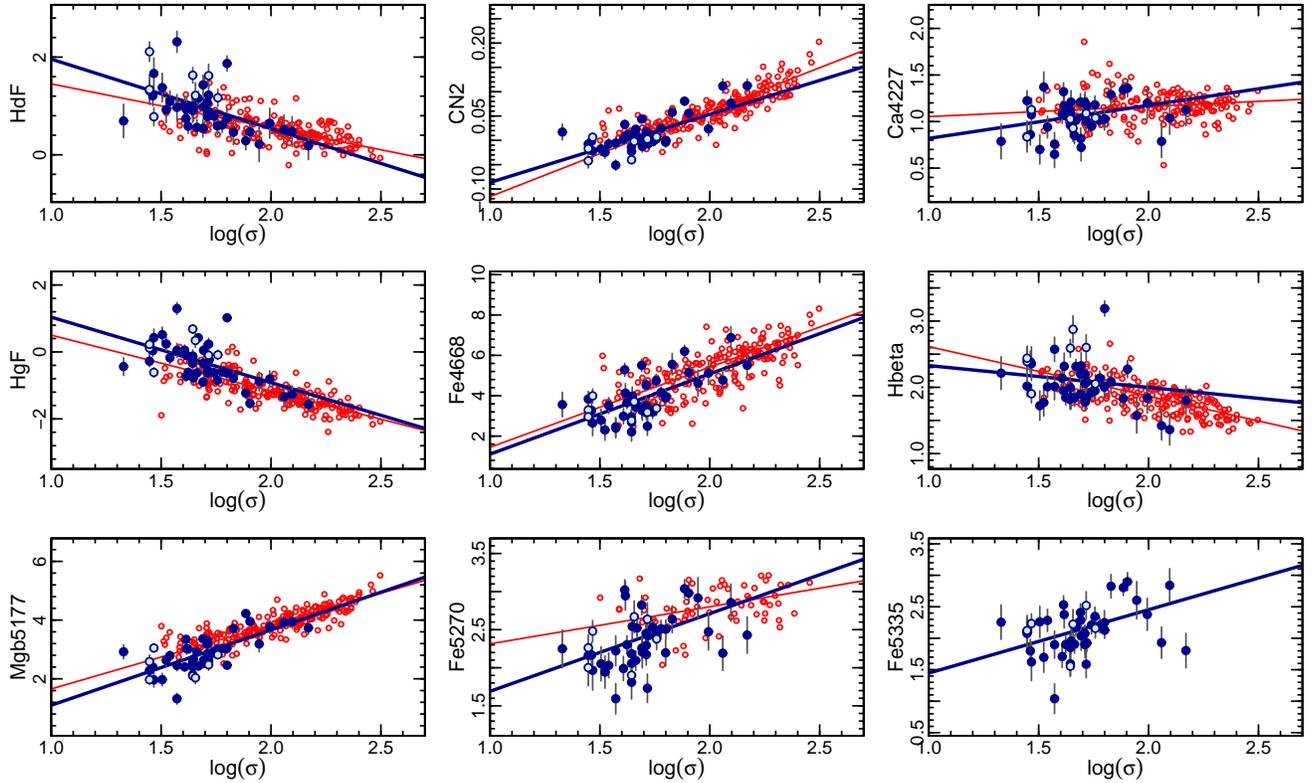}
\caption{Absorption line indices versus velocity dispersion for the subset of galaxies with 
velocity dispersion data from the literature. Symbols are as in Figure~\ref{fig:ixlum}.
}\label{fig:ixsig}
\end{figure*}

Figures~\ref{fig:ixlum}~and~\ref{fig:ixsig} show the most important indices as a function of luminosity and velocity dispersion  for the Coma sample, compared to the more luminous Shapley galaxies. Equivalent figures
for other measured indices are presented in Appendix~\ref{sec:moreixsiglum}. 
In general, the Coma dwarfs fall close to the extrapolation from more massive red galaxies, although
a few differences can be noted: The CN1, CN2 indices are somewhat higher than the extrapolated relation; this is qualitatively
consistent with the apparent curvature in the CN--$\sigma$ relation already identified from the Shapley data alone (Smith et al. 2007). 
The Fe4668 and Fe5015 indices appear somewhat lower than the giant-galaxy trend when plotted against $M_R$, but the difference
is not seen in the index--$\sigma$ relations. Finally the H$\gamma$ and H$\delta$ indices are above the trend lines in the
index--$\sigma$ relations. 

Already from Figure~\ref{fig:ixlum}, there is evidence for environmental dependence in galaxy properties, 
as reported by Smith et al. (2008a). Specifically, galaxies at larger distance from the cluster centre (the figure highlights
those beyond 0.7\,Mpc) have higher Balmer indices (HdF, HgF, Hbeta), and lower Mgb5177, than galaxies of the same luminosity 
nearer to the cluster core. The environmental dependence of the galaxy properties are discussed further in Section~\ref{sec:enviro}.

\section{Stellar population parameters}\label{sec:poppars}

In this section, we use population synthesis models to transform the index measurements into estimates of stellar population age and element abundances. Here, age will refer simply to SSP-equivalent age, $t_{\rm SSP}$, which is the age of a coeval population of stars that best 
matches the observed index data. For more extended star-formation histories, $t_{\rm SSP}$ will be heavily weighted towards the most 
recently-formed stars (e.g. Serra \& Trager 2006).
We explore alternative parametrizations for the formation histories in Section~\ref{sec:allansonsfh}. 

\subsection{Choice of models}

We will analyse the absorption line measurements primarily with reference to 
the stellar population models of Schiavon (2007). We provide a comparison with results
from the TMBK model set in Section~\ref{sec:allansontmbk}. 
The Schiavon models are based on the flux-calibrated spectral library of Jones (1999), and
incorporate the stellar-atmosphere effects of abundance ratio variations in Mg, Ca, C and N. 
Note however that, in common with TMBK, they do not account self-consistently
for the effect of variable abundance ratios in the stellar evolutionary tracks 
(see e.g. Salasnich et al. 2000; Coelho et al. 2007; Dotter et al. 2007). 
The calculations by Coelho et al. suggest these isochrone effects are small for the line-strength indices. 
A useful feature of the Schiavon models is the public availability of the {\sc idl} code, \eza, for estimating stellar population
parameters from a set of measured indices (Graves \& Schiavon 2008). 

\subsection{Index--Index diagnostic diagrams}

This section provides a qualitative commentary on the distribution of the Coma dwarfs
in a number of index--index diagrams, with comparison to the Shapley sample. The discussion follows the 
steps taken by \eza\ to determine the age, metallicity\footnote{We will use the term ``metallicity'', rather than 
the more precise but cumbersome ``iron abundance'', to refer to Fe/H.},  and abundance ratios for each galaxy. 

\begin{figure*}
\includegraphics [angle=0,width=85mm]{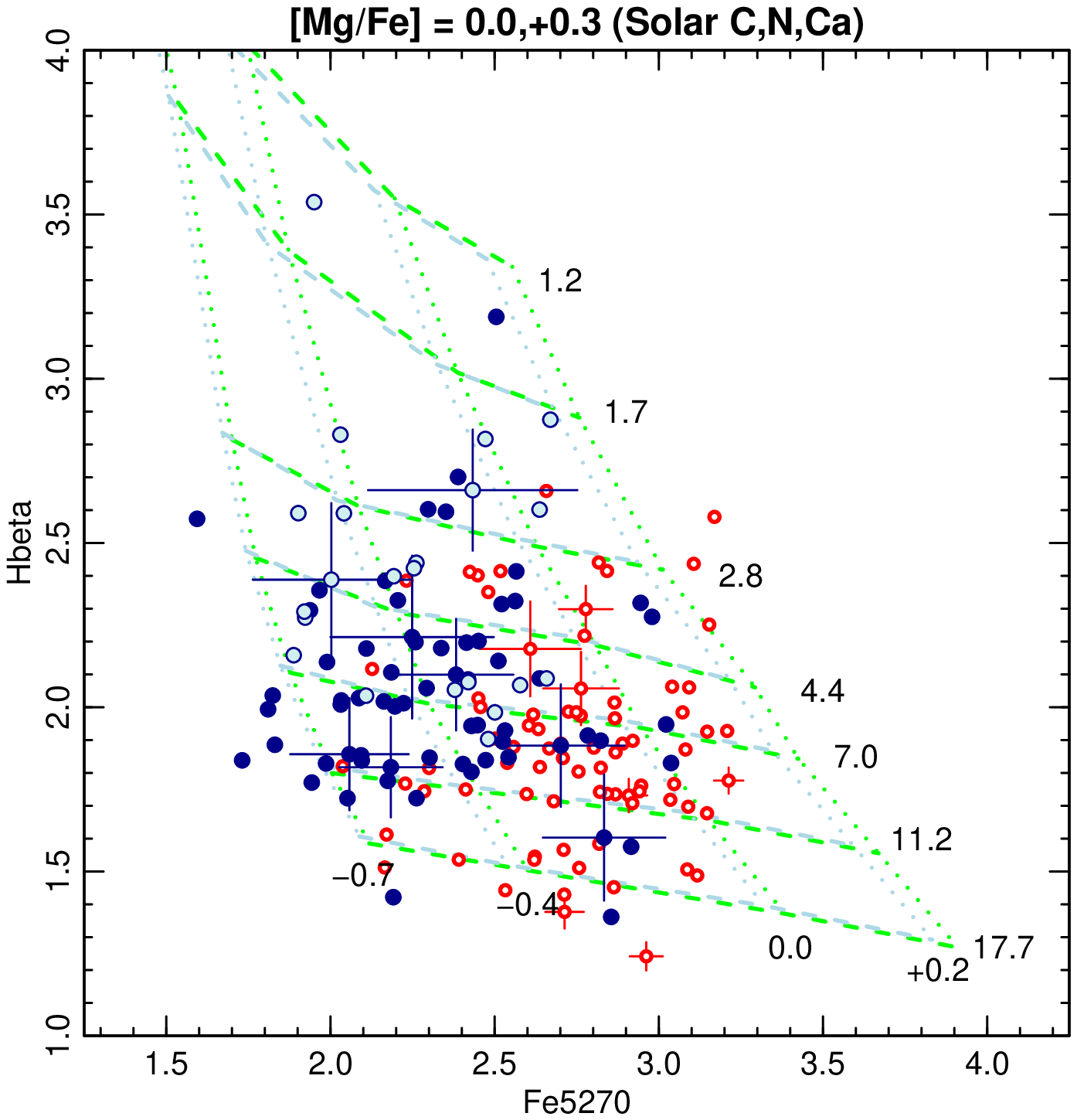}
\includegraphics [angle=0,width=85mm]{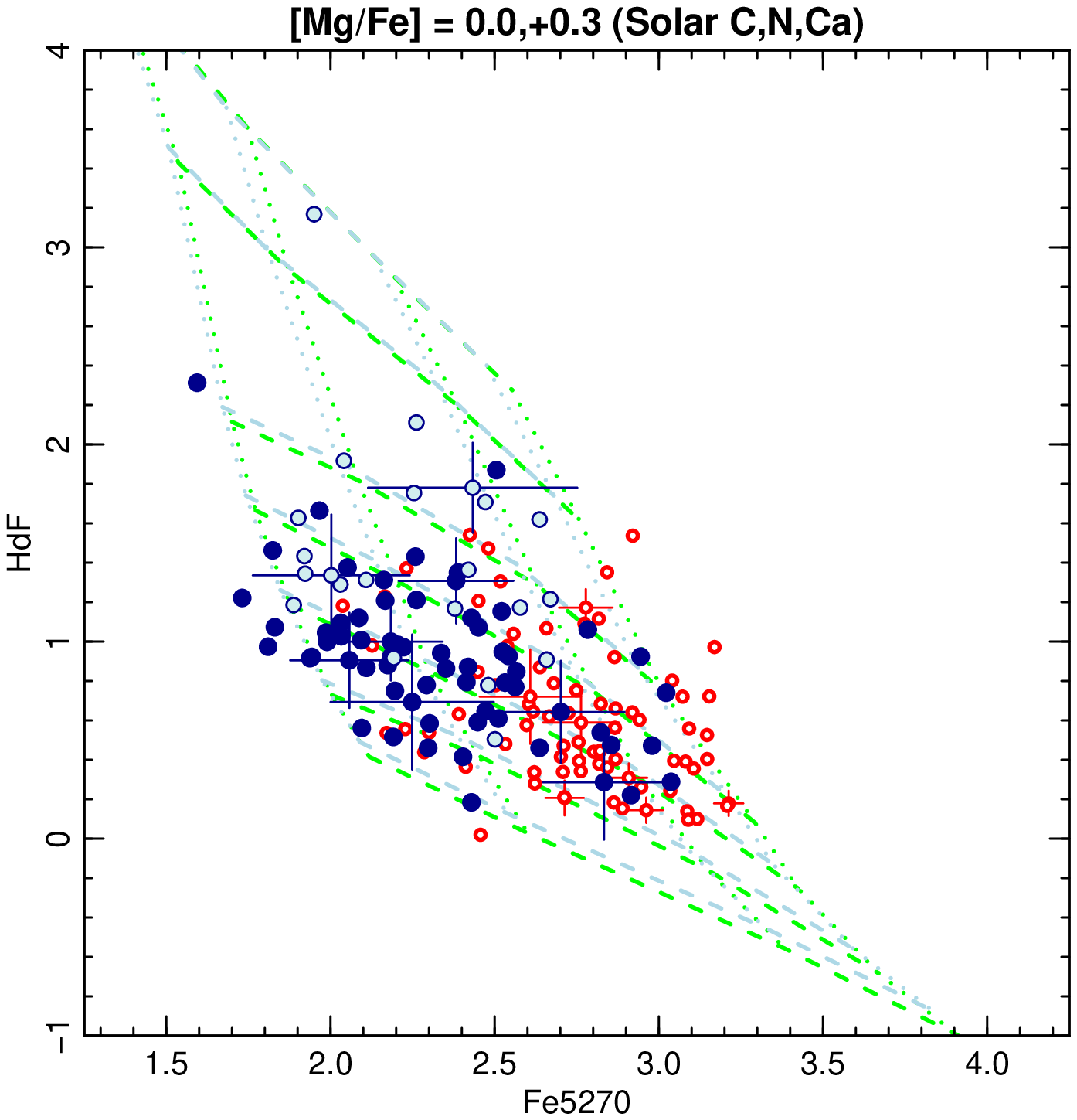}
\caption{Diagnostic diagrams for age and metallicity. The Coma and Shapley galaxies are shown using the same symbol types as in 
Figures~\ref{fig:ixlum} and \ref{fig:ixsig}. For clarity, the error bars are shown on only a subset of points (the same galaxies in each panel). 
The model grids are from Schiavon (2007), with [Mg/Fe]=0.0 (light blue) and [Mg/Fe]=+0.3 (green). The dotted lines are of constant metallicity 
[Fe/H] = --0.7, --0.4, 0.0, +0.2, while the dashed lines indicate constant
age 1.2, 1.7, 2.8, 4.4, 7.0, 11.2 and 17.7 Gyr. 
Note that because the iso-metallicity lines are defined at constant Fe/H (not constant {\it total} metallicity), the 
locations of these grids are fairly insensitive to abundance ratio variations.}\label{fig:hbfe52}
\end{figure*}

From the Hbeta versus Fe5270 diagram (Figure~\ref{fig:hbfe52}, left)
we can approximately read off the SSP-equivalent ages and metallicities, as in the first step of the
\eza\ process. Note that there is only a weak sensitivity to the Mg/Fe ratio,  because the models are defined
with abundances varying at fixed Fe/H, rather than at fixed total metallicity Z/H. We can infer a wide range in both metallicity ($-0.7\la$[Fe/H]$\la{}0.0$)
and age ($2\la{}t_{\rm SSP}\la{}11$\,Gyr) for the Coma sample. Similar ranges are indicated by the HdF versus Fe5270 diagram 
(Figure~\ref{fig:hbfe52}, right), but here the grid is more steeply tilted relative to the axes and error-bars, which hinders a visual 
assessment. The Coma dwarfs are clearly offset from the  brighter sample of Shapley galaxies, in the direction
of lower metallicity. There are also more Coma galaxies with strong Balmer lines, corresponding to 
ages $<3$\,Gyr, than seen among the Shapley sample. 

\begin{figure*}
\includegraphics [angle=0,width=85mm]{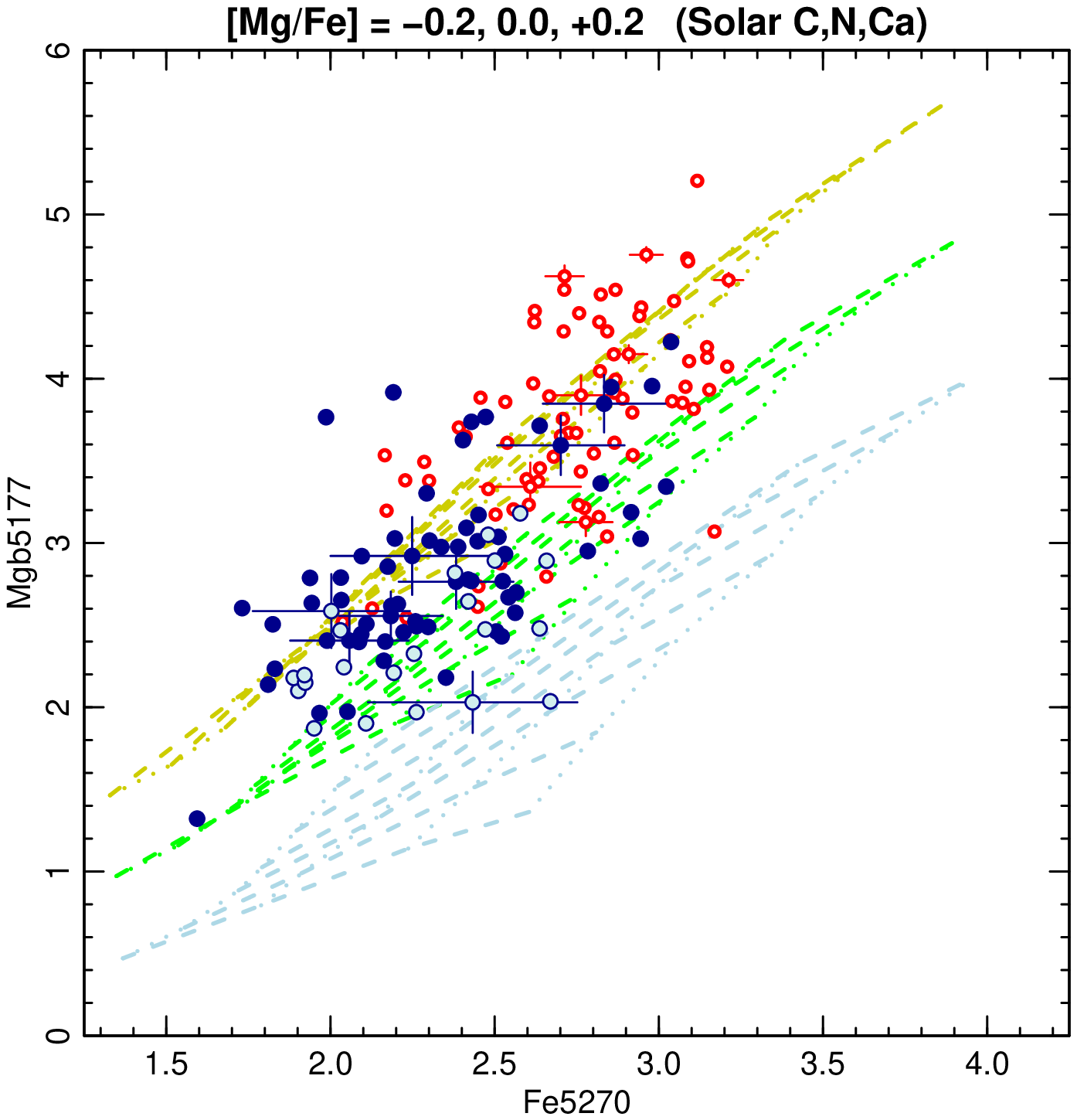}
\includegraphics [angle=0,width=85mm]{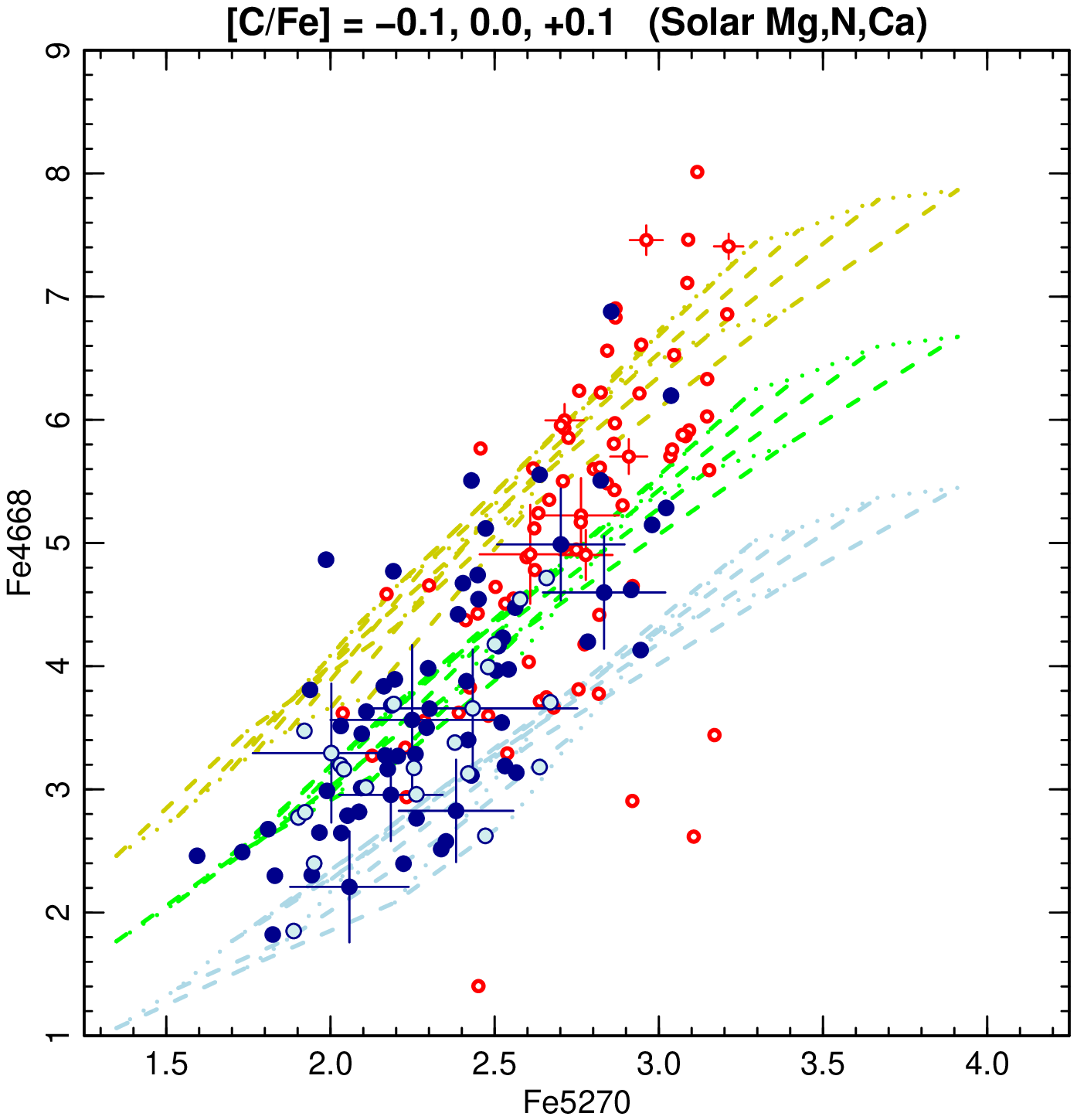}\\
\includegraphics [angle=0,width=85mm]{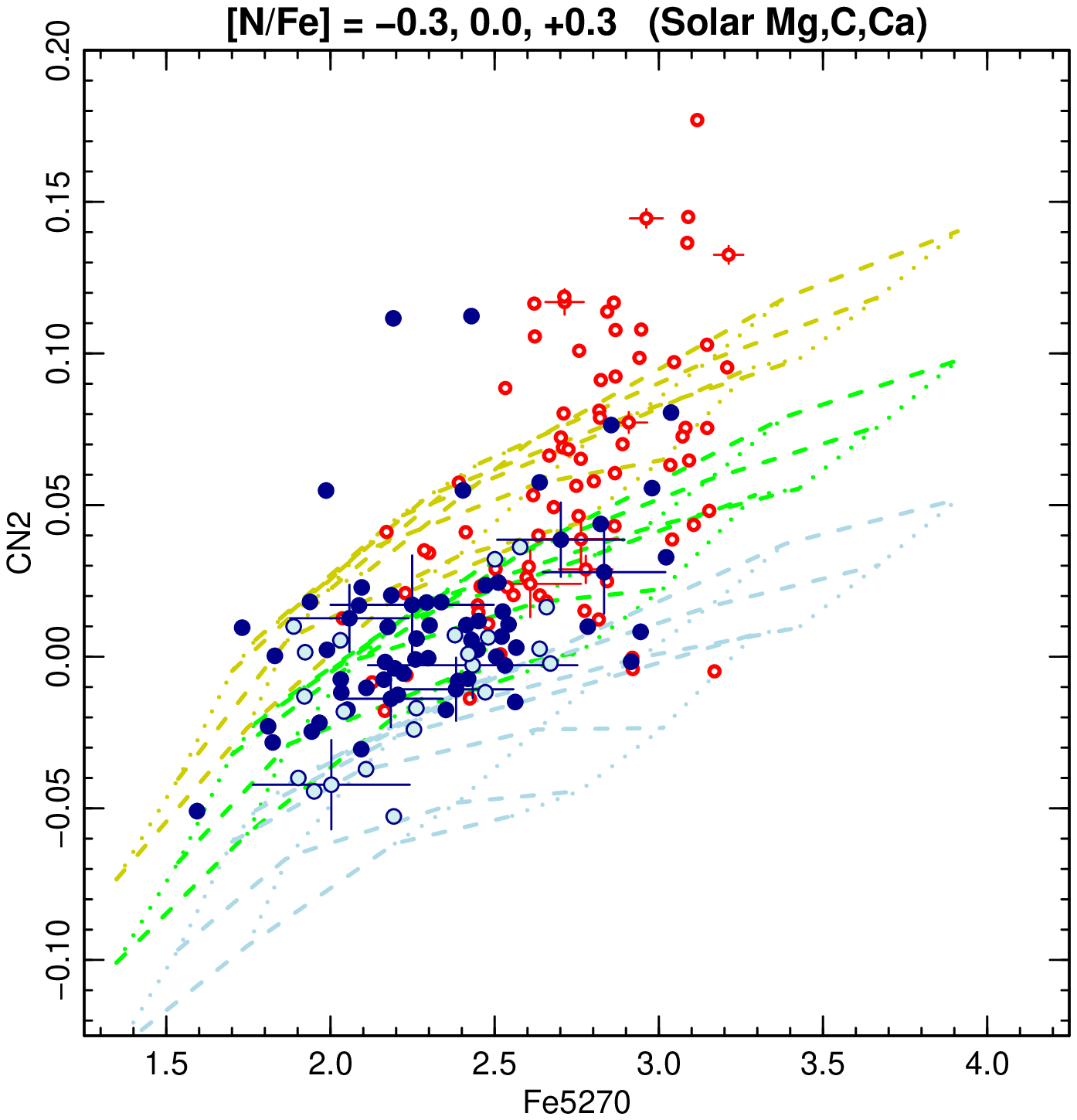}
\includegraphics [angle=0,width=85mm]{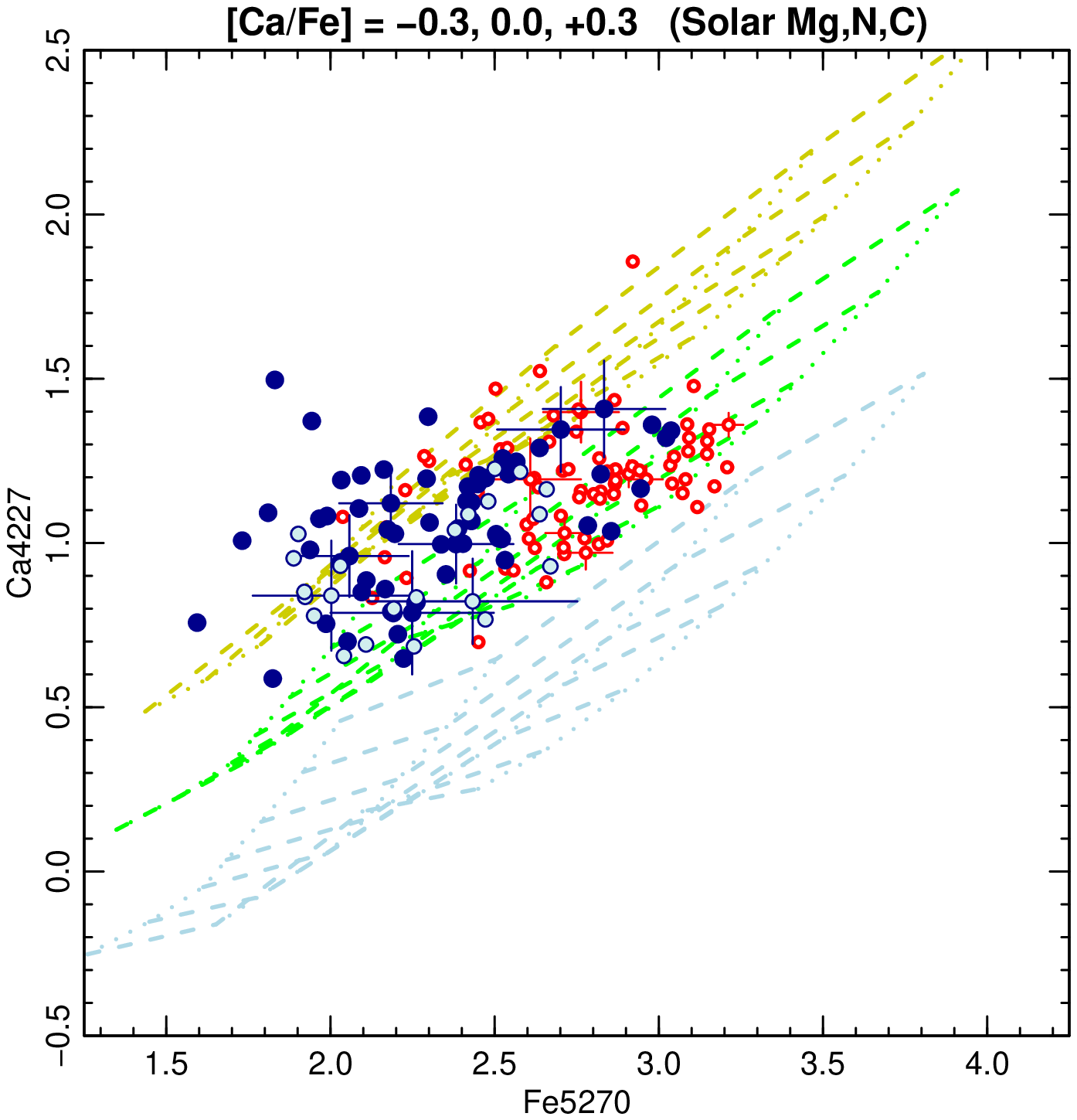}
\caption{Diagnostic diagrams for the abundance ratios Mg/Fe, C/Fe, N/Fe and Ca/Fe. 
The Schiavon (2007) model grids are plotted for enhanced (yellow), solar (green) and 
depleted (blue) abundance ratio X/Fe, where X is the element being probed in each panel. The age and Fe/H lines
for each grid are as in Figure~\ref{fig:hbfe52}.
Coma and Shapley galaxies are shown using the same symbol types as in Figure~\ref{fig:hbfe52}, and
with error bars plotted for the same subset of galaxies. 
}\label{fig:xfediag}
\end{figure*}

The abundance ratios of light elements can be visually estimated from other diagnostic diagrams, as shown in Figure~\ref{fig:xfediag}. 
From the Mgb5177 versus Fe5270  diagram, we see that the Mg abundances are well constrained, with typically
slightly super-solar ratios, [Mg/Fe]\,$\approx$\,+0.1 for the Coma dwarfs. The Mg/Fe ratios are lower than the Shapley sample, on average. 
The Fe4668 versus Fe5270 diagram shows that the C abundances are also well constrained, with dwarfs generally 
subsolar, [C/Fe]\,$\approx$\,--0.05. The average C/Fe ratio is lower than in the Shapley sample. 
The CN2 versus Fe5270 diagram shows that the N abundances can be measured only if the age and metallicity are known, 
since the grids do not collapse into linear tracks as in the previous two figures. Also, the derived N/Fe also depends 
somewhat on the assumed C/Fe. The typical N/Fe ratios are around solar for the Coma dwarfs, and lower than in the Shapley sample on
average. 
The Ca4227 versus Fe5270 diagram can be used to estimate Ca/Fe, but this figure is harder to interpret, since the Ca4227 index 
depends on C and N as well as Ca itself. Assuming solar CN, the typical 
value for our sample is [Ca/Fe]\,$\approx$\,+0.2, but this is probably overestimated by $\sim$0.1 due to the underabundance of C. At face value, 
the average Ca/Fe ratios appear to be higher in the Coma dwarfs than in the Shapley galaxies, but again this is hard to decouple 
from the C and N abundances without formally inverting the models.

\subsection{Inversion method}

The \eza\ code (Graves \& Schiavon 2008) performs a ``sequential'' grid inversion, finding abundance ratios which 
yield the most consistent age and Fe/H estimates across a range of index--index diagrams. 
We adopt the default options for \eza, including solar-scaled isochrones from Girardi et al. (2000), Salpeter stellar initial 
mass function, solar O/Fe ratio, unconstrained $\alpha$ elements (Si, Ti, Na) tracking Mg, and
the Cr abundance tracking Fe. 

The initial estimate of age and Fe/H is made using Hbeta and the iron indices Fe5270 and Fe5335. 
The Mg/Fe ratio is adjusted to obtain the same metallicity and age 
from the Hbeta and Mgb5177 indices. C/Fe is then obtained using the Fe4668 index. 
With C/Fe in hand, the N/Fe abundance is adjusted for consistency 
with the measured CN2. Finally, Ca/Fe is obtained using Ca4227. The procedure is iterated, 
deriving a new age and metallicity estimate from the updated abundance pattern. 
The authors of \eza\ strongly caution against application of \eza\ in the [Fe/H]$<-1$ regime. We will find that 
our Coma and Shapley sample galaxies have $-1.0<$[Fe/H]$<+0.1$, and hence lie in the supposedly ``safe'' range.
Once the abundance pattern has been established, the code generates new age
estimates using the HgF-vs-Fe and HdF-vs-Fe diagrams, in addition to the default estimate from Hbeta-vs-Fe.
We will not use the estimates from the higher-order Balmer indices in this paper; we comment on them briefly in 
Section~\ref{sec:ageresults}. 

By default, the code propagates index errors through the inversion procedure, but does not provide
an estimate of the error covariance matrix for each galaxy. Moreover, propagation of the errors is computationally
slow. We have performed a set of Monte Carlo (MC) simulations on a small number of 
representative spectra, to assess the typical covariance structure and to define an error model for the parameters. 
We assign error estimates to individual parameter measurements according to 
a fit to the MC simulations. 
The estimated errors account acceptably for the scatter in \eza\ parameters estimated from the half-exposure spectra, 
and can be considered reliable to $\sim$10\%.
Full details of the error model are presented in Appendix~\ref{sec:ezerrmod}. 

For the comparison sample of Shapley galaxies, we only use Fe5270 as the Fe/H indicator, since Fe5335 is contaminated
by the sky emission line. To test for any consequences on the derived parameters, the fits for the Coma data were repeated 
with Fe5335 excluded. Comparison with the solutions obtained using both lines confirms that excluding Fe5335 leads to negligible offsets: 
on average the recovered Fe/H is smaller by 0.02\,dex, while Mg/Fe and Ca/Fe are larger by 0.02\,dex, with a 
galaxy-to-galaxy scatter is $\sim$0.1\,dex in all cases. The derived ages are affected by $\la$5\,per cent. 

Meaningful \eza\ inversions could not be obtained for eight of the 89 Coma sample galaxies, including the three
objects with emission lines. The failed cases are described individually in Appendix~\ref{sec:ezafails}. 
Most of these cases affect the galaxies selected as bright comparison objects, rather than 
the dwarf galaxy sample itself. In part this is because the spectra of faint objects result from many separate
observations, and are better cleaned of cosmic rays and other defects in the process of combining the data. 

For the 81 successful fits, the ages are determined with a median error of 0.12\,dex ($\sim$30\%); Fe/H is determined to 0.11\,dex, 
the Mg/Fe and C/Fe ratios to 0.08\,dex, and the N/Fe and Ca/Fe ratios to 0.10\,dex. 

\subsection{Data presentation}

The absorption index measurements for the Coma sample, and the \eza\ inversion results for both the Coma and
the Shapley sample, are tabulated in Appendix~\ref{sec:tables}. The Shapley index data were presented by
Smith et al. (2007). Our reduced spectra for the Coma dwarf sample will be made available via the www upon publication.  

\section{Results}\label{sec:results}

\subsection{SSP-equivalent ages}\label{sec:ageresults}

Our derived SSP-equivalent ages, $t_{\rm SSP}$, range from 1.4 to 13.1\,Gyr, with median 6.5\,Gyr and 
inter-quartile range 3.9--8.3\,Gyr, corresponding to a redshift interval $z=0.35-1.17$. 
These results can be loosely interpreted as the epoch during which star-formation ceased in Coma dwarfs. A detailed
comparison to look-back studies depends on the assumed star-formation history (Section~\ref{sec:allansonsfh} and
Allanson et al. in preparation). 

In this section,we only consider the SSP-equivalent ages estimated self-consistently by \eza\ from the Hbeta index. 
In common with previous work (Schiavon 2007; Graves et al. 2007), we find that the ages obtained by \eza\ from HgF and HdF are 
systematically younger than those found using Hbeta. This can be traced to a clear offset in HgF and HdF to higher values than predicted
by the models at a given value of Hbeta. 
Schiavon and Graves et al. interpret this offset as evidence for composite stellar populations, i.e. a fraction of young 
stars added to an older dominant population, with the younger population having greater impact at blue wavelengths, and hence in the higher-order
Balmer lines.
In this case, we would expect a larger effect in younger galaxies than than in older ones.
For our sample, we find instead a greater discrepancy for galaxies with Hbeta ages above 
6\,Gyr, where the high-order lines yield ages $\sim$30\% younger, while for younger galaxies, consistent ages are recovered. 
The largest differences are apparent at [Fe/H]$<$--0.5, but there is no strong correlation of the offset with Fe/H or any other 
stellar-population parameter. 
We show in Section~\ref{sec:allansonsfh} that parametrized composite SFH predictions, based on TMBK models, do not 
in general provide better fits to the index data (including multiple Balmer indices). 
It appears that a convincing explanation of the younger HgF and HdF ages
in \eza\ is not yet at hand. This discrepancy should clearly 
serve as a caveat that some important spectral characteristics of galaxies
are still not well described by the models, at least in their simplest form. 

\begin{figure*}
\includegraphics [angle=270,width=178mm]{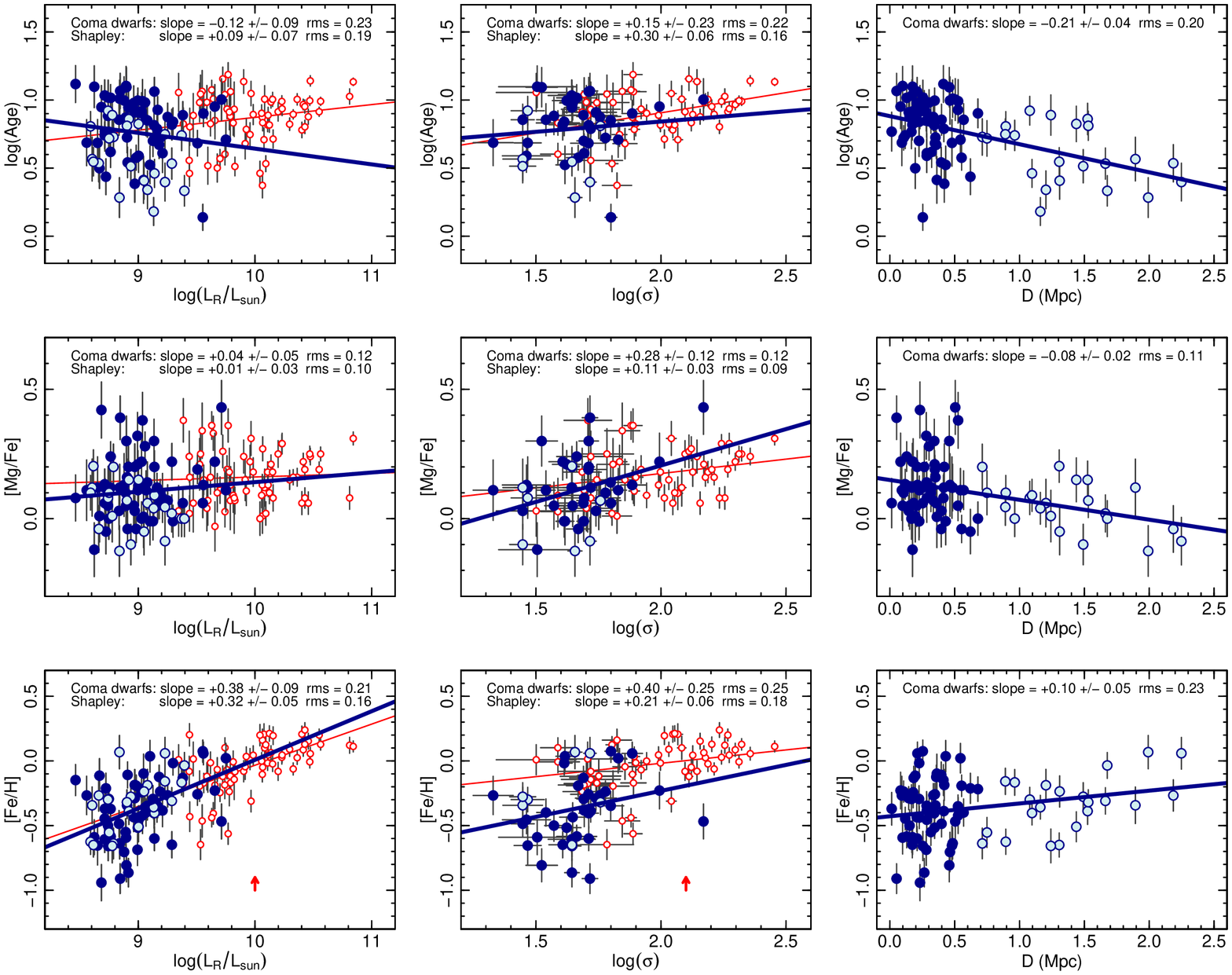}
\vskip -3mm
\caption{Correlations of the SSP-equivalent age, metallicity and Mg/Fe ratio 
with luminosity, velocity dispersion (from the literature compilation) and cluster-centric radius. 
Symbol types are as in previous figures. 
The red and blue lines show separate fits to the Shapley and Coma samples (all radii), respectively. 
The slope and scatter of these fits is reported in each panel. 
The Shapley objects have been corrected upwards by 0.1\,dex in Fe/H to account for 
aperture effects, as indicated by the red arrows.
} 
\label{fig:parvslumsig}
\end{figure*}

Figure~\ref{fig:parvslumsig} (upper row) shows the derived SSP-equivalent ages  as a function of luminosity,
of velocity dispersion (where available), and of distance from the cluster centre. 
We indicate two ranges of cluster radius by different symbol types. The figure shows the Shapley data for comparison, 
but the relations we report below are for the Coma sample only. As already noted, the subset of galaxies
having measured velocity dispersions is not fully representative of the sample at large: within
0.7\,Mpc radius from the cluster centre,  $\sim$60\% of galaxies have measured $\sigma$, 
compared to $\sim$30\,\% of galaxies  beyond this radius. 
There is no significant trend in age with $\sigma$ or with luminosity $L_r$, over the limited range
spanned in these parameters by the Coma sample alone. 
The recovered fits are $t_{\rm SSP} \propto \sigma^{+0.15\pm0.23}$ (rms=0.22\,dex) and  
$t_{\rm SSP} \propto L_r^{-0.12\pm0.09}$ (rms=0.23\,dex). 
The marginal anti-correlation of age with luminosity in part probably reflects the larger stellar light-to-mass ratio
in younger populations, i.e. there is not necessarily an equivalent trend with stellar mass. 
There is a clear difference in the age distribution for the outer galaxies, with a relative absence of the
oldest ($\sim$10\,Gyr) dwarfs beyond 0.7\,Mpc from the cluster centre, and a larger fraction of objects
with $t_{\rm SSP}\approx3$\,Gyr.

Within the limited mass range probed by the Coma sample alone, we do not observe the signature of downsizing, 
i.e. a correlation of stellar age with galaxy mass. Comparing to the Shapley sample provides a wider baseline. 
Although the Coma dwarfs are younger on average than the galaxies in Shapley (median ages 6.5\,Gyr and 7.9\,Gyr respectively), 
the difference disappears when only the inner sample is considered (median age 7.4\,Gyr). The latter comparison
is the more appropriate, since the Shapley sample is dominated by galaxies within $\sim$0.8\,Mpc from rich cluster cores. 
This could be interpreted as supporting the claims (\sanch\ et al. 2006b; Trager et al. 2008) that the core of Coma does not exhibit the 
downsizing effect. The strong dependence of the dwarf galaxy age distribution
on location in the cluster suggests that downsizing and environmental effects 
may be difficult to disentangle.

\subsection{The Mg/Fe ratio}

The $\alpha$-element abundance ratio $\alpha$/Fe in a stellar population is sensitive to the duration of its 
star-formation period, since the production of $\alpha$ elements is dominated by Type II supernovae, 
while most Fe is released by delayed Type Ia supernovae. 
In principle, with suitable calibration from chemical evolution models, a relationship between $\alpha$/Fe and the 
characteristic timescale of star-formation can be inferred (e.g. Thomas et al. 2005). 
This section describes the behaviour of Mg/Fe, which is generally equivalent to previous results for 
$\alpha$/Fe, most of which were based on the Mgb5177 line. 

Figure~\ref{fig:parvslumsig} (middle row) shows the correlations of Mg/Fe with luminosity, velocity dispersion and cluster-centric
radius. 
For the Coma dwarf sample, we recover a wide range in Mg/Fe, from 0.4\,dex above solar (higher than typical for giant 
ellipticals) down to 0.1\,dex below solar. Within the Coma sample alone, there is no significant correlation with luminosity
(Mg/Fe\,$\propto{}L_r^{0.04\pm0.05}$, rms=0.12\,dex), but a weak trend with velocity dispersion is recovered
(Mg/Fe\,$\propto{}\sigma^{0.28\pm0.12}$, rms=0.12\,dex). The Shapley sample shows Mg enhancements up to 0.3\,dex for the 
giant galaxies. As in the case of SSP-equivalent age, the Mg/Fe distribution differs between the inner and outer sample galaxies: 
the median value is similar in each case, but for the outer sample there is an absence of galaxies with [Mg/Fe]$>$0.2, and a greater proportion
of objects with subsolar ratios. At face value, the outer dwarfs appear to have experienced longer periods of star formation
than those in the cluster core, as well as having formed stars until much more recently. 

A  investigation of the systematic behaviour of Mg/Fe, and the other measured abundance ratios, is presented 
by Smith et al. (2008b). 

\subsection{Metallicity and the Z-plane}

Figure~\ref{fig:parvslumsig} (lower row) shows the correlations of Fe/H with luminosity, velocity dispersion and cluster-centric
radius. Although to some degree $L_r$ and $\sigma$ should both serve as proxies for galaxy mass, the correlations of metallicity 
with these different mass tracers are not equivalent. A highly significant (at $>4\sigma$) correlation with luminosity is observed
for the Coma dwarf sample: 
${\rm [Fe/H]}\propto{}L_r^{+0.38\pm0.09}$ (rms 0.21\,dex), while the correlation with velocity dispersion is only marginal:
${\rm [Fe/H]}\propto\sigma^{+0.40\pm0.25}$ (rms 0.25\,dex). 
Contrary to the case for age, the stellar mass-to-light ratio depends only weakly on metallicity. The 
data therefore suggest a primary correlation of Fe/H with stellar mass (traced by $L_r$), rather than with the 
depth of the potential well (traced by $\sigma$). The Fe/H$-L_r$ correlation for the Coma galaxies is consistent
in slope and normalisation with that derived for the Shapley sample at higher luminosity. By contrast the metallicity
of the Coma dwarfs is lower than extrapolated from the Shapley Fe/H$-\sigma$ relation.

\begin{figure}
\includegraphics [angle=270,width=85mm]{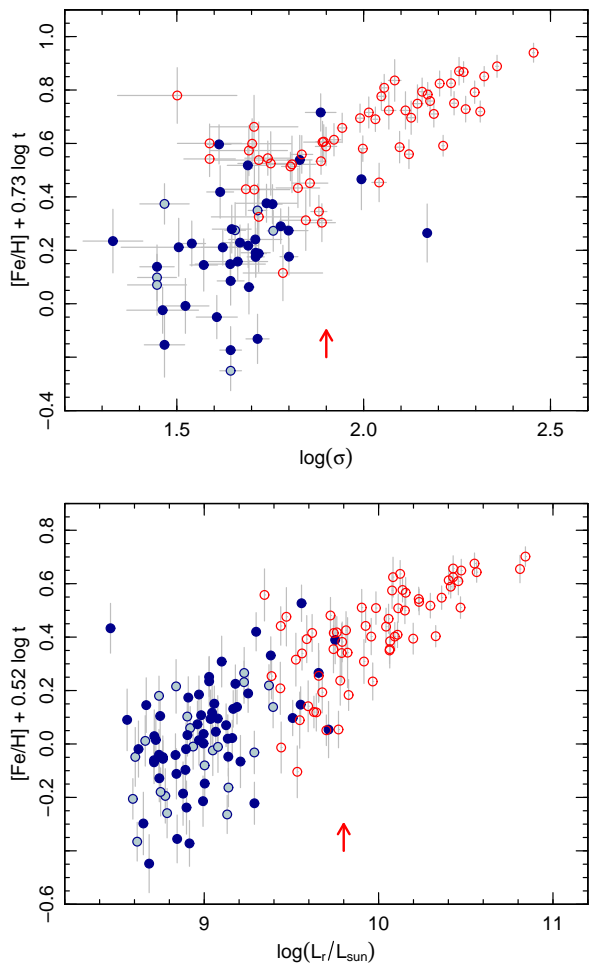}
\caption{Edge-on projection of the Trager et al. (2000b) Z-plane (upper panel) and the equivalent relation substituting luminosity
instead of velocity dispersion (lower panel). 
Symbol types are as in previous figures. 
The Shapley objects have been corrected upwards by 0.1\,dex in Fe/H to account for 
aperture effects, as indicated by the red arrows.}
\label{fig:zplane}
\end{figure}

An anti-correlation of age and metallicity, among galaxies of given velocity dispersion,
was noted for giant galaxies by Trager et al. (2000b), who obtained the planar relationship
\[
{\rm [Fe/H]}=(0.76\pm0.13)\log\sigma - (0.73\pm0.06)\log t_{\rm SSP} - 0.87\ ,
\]
with a scatter of 0.09\,dex. 
A simple least-squares fit to the Coma dwarf sample, not accounting for the correlated errors, yields 
\[
{\rm [Fe/H]}=(0.51\pm0.20)\log\sigma - (0.74\pm0.14)\log t_{\rm SSP} - 0.63
\]
with rms 0.19\,dex. Our plane has a marginally shalower slope with $\sigma$, and a larger scatter, 
but the age--metallicity anti-correlation at fixed $\sigma$ is very similar. 
The scatter around the plane is primarily intrinsic, rather than due to the measurement errors.  
If we use luminosity instead as the mass tracer, we obtain
\[
{\rm [Fe/H]}=(0.32\pm0.07)\log L_r - (0.52\pm0.09)\log t_{\rm SSP} - 2.84 
\]
with an rms of 0.18. The edge-on projections of these planes are shown in Figure~\ref{fig:zplane}. 

To confirm the reality of the age--metallicity anti-correlation, it is essential to consider the effects of correlated errors
in the measured parameters (Kuntschner et al. 2001). 
To this end, we constructed Monte-Carlo realisations in which there is no intrinsic age--metallicity correlation. 
In each realisation we assume the ``true'' age of each galaxy is as observed, and a ``true'' metallicity is assigned 
which is linearly related to $\sigma$ but not related to age. Then we perturb the age and metallicity according to the correlated 
error on each observed data point. Fitting the age--metallicity--mass plane resulting from each simulation, we obtain a distribution 
of coefficients
for $\log t_{\rm SSP}$ which is biased negative, as expected, due to the anti-correlated errors. The median coefficient is $-0.15$, 
and 99\% of simulations yield a coefficient between $-0.30$ and $0.00$. Thus the error anti-correlation can account for only one fifth
of the observed dependence of Fe/H on age at fixed $\sigma$, and only one third of the correlation at fixed luminosity.
For simulations where the input model {\it includes} an intrinsic 
anti-correlation as strong as that observed, the error correlations have negligible impact on the recovered parameters. 

We conclude that the ``Z-plane'' noted by Trager et al. for giant galaxies is followed also in the dwarf galaxy regime.
However, the dwarfs appears to exhibit a larger intrinsic scatter around the plane, than found for the giant galaxies.

\subsection{Correlations with environment}\label{sec:enviro}

Based on a preliminary analysis of the data presented here, 
Smith et al. (2008a) presented evidence for strong correlations of dwarf galaxy properties with cluster-centric radius. 
These effects can be seen  in the different distributions of light and dark blue points in the first column of 
Figure~\ref{fig:parvslumsig},
and are emphasised by the panels in the third column which show the stellar population parameters directly as a function
of radius. 

Among the Coma dwarfs, 
radius from the cluster centre is a better predictor of SSP-equivalent age than either luminosity or 
velocity dispersion, with a correlation that is formally significant at the $\sim$5$\sigma$ level. 
Given our limited azimuthal coverage of the outer parts of the cluster, we cannot distinguish between a true cluster-centric
gradient, or a population of young galaxies localised in the south west. Nor can the data be said to favour
a continuous gradient, rather than a ``step'' between inner and outer samples. The latter case might result from a 
radially-changing mixture between two distinct galaxy populations. 

The Mg/Fe ratios show a similar decline with increased cluster-centric radius, suggesting that star-formation histories
were more extended among the outer galaxies, resulting in increased incorporation of Fe-rich Type Ia supernova ejecta. Consistent with
this, we find a marginally-significant increase of Fe/H with radius, such that the Mg abundance itself (i.e. Mg/H)
is consistent with no correlation with distance from the cluster centre. 

Note from Figure~\ref{fig:zplane} that the dwarf galaxies in the inner and outer 
parts of Coma fall in the same part of the Z-plane\footnote{At least when considering the luminosity version of the plane. 
For the traditional Z-plane using velocity dispersion, the limited $\sigma$ data in the outer region prevents a meaningful test of 
this result.} 
Introducing a cluster-centric radius term into the Z-plane, its coefficient is consistent with zero.
This confirms that the environmental differences in age and metallicity are anti-correlated following the slope of the plane itself. 

\subsection{Robustness against choice of models}~\label{sec:allansontmbk}

\begin{figure*}
\includegraphics [angle=270,width=180mm]{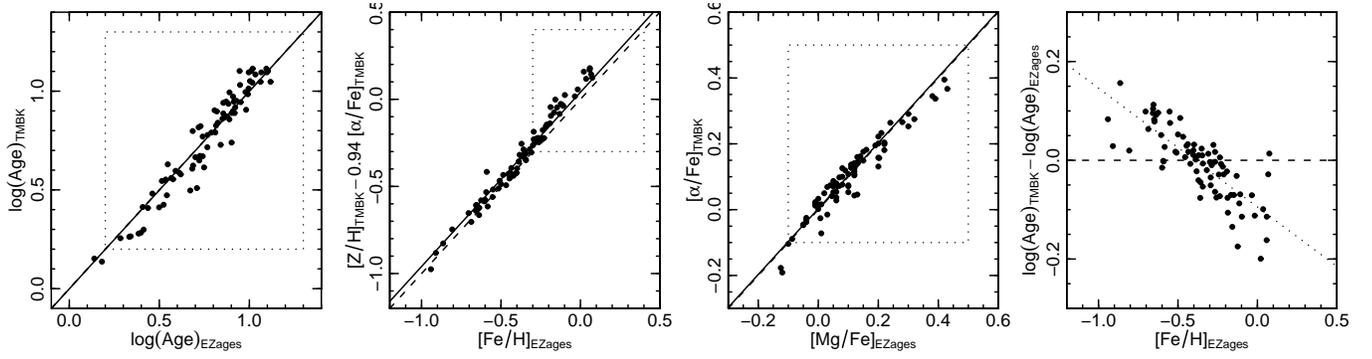}
\caption{Comparison between stellar parameters, estimated for the Coma dwarf sample, using \eza\  
and using the Thomas et al. (2003) models (TMBK). 
The dotted box indicates the axis range in Figure 10 of Graves \& Schiavon (2008) which shows an equivalent comparison for more luminous galaxies.}
\label{fig:allansonssp}
\end{figure*}

Graves \& Schiavon (2008) have made a comparison between stellar population parameters derived from 
\eza\ with results from fitting to the TMBK models, using the Thomas et al. (2005) sample of giant early-type galaxies. 
For a similar test, we have performed $\chi^2$
fits to the TMBK SSP predictions, using a restricted subset of indices (Hbeta, Fe5270, Fe5335, Mgb5177), matching
those used in our \eza\ fits. These indices are quite insensitive to the abundances of C, N and Ca, the more flexible abundance 
mixture fit by \eza\ should not affect the comparison. Rather, we are comparing the underlying models, i.e. Thomas et al. (2003) versus
Schiavon (2007). Figure~\ref{fig:allansonssp} shows the results of this test, in a format readily comparable to
Figure 10 of Graves \& Schiavon (2008). 

For SSP-equivalent age, we find no offset (0.001$\pm$0.008\,dex) between
the two methods, and a scatter of 0.07\,dex. 
(For comparison, Graves \& Schiavon observed a 0.13\,dex offset with younger ages derived from TMBK than from \eza.)
At face value, the scatter in the age comparison is surprisingly large, given that exactly the same data are being used in each fit method. 
Fitting the C, N and Ca abundances in \eza\ does not seem to be the cause of the scatter, since the age differences are 
uncorrelated with C/Fe, etc. However, the deviations
in age are strongly anti-correlated with metallicity (rightmost panel of Figure~\ref{fig:allansonssp}). 
The scatter around the $\Delta{}t_{\rm ssp}$--[Fe/H] relation is only 0.04\,dex.
We conclude there is a small metallicity-dependent relative bias in the ages obtained from the two methods. This effect corresponds
to a relative ``tilt" of the Hbeta versus Fe5270 grid, between the two sets of models. 

For metallicity, we compare Fe/H from \eza\ to the same quantity computed from TMBK parameters Z/H (total metallicity) 
and $\alpha$/Fe, following
the relation [Fe/H] = [Z/H] -- 0.94 \afe, as given by Thomas et al. (2003). We find a systematic offset of $0.040\pm0.005$\,dex, 
with TMBK yielding higher Fe/H, and a scatter of 0.04\,dex. Graves \& Schiavon found a larger offset, 0.08\,dex in the same sense.  
However, our metallicity range is much broader than that of the Thomas et al. (2005) sample; restricting
our comparison to [Fe/H]$ > -0.2$ as in  Graves \& Schiavon, we recover an of offset 0.09\,dex, similar to their result. 
Finally, comparing \eza\ Mg/Fe to TMBK $\alpha$/Fe (which are equivalent, since Mgb5177 is the only $\alpha$ sensitive
feature used here), we find no significant offset ($0.003\pm0.004$) and a scatter of 0.03\,dex. Again, Graves \& Schiavon obtained
a larger offset (0.03\,dex) in the same sense.

To summarize, the use of the Schiavon models results in SSP parameter estimates which differ only slightly, and {\it predictably},
from what would be derived using the TMBK models. For reference, the empirically-determined translation from 
\eza\ to TMBK-based parameters is: 
\[
\log t_{\rm SSP}^{\rm TMBK} = 0.95 \log t_{\rm SSP}^{\rm EZ} - 0.27{\rm [Fe/H]}^{\rm EZ} - 0.07
\]
\[
[{\rm Z/H}]^{\rm TMBK} = 1.10 {\rm [Fe/H]}^{\rm EZ} + 0.77 {\rm [Mg/Fe]}^{\rm EZ} + 0.10
\]
\[
[\alpha/{\rm Fe}]^{\rm TMBK} = 0.99 {\rm [Mg/Fe]}^{\rm EZ} + 0.06 {\rm [Fe/H]}^{\rm EZ} + 0.02 \ \ ,
\]
with rms scatters of 0.04 in $\log t_{\rm SSP}$ and [Z/H] and 0.03 in $[\alpha/{\rm Fe}]$. These translations strictly
apply only to estimates based on the specific set of indices used here, i.e. Hbeta, Mgb5177, Fe5270 and Fe5335.

\begin{figure}
\includegraphics [angle=270,width=85mm]{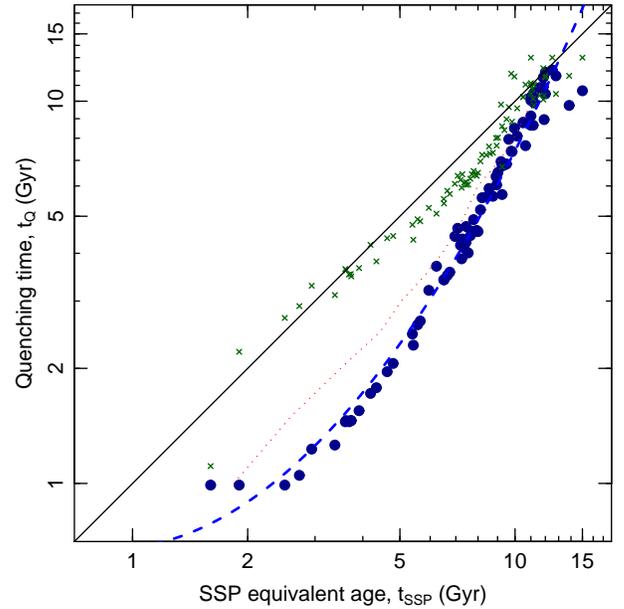}
\caption{The relationship between quenching time and SSP age for the Coma sample, derived from fits by Allanson et al. (in preparation).
The figure shows that the an SSP-equivalent age of, say, 5\,Gyr can alternatively be interpreted as a constant star-formation history, 
truncated $\sim$2\,Gyr ago.  The filled symbols are standard abrupt-quenching models, with the fit line given in the text. 
Small crosses indicate models in which the quenching is followed by a exponential decline with time-constant of 1\,Gyr.
The dotted red line shows the similar relation derived by Trager et al. (2008) for quenched constant-SFR models, with slightly
different assumptions. 
}
\label{fig:sspquench}
\end{figure}

\begin{figure*}
\includegraphics [angle=270,width=180mm]{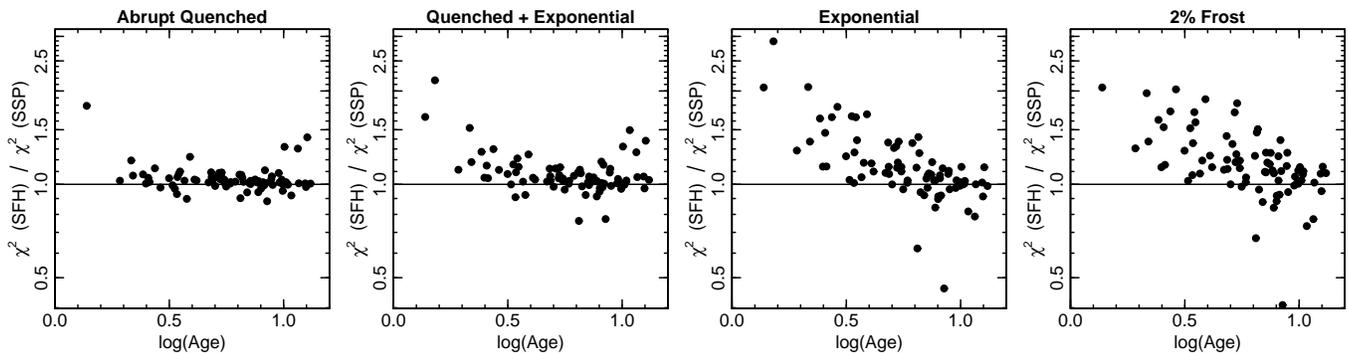}
\caption{Comparison between $\chi^2$ obtained for Allanson et al. complex SFH models and for SSP fits to the same galaxy. 
The ordinate is the \eza\ SSP-equivalent age as derived previously. The abrupt-quenching models perform as well as the 
SSP; the other models have very recent star formation, and generally yield higher $\chi^2$, i.e. the 
fail to reproduce the indices so well. See text for details of the models.}
\label{fig:allansonchi}
\end{figure*}

\subsection{Alternative star-formation history models}\label{sec:allansonsfh}

We have so far employed only the SSP-equivalent age, defined as the age of a single-burst model which 
best reproduces the observed index set for a given galaxy. 
This is the simplest possible representation of the 
star-formation history (SFH), and is useful for comparing trends with luminosity, environment or other parameters. 
In this section, we consider alternative parametrizations for the SFH, to determine whether we
can distinguish SSPs from more complex formation histories, and how our conclusions regarding galaxy ``age'' 
are affected by our SFH assumptions.  

Allanson et al. (in preparation) have computed line-strength indices for various one-parameter SFH families, using 
the TMBK SSP models as the building blocks. The models are: 
(a) the default SSP model characterised by age of a single burst;
(b) a constant-SFR model parametrized by the time since quenching, with an abrupt cut-off; 
(c) a constant-SFR model with an exponential post-quenching decline;
(d) an exponentially-declining SFR parametrized by e-folding time; 
(e) a two-burst or ``frosting'' model, parametrized by burst age (the burst mass fraction is fixed at 2\%). 
In each case, a single metallicity and $\alpha$-element abundance ratio are imposed, representing 
average values over the galaxy enrichment history. (Note that recovery of these parameters
is not our primary goal here.)
We fit each of the Coma sample galaxies using the above five families of models, using 
indices  HdF, HgF, Fe4383, Hbeta, Fe5015, Mgb5177, Fe5270, and Fe5335. (Note that this is 
a slightly larger set of indices than used in the \eza\ fits, in particular including the higher-order Balmer indices.)

One of the aims of this paper is to determine when the faint end of the red sequence was populated, based on the
ages of today's passive dwarf galaxies. For a given $t_{\rm SSP}$, different star-formation histories imply different times since
joining the red sequence.  We focus mainly on SSP versus quenched models.  Although these are both highly idealized, 
many more general SFHs (e.g. quenching combined with a final burst) would yield results intermediate between these cases. 
For a true SSP, the galaxy exists only after its formation time $t_{\rm SSP}$, and immediately begins to fade onto the 
red sequence. In the quenching model, the galaxy begins to move
onto the red sequence when star-formation ceases at $t_{\rm Q}$. For a given set of index measurements, 
$t_{\rm Q} < t_{\rm SSP}$, since the most recent stars must overcome the contribution from earlier
generations to yield the same Balmer indices as in the SSP case. 

Figure~\ref{fig:sspquench} compares SSP and quenching ages derived for the Coma dwarf galaxy sample. 
The latter are derived from the Allanson et al. abrupt-truncation model which best matches the observed indices. 
A quadratic fit yields a good approximation to the results: 
\[
\log t_{\rm Q} =  0.92 (\log t_{\rm SSP})^2 + 0.13 \log t_{\rm SSP} - 0.17\, .
\]

This relation is for {\it abrupt} complete truncation of the SFH at $t_{\rm Q}$ (i.e. model ``b'' above). 
This might be appropriate for truncation by ram-pressure stripping of cold gas from the disk. For the case
of strangulation/suffocation, i.e. the removal of the hot gas halo and subsequent exhaustion of remaining
cold gas, a more realistic model would include a gradual decline after the quenching event. In model ``c'', we allow
a post-quenching exponential fall-off with a 1\,Gyr time-constant. In this case, the quenching times become almost identical to the 
SSP ages, because the youngest stars in the exponential tail compensate for the older stars formed prior to quenching. 

Next, we test whether, within the range of models considered by Allanson et al., 
the index patterns favour particular SFH models. Figure~\ref{fig:allansonchi} shows the 
$\chi^2$ statistic for each model relative to the $\chi^2$ for the SSP case. 
For a large majority of galaxies we find that the complex SFH models in fact perform {\it less} well than the SSP model. In particular, 
the exponential and frosting models fail to reproduce the indices for the ``young'' galaxies, where the form of the SFH has greatest
impact on the spectrum. This arises because such models include substantial very recent ($<2$\,Gyr) star-formation, which 
would generate stronger high-order Balmer lines than observed, as the very young stars make greater contributions in the blue. 
The frosting model adopted here is rather artificial, with a dominant old (13\,Gyr) population, plus a 2\% (by mass) burst at a later
epoch which is fit to each galaxy. None the less, we can conclude robustly that frosting of uniformly-old dominant populations
cannot reproduce the observed spectra unless the mass in the later burst is much greater than 2\%. Thus, the young ages
are {\it not} caused by a late trickle of star-formation which contributes insignificant mass.  
The abruptly-quenched models are similar to SSPs in not hosting very young stars; these models fit the data almost as
well as SSPs, but even here the $\chi^2$ ratio prefers SSP fits for the younger galaxies. Finally, the exponentially-quenched 
models are disfavoured because, like the pure exponential models, they include a trickle of star formation to very late times. 

In summary, to the extent that the index data are capable of distinguishing between alternative star-formation histories, they tend
to favour models with very limited star formation in the past 1-2\,Gyr. Models with little such recent activity, in particular
the SSP and abrupt-quenching cases, perform equally well. Of course, these models have different implications for the
growth of the red sequence, and  thus make different predictions for studies at higher redshift. In 
Section~\ref{sec:redseq}, we consider the difference between SSP and abrupt-quenching cases as indicative of this 
ambiguity.

\section{Discussion}\label{sec:discuss}

\subsection{Ages and metallicities for dwarf galaxies}

In this section, we compare our results to previous spectroscopic work on dwarf galaxies in general. 
We defer detailed discussion of environmental correlations in Coma until Section~\ref{sec:comaenv}.

In recent years several groups have used line-strength indices to estimate ages and metallicities for dwarf elliptical galaxies 
in the Virgo cluster, and other systems at comparable distance (Geha et al. 2003; van Zee et al. 2004; Michielsen et al. 2008; 
Sansom \& Northeast 2008). All four of these studies focused on a luminosity range similar to that in to our work (broadly $-19.5\la{}M_r\la{}-16.5$),
but have been limited to much smaller samples (typically 10--20 galaxies, compared to the $\sim80$ here). With the exception of Sansom \& Northeast, 
these studies report a broad distribution of SSP-equivalent age, 2--10\,Gyr, and Fe/H ranging from solar down to one-tenth of solar. 
These ranges agree well with the
span of values we find for the Coma dwarf galaxies (ignoring the distinction between inner and outer objects for now).  
The $\alpha$/Fe  reported
in these papers are generally centred near solar, but again with a large range, between half and twice the solar ratio. 
For our Coma dwarfs, we find $-0.1\la$[Mg/Fe]$\la{}+0.4$. There are no very Mg-underabundant galaxies in our sample. 
Sansom \& Northeast reported somewhat different results, with their sample of dEs in low-density groups {\it all}
having young ages ($<3$\,Gyr) and 
subsolar $\alpha$/Fe.  They find a wide range in the metallicities, and an apparent bimodality, with 
around half of the galaxies having [Fe/H]$\approx$--1.2. Their higher-metallicity dwarfs lie approximately on the 
index-$\sigma$ relations that we find from the Coma sample. In contrast their very low-metallicity 
dwarfs would fall well below our index-$\sigma$ relations; no comparable population is seen in our sample. 

A potential concern with all of the nearby dwarf studies is that many dwarf ellipticals harbour a compact stellar nucleus, 
which could have a formation history different from that of the bulk of the galaxy (e.g. Lotz, Miller \& Ferguson 2004). 
For nearby galaxies, e.g. in Virgo, where the dwarfs have typical effective radii $\ga$10\,arcsec,  
an extracted $\sim$2\,arcsec central spectrum can be significantly affected by the nucleus, 
contributing 20--40\% per cent of the light, and thus may be unrepresentative of the diffuse stellar population. This is much less of a concern for more 
distant galaxies, where a greater fraction of the galaxy light is sampled by the extracted spectrum, and the contribution from the
nucleus correspondingly reduced.  In particular, the nuclear contamination is an order of magnitude
smaller at the distance of Coma than it is in Virgo. 

Poggianti et al. (2001) analysed stellar populations in a sample of $\sim$150 dwarfs in the Coma cluster, 
based on spectra from Mobasher et al. (2001). Their sample extends to slightly lower luminosities  
($M_r\approx-16$) than our Hectospec observations, but their typical signal-to-noise ratio is very much
lower than for our data (they report a mean $S/N{}\approx{}9$, compared to $S/N{}\approx{}45$ here). As such, their
ability to recover meaningful ages and metallicities for individual galaxies was severely limited.
Their reported ages span the 1--20\,Gyr extent of the models they used, and indeed many galaxies lie
beyond the grid limits, in part due to the large measurement errors. The metallicities similarly span a very wide range. 
Poggianti et al. claimed to recover the signature of an age--metallicity anti-correlation in the dwarfs, but this may have been  
spurious, given the large and correlated errors. Poggianti et al. also reported a bimodality in Fe/H 
for the youngest galaxies in their sample. Given the low signal-to-noise, this result
(which is  curiously similar to the later claim of Sansom \& Northeast 2008) may have been an artifact of the 
fitting process. In any case, we confirm neither the bimodality in Fe/H, nor the presence of young, very metal
poor galaxies in Coma. 

More recent studies of Coma galaxies have been made by \sanch\ et al. (2006a,b) and 
Trager et al. (2008).  Based on these samples, Trager et al. claim that no signal of 
``downsizing" is seen in Coma, in apparent contradiction to results obtained in other environments. 
Our conclusion that many dwarf galaxies in the cluster core have ages $\sim$10\,Gyr at face value supports this 
picture, although we do also find a minority of galaxies with $t_{\rm SSP}<5$\,Gyr, even in the core. 
In fact, however, the absence of downsizing in the Trager et al. sample results largely from an absence of old high-mass
galaxies, since they recover an intermediate age of $\sim6$\,Gyr for all galaxies.  
Moreover, the results of \sanch\ et al and Trager et al. are not readily comparable to 
our results, since they are based on small samples and dominated by much more luminous galaxies. 
In particular, fainter than $M^\star+2$ (corresponding to the {\it brightest} galaxies in our sample) Trager et al. 
observed only two galaxies and \sanch\ et al. only six. 

Perhaps the most similar study to our own, in terms of sample size, luminosity range, and signal-to-noise ratio, is that
of Chilingarian et al. (2008), who observed 46 galaxies in the rich cluster Abell 496 ($z=0.033$). 
Of these, 30 galaxies are in the luminosity range $-17.0<M_r<-19.0$. Their observations were made at high 
spectral resolution, and consequently have a  
very restricted wavelength coverage. Their recovered distributions of age, metallicity and Mg/Fe are in 
excellent agreement with our results for the Coma dwarfs, and they demonstrate the continuation of the Z-plane into the 
dwarf galaxy regime, as also found here. 

In summary, our results on the typical stellar populations in dwarf galaxies are generally similar to those of previous work 
that was based on smaller samples of objects, and/or on spectra of much lower signal-to-noise.

\subsection{Recent quenching in the south west of Coma}\label{sec:comaenv}

\begin{figure}
\vskip 3mm
\includegraphics [angle=270,width=170mm]{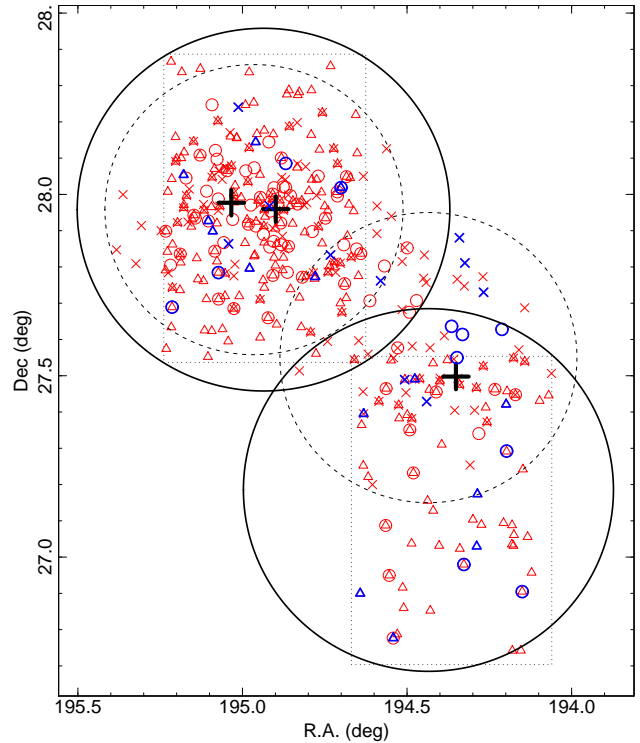}
\caption{Our Hectospec fields (bold circles) compared to the spatial coverage of spectroscopic studies by Caldwell et al. (2003) 
(smaller dashed circles) and Mobasher et al. (2001) (dotted rectangles). 
Observed galaxies are shown by circles (this work), crosses (Caldwell et al.), and triangles (Mobasher et al., red galaxies only). 
The blue symbols highlight candidate recently-quenched galaxies, under definitions adopted by the original papers: 
abnormal spectra (no emission) from Caldwell et al., and k+a galaxies from Poggianti et al. (2004). For our Hectospec sample, 
we highlight objects with $t_{\rm SSP}<3$\,Gyr. 
The positions of the three D galaxies (NGC\,4889, NGC\,4874 and NGC\,4839) are indicated with large black crosses for reference.}
\label{fig:comasw}
\end{figure}

Next, we discuss our results on the spatial dependence of galaxy properties in Coma, in the context of previous spectroscopic studies
by Caldwell et al. (1993) and Mobasher et al. (2001). 

Caldwell et al. (1993) were the first to show that the outer south-west part of Coma harbours a population of 
early-type galaxies displaying clear signs of recent star formation, which are not found in the cluster 
core. These galaxies were identified by their large high-order Balmer absorption and absence of emission lines.
Their data were drawn from two 45\,arcmin diameter fields, with a spatial coverage slightly different than in our work
(see Figure~\ref{fig:comasw}). 
The Balmer-enhanced objects from Caldwell et al. are generally brighter than our dwarf galaxy sample, with $-20\la{}M_r\la{}-18$. 
There are only four galaxies in common between Caldwell et al. and our Hectospec sample, all of which are fairly old
($t_{\rm SSP}=5-10$\,Gyr) and were not flagged by Caldwell et al. as abnormal. 

The discussion by Caldwell et al. of the spatial distribution of the Balmer-enhanced galaxies was influenced by the 
contemporaneous discovery of a south-west substructure in the X-ray emitting gas (Briel et al. 1992). The substructure is
interpreted as an ongoing merger of a subcluster associated with the third D galaxy, NGC\,4839. Caldwell et al. report that
the abnormal spectra galaxies lie mainly between the main part of Coma and the secondary X-ray peak, although in fact
they did not observe galaxies at larger radii than the X-ray substructure. Caldwell et al. emphasised the need 
to extend the spectroscopic coverage of Coma to outer fields other than the south west, to determine whether the 
Balmer-strong phenomenon is spatially localized or part of a more general cluster-centric trend. (Indeed they intended to observe
additional fields in the same observing run but were prevented from doing so by poor weather.)

The next major spectroscopic survey of Coma, hereafter the William Herschel Telescope (WHT) survey, was based on wide-field
photometry of Komiyama et al. (2002). This project targeted five CCD mosaic fields in Coma but only two, in the
cluster centre and in the south west, were fully observed, again due to poor weather. Subsequent spectroscopic observations
(Mobasher et al. 2001) were accordingly restricted to these two regions, which extend further out into the south
west than the Caldwell et al. sample. 
As shown in Figure~\ref{fig:comasw}, our coverage in the south west is similar to that of the WHT survey, because we selected
only known members for follow-up, and the Mobasher et al. sample was the only suitable redshift source for faint galaxies in the 
outskirts of Coma. Compared to our sample, the WHT spectroscopic survey actually covers fainter targets ($r<20.0$, with stellar
 population analysis
generally restricted to $r<19.0$), but at much lower typical signal-to-noise ratio than our Hectospec observations. 

Two papers from the WHT group discuss the environmental dependence of galaxy properties in Coma. Carter et al. (2002) approach
this issue based on Lick indices, and on ages and metallicities derived by Poggianti et al. (2001). They find a strong dependence of the
Mg$_2$ index on radius from the cluster centre, after removing the first-order correlation with luminosity. Of course, as with Caldwell
et al. and the present work, it was not possible to distinguish a cluster-centric gradient from a localised effect in the south west. 
Carter et al. interpreted the Mg$_2$ trend as a gradient in metallicity, but by considering also the corresponding trends in 
Hbeta and Fe5270+Fe5335, we have found that the Carter et al. trends are more compatible with an age gradient. As emphasised by Smith et al. 
(2008a), the implied age gradient is much steeper than the equivalent trend for giant galaxies in the NOAO Fundamental Plane 
Survey (Smith et al. 2006), but is very similar to the result from our Hectospec observations.  

Taking a different approach, Poggianti et al. (2004) classified WHT survey galaxies according to the Dressler et al. (1999) 
scheme, to identify post-starburst ``k+a'' galaxies. In practice, this classification is similar to the 
Caldwell et al. definition of abnormal spectra. Poggianti et al. distinguish between red and blue k+a galaxies, and propose that 
the blue objects (which
are thought to have been quenched less than 500\,Myr ago) are co-located with small scale substructures in the X-ray emission. 
By contrast, the 
red k+a's, which might be expected to match the youngest galaxies in our Hectospec sample, appear  uniformly distributed, 
and present in similar proportions in both the core and the south-west region. 

The relationship of k+a classification to SSP-equivalent age is not trivial, since the former is based on H$\delta$ absorption which
is more sensitive than H$\beta$ (and hence $t_{\rm SSP}$) to very recent star-formation. Moreover, the classification does not take 
account of metallicity effects, which also affect the high-order Balmer lines. The latter may be important given that the classification
scheme was devised for much more luminous (hence more metal-rich) galaxies than those to which it was applied by Poggianti et al. 
None the less, we might expect to recover young SSP-equivalent ages for most k+a galaxies.
Our Hectospec sample includes three of the Poggianti et al. red k+a galaxies. Of these, 
we confirm a fairly young age only for 
GMP4003  
($t_{\rm SSP}=2.4^{+0.9}_{-0.7}$\,Gyr). 
For GMP2692 
(an ``uncertain'' k+a), we obtain a moderately old age ($t_{\rm SSP}=5.0^{+1.6}_{-1.2}$\,Gyr).
The third galaxy, 
GMP4453, 
is classed as a ``secure'' k+a, but we recover 
an intermediate age from the Hectospec data, $t_{\rm SSP}=3.4^{+1.3}_{-0.9}$\,Gyr. 
There are also five WHT survey galaxies in our sample for which 
we obtain young ages ($t_{\rm SSP}=2-3$\,Gyr) but Poggianti et al. did {\it not} identify as k+a. 
This is less surprising, since galaxies with $t_{\rm SSP}>2$\,Gyr need not contain any of the A-stars (lifetime $\la$1\,Gyr) 
which supposedly drive the k+a classification. 

In summary, our results agree with Caldwell et al. (1993) in suggesting a larger fraction of young or recently-quenched
galaxies in the south-west region of Coma than in the cluster core. Similarly, there is agreement with the results of Carter et al. (2002), 
if their gradients are reinterpreted as a trend in age, rather than metallicity. The work of Poggianti et al. (2004), by contrast, found
no environmental correlation for {\it red} post-starburst galaxies. Outside the Coma cluster, other recent work suggests that 
cluster-centric gradients in 
dwarf galaxy properties may be a general phenomenon. Among the Virgo dwarfs studied by Michielsen et al. (2008), 
there are only four galaxies beyond $\sim$1\,Mpc from the cluster centre, but all of them appear young ($\la$3\,Gyr), 
while most of their dwarfs in the Virgo core have ages 4--8\,Gyr. 
There is also an hint of radial age variation in the Abell 496 sample of 
Chilingarian et al. (2008), although the radial extent is more limited, with all galaxies within $\sim$0.5\,Mpc.

\subsection{Growth of the faint red sequence}\label{sec:redseq}

\begin{figure}
\includegraphics [angle=270,width=85mm]{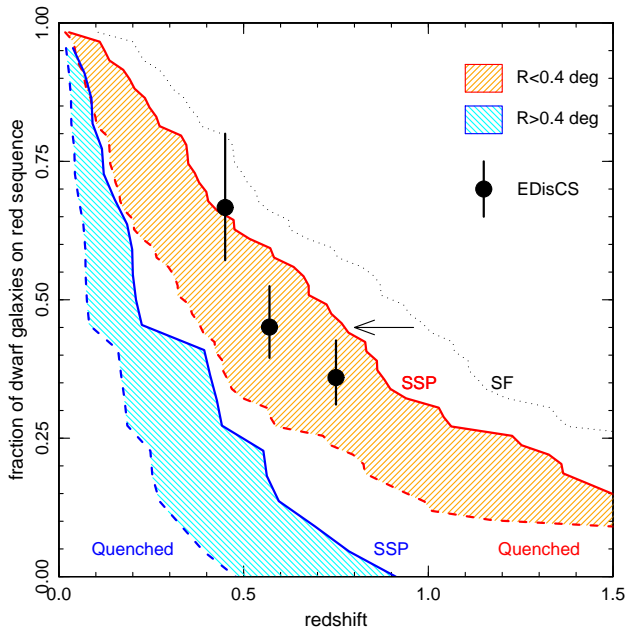}
\caption{Predictions from the galaxy ages for the build-up of galaxies at the faint end of the red sequence.
 The shaded regions show the cumulative distribution of galaxies which had joined the red sequence by a given redshift, 
 assuming SSP (solid boundary) and abruptly-quenched constant SFR models (dashed boundary). 
The red hatched region indicates galaxies within 0.4\,deg (0.7\,Mpc) 
of the cluster centre, while the blue hatched region is for galaxies outside this radius.
We take into account a reddening delay time of 1\,Gyr for the SSP models, and 0.5\,Gyr for the quenched models. The star-formation epoch
for the inner galaxies, under the SSP assumption, is shown by the black dotted line, with an arrow to illustrate the reddening delay. 
The large black points
show the observed fraction of red-sequence dwarfs in EDisCS clusters at $z=0.5-0.8$ (De Lucia et al. 2007),
converted onto the same scale. }
\label{fig:agedist}
\end{figure}

In this section, we consider how our stellar age measurements translate into expectations for the
luminosity function of galaxies on the red sequence (RSLF hereafter) at significant look-back times. Whereas previous comparisons of this kind 
(Nelan et al. 2005; Smith 2005; Smith et al. 2007) were based on a fit to the age--mass relation and its scatter, in this paper
we use the age distribution as measured directly for the red sequence dwarfs. 

Evolution in the RSLF out to $z\sim0.8$ was 
reported by De Lucia et al. (2004, 2007), from the ESO Distant Cluster Survey (EDisCS), 
in the form of a relative paucity of faint red galaxies, relative to the most luminous red galaxies, in distant clusters.
The dwarf galaxies in our Coma sample are representative of the luminosity range used by De Lucia et al. 
to define the giant-to-dwarf galaxy ratio.  By estimating the fraction of our sample galaxies which were on
the red sequence at a given redshift, we can directly compare the ``look-back'' and ``archaeological'' evidence for
downsizing. 

As discussed in Section~\ref{sec:allansonsfh}, the SSP-equivalent age, for a given galaxy, is compatible with a wide 
range of star-formation histories, which in general make different predictions for the epoch at which the galaxy became red. 
In this section, we work with the SSP and abruptly-quenched SFH models, defined respectively by $t_{\rm SSP}$ and $t_{\rm Q}$.
(To avoid mixing different model sets, we use quenching times derived from \eza\ $t_{\rm SSP}$ via the quadratic relation provided in 
Section~\ref{sec:allansonsfh}, rather than the extended TMBK model fits.) Many alternative SFH models, e.g. a quenching 
with a final starburst, can be thought of as intermediate between the SSP and quenching cases. 

Each galaxy is assumed to join the red sequence a short time after ceasing to form stars. The reddening delay time 
is not computed self-consistently from the SFH here; instead we simply assume a constant 1.0\,Gyr for the SSPs and
0.5\,Gyr for the quenched models (where the youngest stars make a smaller contribution to the total light). 
The cumulative distribution of redshifts corresponding to arrival on the red sequence, $z_{\rm red}$, can be converted 
into an estimated deficit of faint red sequence galaxies in distant clusters, relative to the giants which are assumed to
be constant with redshift (Figure~\ref{fig:agedist}). We estimate the epoch at which 50\% of the low-luminosity
galaxies joined the red sequence is 3.9--6.4\,Gyr ago ($z=0.35-0.72$) for the inner galaxies and 1.0--2.6\,Gyr ago 
($z=0.08-0.22$) for the outer sample. The ranges here indicate the spread between SSP and quenched formation assumptions, 
with the higher redshift corresponding to the SSP model. 

We compare our results to the RSLF evolution in EDisCS observed by De Lucia et al. (2007),  after transforming their  values
to a dwarf-to-giant ratio normalised at $z=0$. 
Other studies of large cluster samples (e.g. Stott et al. 2007; Gilbank et al. 2008) agree with the evolution seen in EDisCS,
to first order, although contrary results have been obtained by Andreon (2008).
The EDisCS points fall close to the evolution predicted from the inner sample, with the number of red galaxies
doubling since $z\sim0.5$.  
Thus, the build-up in the faint red sequence observed in distant clusters occurs at a rate which correctly ``predicts'' the 
observed age distribution for dwarf galaxies in the core of Coma. By contrast, the implied very rapid build-up of the 
red sequence in the outer part of Coma is not consistent with  the EDisCS red-sequence evolution. Since the EDisCS
results are limited to distant cluster cores, this difference is not surprising. If the outer dwarfs were indeed accreted only
recently then the appropriate comparison might be with the RSLF of field galaxies, which appears to evolve even more rapidly
than the cluster RSLF (Gilbank \& Balogh 2008). More sophisticated modelling is clearly required to link these different
probes of the environmental dependence of downsizing. 

We can not yet discriminate, based on this test, between SSP and quenching models for the typical star-formation history 
of dwarf galaxies. However, the observed RSLF 
evolution falls within the range spanned by these two rather extreme assumptions. This first-order agreement is in itself
strong evidence against pathological SFHs, e.g. many repeated bursts punctuated by quiescent periods, in which the RSLF 
evolution could be largely decoupled from the distribution of (SSP-equivalent) ages. Such a decoupling has been 
suggested by Trager et al. (2008) based on the supposed absence of downsizing in Coma. Instead, we find evidence
for star-formation histories which may be extended, but are not subject to widespread irregular rejuvenation
events.

\subsection{Comparison to semi-analytic models}

\begin{figure}
\includegraphics [angle=0,width=80mm]{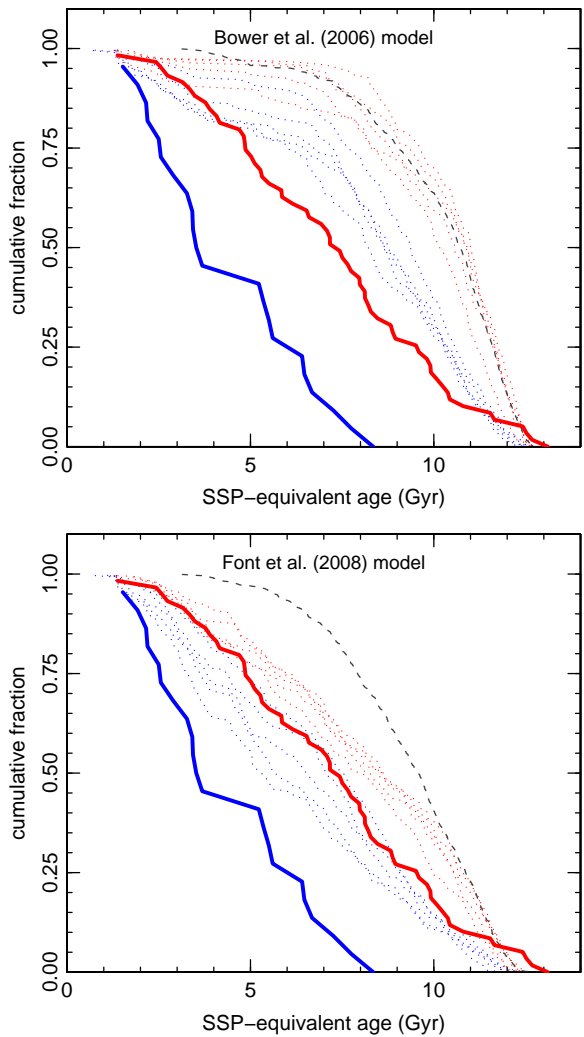}
\caption{The cumulative SSP-equivalent age distribution of Coma dwarfs (thick solid lines), 
compared to predictions from {\sc galform} semi-analytic models (dashed lines). 
The two panels indicate different versions of the models, with different prescriptions for stripping gas from satellite haloes (Bower et al. 2006; Font et al. 2008).
The red and blue lines correspond to galaxies inside and outside 0.7\,Mpc radius from the cluster centre, for observed and model galaxies alike. 
For the model galaxies, the age distribution for five $10^{15}\,M_\odot$ clusters are shown separately to illustrate the
variance between clusters. The grey dashed lines show
the luminosity-weighted ages (inner sample, all five cluster combined), which do {\it not} correspond to an SSP-equivalent measurement (see text).
}\label{fig:galform}
\end{figure}

Semi-analytic models for galaxy formation can be used to extract star-formation histories
for individual simulated galaxies. The model SFHs can be much more complex than can be parametrized by 
toy models such as SSPs or abrupt quenching. However, it is possible that the semi-analytic models may not be
any more successful in reproducing the formation histories of real galaxies, since the predictions depend on many 
complex inputs and assumptions.  

In this section, we present a preliminary comparison of the observed age distribution for the Coma dwarf galaxy sample against predictions
from two recent versions of the Durham {\sc galform} models. The models are based on halo merger trees from the Millenium Simulation
(Springel et al. 2005), and are described by Bower et al. (2006) and Font et al. (2008). 
The main difference between the two sets of predictions, in the current context, is that Bower et al. abruptly remove the hot gaseous haloes of 
galaxies when they are accreted into clusters, while Font et al. include a more physically-motivated prescription for stripping,
which permits more extended star formation in accreted cluster galaxies. 

To connect model and observed quantities, it is necessary to predict the SSP-equivalent age, rather than a luminosity- or mass-weighted
age, for complex star-formation histories. Although weighting the contributions by their luminosity is the correct scheme for predicting 
line-strength indices, the Balmer indices in particular depend non-linearly on age, so that luminosity weighting under-represents the 
Hbeta contribution from the youngest stars (Trager \& Serra 2006; Trager et al. 2008). 
The public  {\sc galform} galaxy catalogues contain only luminosity-weighted ages. To generate correctly weighted SSP-equivalent
values, for valid comparison to the data, we have post-processed a sample of {\sc galform} merger trees as follows. For
each descendant galaxy at $z=0$, we compute the combined star-formation history including all of its progenitors. 
From the stellar mass formed at each timestep, we obtain the luminosity contributions in the descendant, and the V-band
luminosity-weighted contributions to the Hbeta index, using the Schiavon (2007) SSP models. The total luminosity-weighted Hbeta 
is then translated back to an ``observed'' 
age using the same models\footnote{We assume solar metallicity and abundance ratios when converting from age to Hbeta and vice
versa. To include metallicity effects consistently would require computing luminosity-weighted metal line indices and 
using \eza\ to convert back to ``observed'' parameters.}.
As expected, this weighting scheme yields significantly younger ages than the luminosity-weighted values provided 
by the Millennium Simulation database. 

The simulated sample is selected from five clusters with halo mass $M\approx10^{15}\,M_\odot$, with predicted
galaxy luminosity $-18.5 < M_r < -17.0$ and low H$\alpha$ emission. 
The objects are assigned to inner (within 0.7\,Mpc of the cluster centre) and outer (0.7--2.0\,Mpc) samples, 
based on two-dimensional projected radius.
Figure~\ref{fig:galform} shows the {\sc galform} predictions in comparison to the observed age distribution. 
The SSP-equivalent ages of dwarf galaxies in the Bower et al. models are much too old, with almost no objects inside 0.7\,Mpc 
having $t_{\rm SSP}$ less than 8\,Gyr (which is roughly the median of the observed distribution). In the Font et al. version
the more extended star-formation histories yield a distribution of  $t_{\rm SSP}$  which is closer to that observed in the inner region. 
(Note that the difference between the model set is less pronounced in the luminosity-weighted ages than in the SSP-equivalent ages.)

Both models show a difference in the age distribution between inner and outer samples, with younger galaxies in the latter as expected. 
However even in the Font et al. version, the outer galaxies do not show such a dramatic deficit in old dwarf galaxies in the outskirts
as seen in our Coma sample. This may be due to shortcomings in the statistical assignment of orbits to satellite galaxies in the Font et al. 
model, which mean that the position of galaxies within the cluster may not be consistent with the stripping history they 
experience. The disagreement should not be overinterpreted since,  as already stressed, the south west part of Coma 
may be unrepresentative of the outskirts of massive clusters in general.

\subsection{Formation of cluster dwarfs}

Many studies have discussed the origins of passive dwarf galaxies. 
A key issue is whether cluster dwarfs are ``primordial'', having formed from the highest density peaks in proto-clusters (Tully et al. 2002), 
and avoided subsequent merging,
or instead are the descendants of disk galaxies or dwarf irregulars which were quenched and transformed through interaction
with the cluster environment (e.g. Boselli \& Gavazzi 2006). This distinction may be too simplistic and it is likely that both 
scenarios operate at some level: the question is how to distinguish which processes dominate in which mass regimes, 
in which environments, and for which morphological subclasses of dwarf. 

Our results provide little support for the presence of primordial dwarfs in the luminosity range we studied: 
Only $\sim$10\% of our sample galaxies,  all located in the central part of the cluster, are compatible with a 
rapid star-formation event ([Mg/Fe]$\ga$0.3) at $z>1$. 
There is much stronger evidence in favour of transformation from infalling field galaxies: The typical SSP-equivalent ages of 
4--8\,Gyr indicate that
star formation ceased at $z<1$ even for many dwarfs in the cluster core. The typical Mg/Fe ratios suggest extended star-formation histories, 
with timescales $\sim$5\,Gyr (using the relation given by Thomas et al. 2005), prior to quenching. Perhaps most convincingly, our results
demonstrate that the build-up of the passive dwarf population is an ongoing process 
in the outskirts of the cluster, with significant growth occurring even since $z\sim0.2$. 

The physical mechanisms responsible for quenching the infalling galaxies are not easy to determine from stellar population properties, since
the red galaxies are observed $\ga$1\,Gyr after star-formation ceased. We have seen that little discrimination can be made between prior
star-formation histories, beyond the constraint of no significant star formation in the past $\sim$1\,Gyr. Thus we cannot 
directly distinguish between mechanisms involving only the {\it removal} of gas, such as ram-pressure stripping of the
cold interstellar medium (Quilis, Moore \& Bower 2000; Tonnesen, Bryan \& van Gorkom 2007), 
or ``strangulation'' by stripping a hot gas reservoir (Larson, Tinsley \& Caldwell 1980; Balogh, Navarro \& Morris 2000; 
McCarthy et al. 2008), as opposed to those driven by {\it exhaustion} 
of gas in starbursts, either driven by mergers in the cluster outskirts (Moss 2006) or by tidal shocking in sub-cluster mergers (Bekki 1999). 
Our comparison with {\sc galform} semi-analytic models, though preliminary, suggests that the strangulation mechanism as
implemented by Font et al. (2008) may be effective in reproducing the age distribution, 
without quenching star-formation too rapidly as in previous versions (e.g. Bower et al. 2006). On the other hand, even the 
Font et al. (2008) model does not reproduce the prevalence of young dwarfs 
in the outer regions. If the correspondence of recent quenching with the merging NGC\,4839 group in the south west can be confirmed, it would 
suggest that substructure-related mechanisms such as that described by Bekki (1999) may be important processes in driving 
quenching more generally. 

We are pursuing several observational tests that will help to determine the most important quenching processes. One 
avenue for future work is to combine stellar population information with morphological and structural information, from 
ground-based imaging and from the HST/ACS Treasury Survey (Carter et al. 2008). Lisker et al. (2007) have proposed 
that multiple channels may be necessary to explain the diversity of properties among morphologically different 
sub-classes of dwarf. In this case, we might expect objects with disk features (e.g. residual spiral arms) to have 
young populations, while the oldest galaxies might be of classical dE morphology. 
Moreover, although dE nuclei make little direct contribution to the spectra, there could be indirect correlations between 
stellar age and nucleus luminosity (or nucleus colour), if the process of nucleation is linked to the formation history of the
galaxy as a whole. 

In this paper, we have been limited to fibre spectroscopy to probe a statistically meaningful sample of dwarf galaxies, but the absence of 
spatial resolution hampers interpretation of recently quenched galaxies. In particular, if quenching is accompanied by a 
tidally-triggered starburst, the young stars are expected to be centrally concentrated, but cessation by
simple ram pressure stripping would leave a uniformly young disk. To distinguish such cases we are pursuing 
integral field spectroscopy of selected targets using 8m-class telescopes. 

Finally, as already noted, additional outer Hectospec fields will be analysed to determine whether the young ages in the 
south west field are a widespread characteristic of the cluster outskirts, or instead a localised phenomenon. 
The latter would clearly favour mechanisms associated with the infalling group, whether due to interaction with the 
intra-cluster medium or to the tidal influence of the subcluster merger.

\section{Conclusions}\label{sec:concs}

In this paper, we have presented an extensive study of passive dwarf galaxies in the Coma cluster, based on new high signal-to-noise
spectra for a large galaxy sample. We have used line-strength measurements to make a detailed investigation of the stellar 
populations in cluster dwarfs, in a luminosity range previously studied only with much smaller samples or at much lower
signal-to-noise. 

We have shown that red-sequence dwarf galaxies in the Coma cluster have a broad age distribution, which is strongly dependent
on their location within the cluster. Dwarfs in the cluster outskirts, at least in the south west region, overwhelmingly exhibit
signs of star formation within the past few Gyr, and of extended chemical enrichment histories. This supports the scenario
in which a large proportion of the cluster dwarf population is formed by environment-driven quenching of star-forming galaxies 
which fall in from the surrounding field. However, the physical mechanism responsible for quenching cannot be determined
from the current data alone. In the cluster core, by contrast, most dwarf galaxies are older (5-10\,Gyr), with some $\sim$10\% 
being compatible with formation at $z\ga1$ in a rapid starburst. These objects are {\it candidates} for truly primordial dwarfs, surviving
from the earliest epoch of galaxy formation. However, it is also possible that most of the core dwarfs were formed in  a manner 
 qualitatively similar to the outer dwarfs, i.e. through environmental quenching, but simply at earlier epochs. 

We have shown that, whatever the mechanisms driving the shut-down of star formation in dwarfs, 
the archaeological evidence from stellar ages is compatible with the observed evolution in the red sequence luminosity function in
distant clusters. The latter yields a doubling in the number of galaxies on the faint red sequence since $z\sim0.5$. 
This is well matched by the fraction of today's dwarfs, in the cluster core, that we estimate had ceased forming stars and
faded onto the red sequence by that epoch. The age distribution of dwarfs in the cluster core can be approximately
reproduced by recent variants of semi-analytic galaxy formation models, which include improved treatment of stripping effects 
on cluster members. For the outer galaxies in the south-west region, the red-sequence evolution implied by their stellar
ages is both more rapid than seen in (the cores of) distant clusters, and more rapid than predicted by semi-analytic models. 
The interpretation of our results in the south-west region is highly dependent on whether a similar age distribution is
found in other outer parts of the cluster. 

Caldwell et al. (1993) commented that the ``most pressing need is 
to extend the work ... to study other regions surrounding the Coma cluster''. Fifteen years later, this remains the highest
priority. We recently obtained Hectospec observations of 
additional outer fields, to be reported in a future paper, which should finally establish whether recent quenching is
a general effect in the outskirts of Coma (and hence presumably also in other clusters), or is specific to the ongoing substructure 
merger in the south-west region.  

\section*{Acknowledgments}

RJS is supported by STFC rolling grant PP/C501568/1 `Extragalactic Astronomy and Cosmology at Durham 2005--2010'.
MJH is supported by NSERC. 
ROM is supported by NSF grant AST-0607866. 
We thank Richard Cool for providing {\sc hsred} and for advice on its use
and Richard Bower for helpful discussions about {\sc galform}.
The Millennium Simulation databases used in this paper and the web application providing online access to them were
constructed as part of the activities of the German Astrophysical Virtual Observatory.

{}

\appendix

\section{Index data comparisons}\label{sec:appixcomp}

As a test for reproducibility in the measurements, and to validate our error estimates, 
we use internal comparisons between indices measured on the half-exposure spectra. 
These spectra were generated by randomly dividing the individual observations of a galaxy
into two groups, with approximately equal total exposure time in each group. The data for each group 
were combined and the two resulting spectra were propagated through all measurement routines in a manner identical
to that used for the fully-combined spectra. Since the multiple observations of each galaxy were obtained with
different fibres, on different fibre configurations, the half-exposure spectra allow for a characterisation
of additional error sources that are not included in the purely statistical formal error estimates. 

\begin{figure*}
\includegraphics [angle=0,width=180mm]{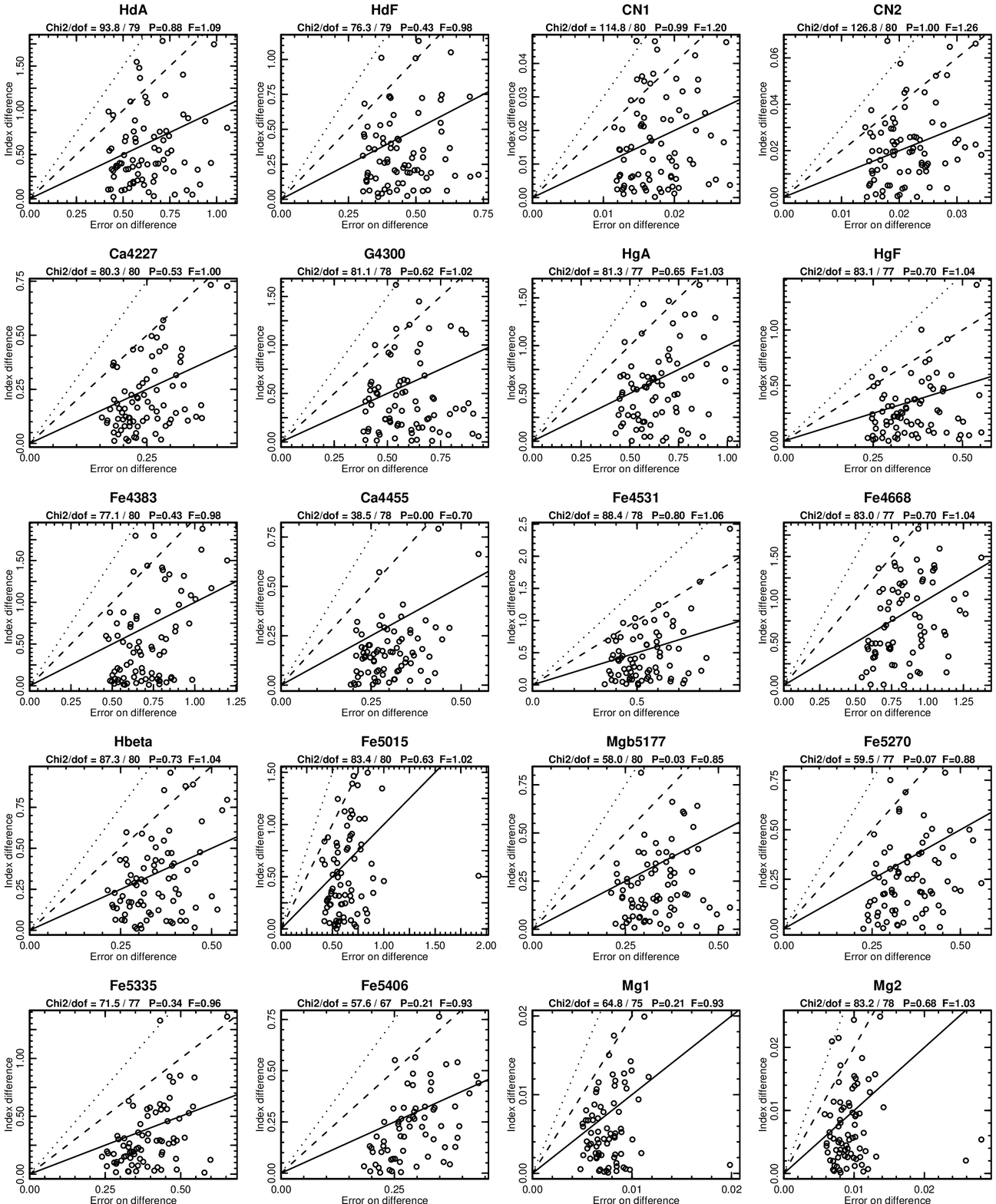}
\caption{Reproducibility test for the index measurements. The indices are measured on two independent spectra for 
each galaxy, each formed from approximately half the total exposure obtained. Each panel plots the 
difference between the two measurements, against the error on the difference as calculated from the 
nominal uncertainties. Solid, dashed and dotted lines indicate deviations at 1$\sigma$, 2$\sigma$ and 3$\sigma$. In the title line, 
$P$ is the probability of observing a $\chi^2$ smaller than that in the real data, given the errors, and 
$F$ is the factor $\sqrt{\chi^2/\nu}$ by which the errors must be artificially inflated to recover a 
reduced $\chi^2$ of unity.}
\label{fig:divabix}
\end{figure*}

Figure~\ref{fig:divabix} confirms that for most indices the expected statistical errors reproduce the 
scatter between repeated measurements: the index errors appear to be accurate to better than 10\%. 
We highlight a number of exceptions. For CN1 and CN2, the errors would need to be increased by 20--25\% to 
obtain an acceptable $\chi^2$. For Mgb5177 and Fe5270 the errors are 10--15\% underestimated, while
for Ca4455 the errors are overestimated by a larger factor ($\sim$30\%). We speculate that the additional 
scatter in the CN indices is due to flux calibration uncertainty in the blue region, combined with
the fairly wide ($\sim$200\,\AA) extent of these index definitions. Note that wide indices further to 
the red are reproduced very well; even Mg1 and Mg2, which span $\sim$500\,\AA, agree acceptably within
the formal errors. 

\begin{figure*}
\includegraphics [angle=0,width=170mm]{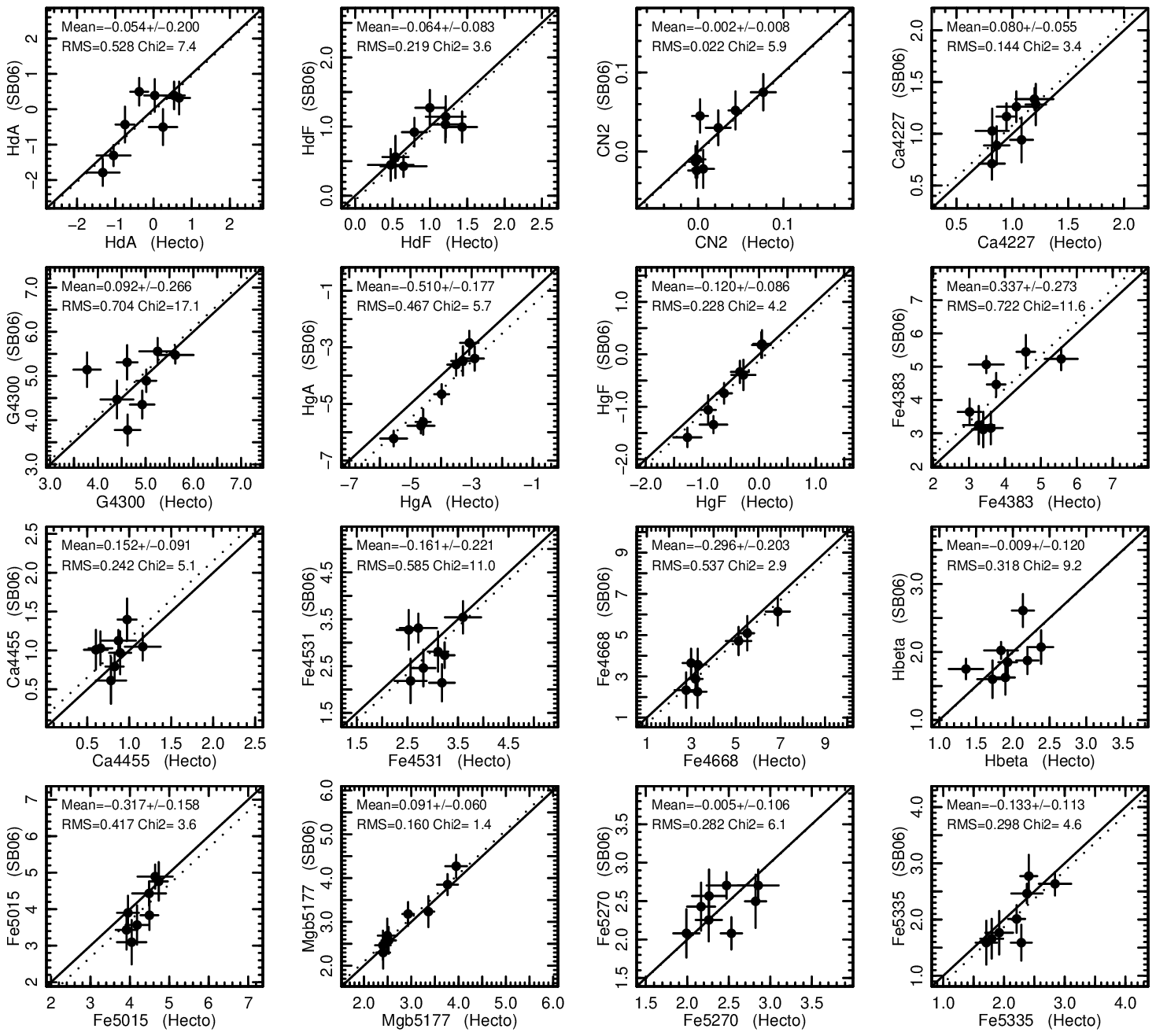}
\caption{Comparison of index measurements with \sanch\ et al (2006a) for eight galaxies in common. 
In each panel we report the offset, rms scatter and
$\chi^2$ around the offset.  Most of the comparisons have $\chi^2$ within the 95\% interval for seven degrees 
of freedom ($\chi^2=1.7-16.0$).}\label{fig:sanch_ixcomp}
\end{figure*}

An external comparison of our index measurements to the data of \sanch\ et al. (2006a), for the eight galaxies in
common, are shown in Figure~\ref{fig:sanch_ixcomp}. 
The \sanch\ et al. data are tabulated after correction to the Lick instrumental response. 
To match our  measurements, the comparison is made after correcting their data back to the 
flux-calibrated system using the offsets provided in their Table A.1. 
The \sanch\ et al. measurements are 
from an aperture 2.0$\times$2.7\,arcsec$^2$, i.e. somewhat larger than our fibre apertures.
Overall, excellent agreement is seen between the two data sources. Formally significant offsets (at the 2--3$\sigma$ level) are found only
for HgA and Fe5015. After allowing for the offsets, the $\chi^2$ statistic for the comparison is acceptable 
(i.e. within the 95\% interval for seven degrees of freedom) in most cases. The exceptions are Mgb5177, where $\chi^2$ is slightly too low, 
and G4300, where it is too high. Note that HgA, Fe5015 and G4300 are not used for the analysis in this paper.

\section{Additional index scaling relations}\label{sec:moreixsiglum}

\begin{figure*}
\vskip -2mm
\includegraphics [angle=0,width=150mm]{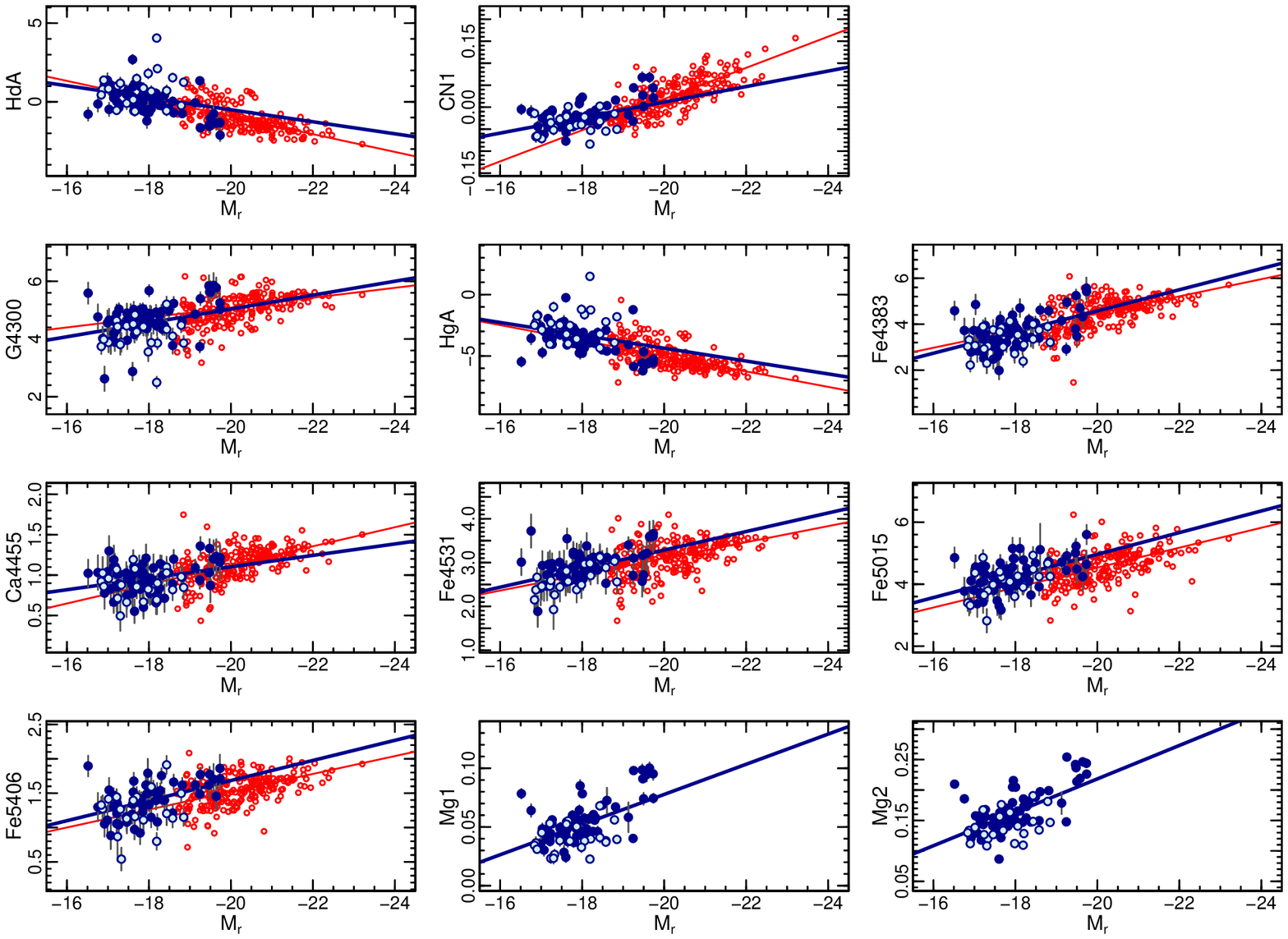}
\vskip -2mm
\caption{Absorption line indices versus luminosity for supplementary indices. Symbols are as in Figure~\ref{fig:ixlum}.}\label{fig:ixlum_more}
\end{figure*}

\begin{figure*}
\vskip -2mm
\includegraphics [angle=0,width=150mm]{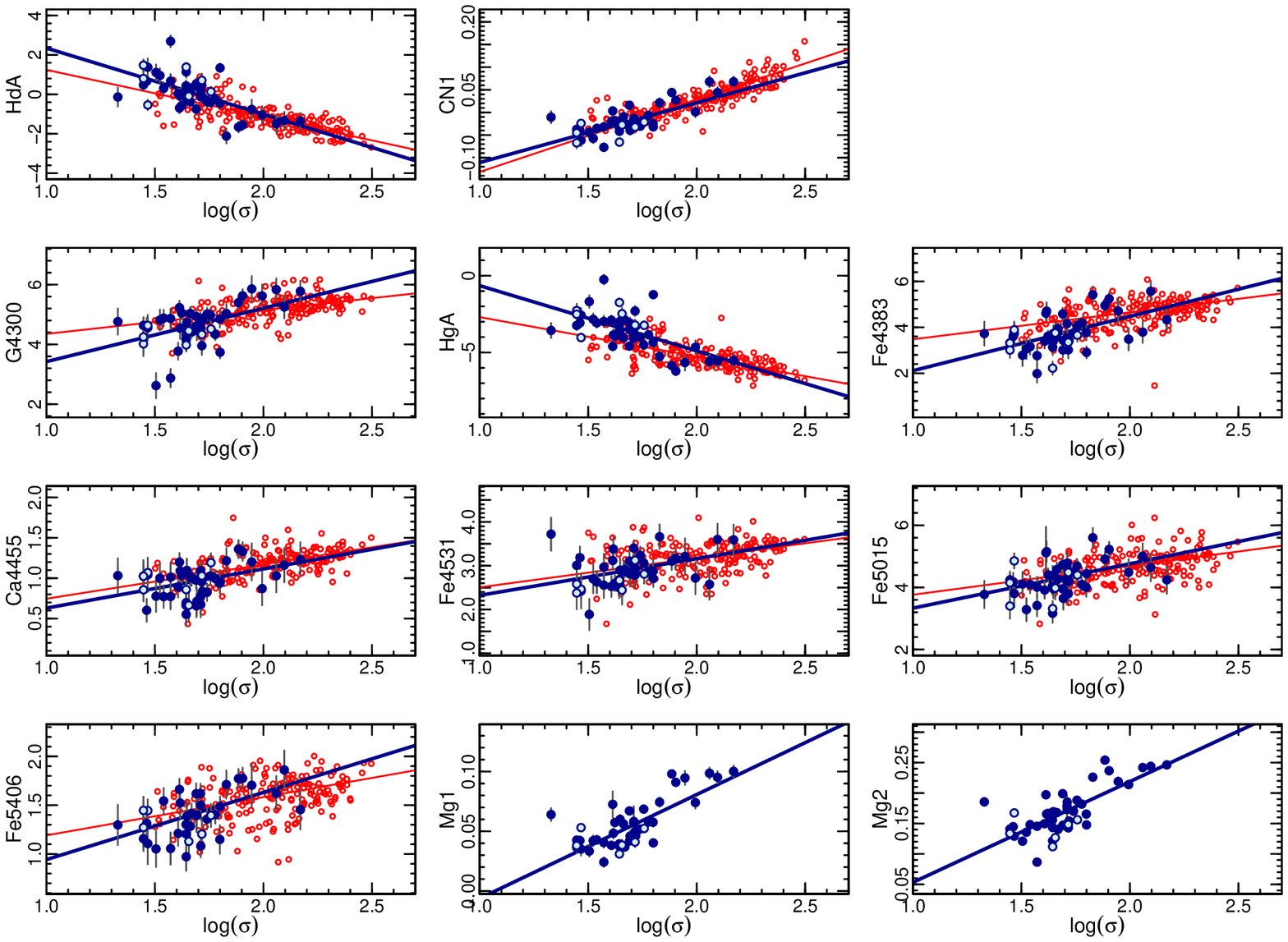}
\vskip -2mm
\caption{Absorption line indices versus velocity dispersion for supplementary indices. Symbols are as in
Figure~\ref{fig:ixlum}.}\label{fig:ixsig_more}
\end{figure*}

Figures~\ref{fig:ixlum_more} and \ref{fig:ixsig_more} provide supplementary index--luminosity and
index--$\sigma$ relations. These figures show the relations for indices which are not used directly in the present paper, 
and are provided for completeness only. 

\section{\eza\ error model}\label{sec:ezerrmod}

\begin{table*}
\caption{Errors as function of $S/N$ for the stellar population parameters derived from  \eza.
For each parameter, the quantities $\varepsilon$ give the standard error
at $S/N=30,40,50,60$ per \AA\ (measured over 4400--5400\,\AA), based on Monte-Carlo simulations.
Over this range in $S/N$, the errors can be described by linear relationships $\varepsilon=a(S/N)^{-1}+b$. 
}\label{tab:ezerrmod}
\begin{tabular}{l|cccccc|c|cccccc|}
\hline
 & \multicolumn{6}{c}{HECTOSPEC} & \ \ \ & \multicolumn{6}{c}{AAOMEGA} \\
parameter & 
$\varepsilon(30)$ & 
$\varepsilon(40)$ & 
$\varepsilon(50)$ & 
$\varepsilon(60)$ & $a$ & $b$ &&  
$\varepsilon(30)$ & 
$\varepsilon(40)$ & 
$\varepsilon(50)$ & 
$\varepsilon(60)$ & $a$ & $b$ \\ 
\hline
$\log(t_{\rm SSP})$      & 0.163  & 0.127  & 0.110  & 0.093  &  4.144 & +0.025   && 0.167  & 0.129  & 0.102  & 0.086  &  4.889 &  +0.005  \\
$[$Fe/H$]$        & 0.144  & 0.112  & 0.099  & 0.084  &  3.496 & +0.026   && 0.134  & 0.110  & 0.089  & 0.082  &  3.211 &  +0.028  \\
\hline
$[$Mg/Fe$]$       & 0.114  & 0.077  & 0.069  & 0.055  &  3.436 & --0.003  && 0.114  & 0.096  & 0.072  & 0.062  &  3.163 &  +0.011  \\
$[$C/Fe$]$        & 0.108  & 0.082  & 0.074  & 0.058  &  2.845 & +0.013   && 0.128  & 0.097  & 0.069  & 0.063  &  4.090 & --0.008  \\
$[$N/Fe$]$        & 0.140  & 0.110  & 0.094  & 0.077  &  3.678 & +0.018   && 0.153  & 0.125  & 0.091  & 0.076  &  4.688 & --0.000  \\
$[$Ca/Fe$]$       & 0.138  & 0.108  & 0.087  & 0.079  &  3.650 & +0.016   && 0.147  & 0.125  & 0.094  & 0.078  &  4.235 &  +0.010  \\
\hline
\end{tabular}
\end{table*}

We have not used the errors estimated by the \eza\ model grid inversion code, since these
are computationally intensive to obtain, and do not provide a full description of the error correlations. Instead, 
we use a set of Monte Carlo experiments to derive an error model which is applied to the whole 
dataset. 

The simulations start with the measured indices for two of the galaxies in our sample 
(GMP4175 and GMP3719).
We apply gaussian perturbations to each index Ix, based on an assumed 
signal to noise ratio $S/N$ (at 4400--5400\,\AA), according to 
$\varepsilon_{\rm Ix} = f_{\rm Ix} (S/N)^{-1}$. 
The factors $f_{\rm Ix}$ are derived from the index error versus $S/N$ 
relations for the complete Hectospec sample. The perturbed indices are used as inputs to the 
\eza\  code. 
Simulations were performed for four choices of $S/N = 30, 40, 50, 60$\,\AA$^{-1}$, spanning the 
range covered by our data. A set of 200 realizations were produced for each $S/N$. 
Equivalent simulations were performed using one of the Shapley galaxies (NFPJ132914.7-314934), 
to allow for any effect of different spectrograph response, and hence different $\varepsilon_{\rm Ix}$ vs $S/N$ relations. 

\begin{figure*}
\includegraphics [angle=270,width=100mm]{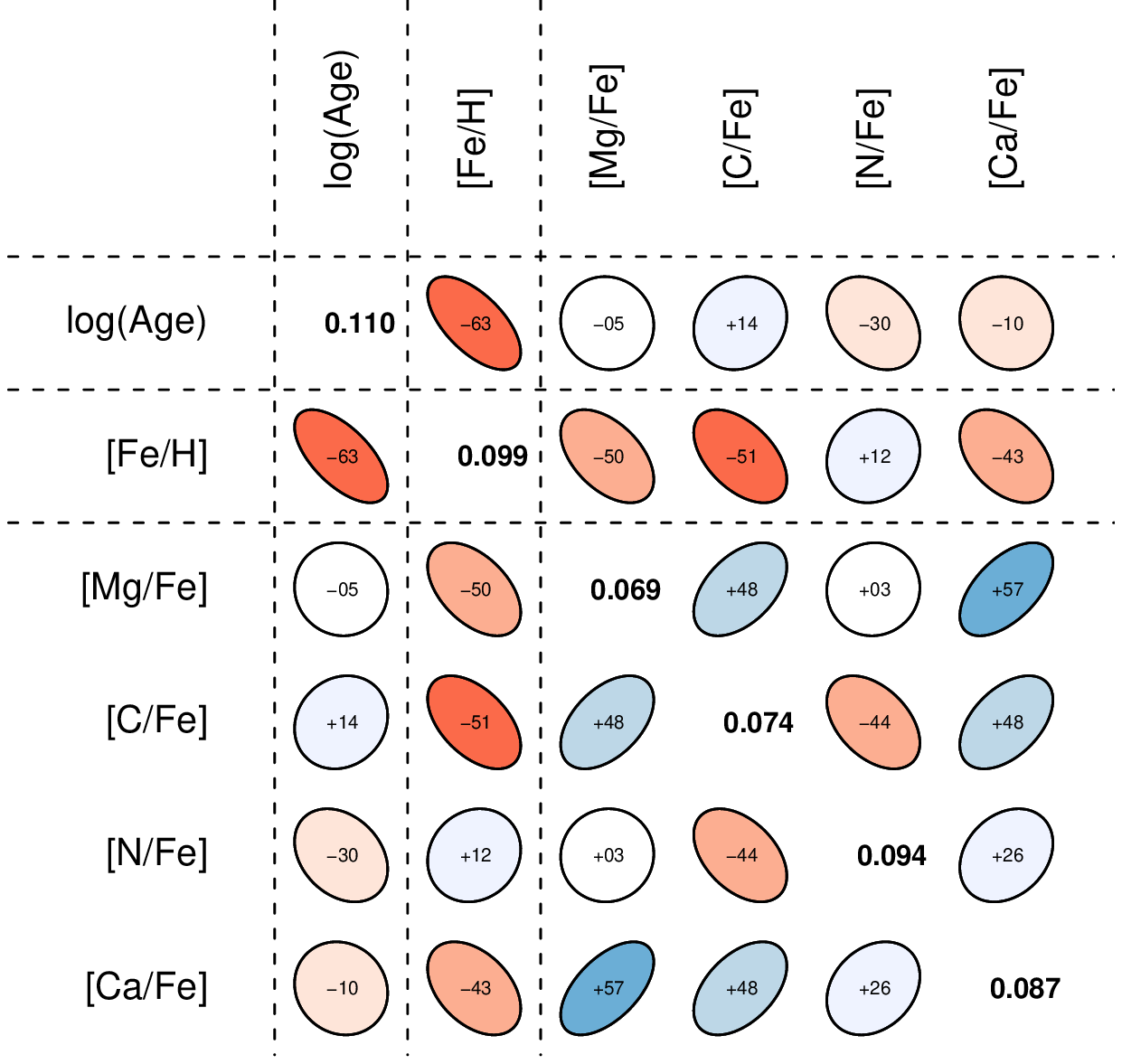}
\vskip -3mm
\caption{Error correlation matrix for the stellar population parameters estimated using \eza. This matrix
is as derived for $S/N=50$\,\AA$^{-1}$. Each off-diagonal cell is labelled with the correlation coefficient between the corresponding
parameters, expressed as a percentage. The shapes of the ellipses indicate the correlation coefficients, and 
are colour-coded red for negative and blue for positive correlation. Diagonal cells are labelled with the 
standard deviation for the corresponding parameter, i.e. the error marginalized over all the other parameters.
These errors are as given in the column $\varepsilon(50)$ of Table~\ref{tab:ezerrmod}.}
\label{fig:ezcovmats}
\end{figure*}

Figure~\ref{fig:ezcovmats} shows the correlation matrices obtained from these simulations. 
Among the prominent features are the age--metallicity degeneracy, i.e. an anti-correlation of the 
errors in $\log t_{\rm SSP}$ and [Fe/H], with correlation coefficient $\sim$0.6. 
There are also error correlations among the abundance ratios, with Mg/Fe, C/Fe and
Ca/Fe all positively correlated. This arises because perturbations to the Fe indices
change all the abundances relative to Fe. The correlations for N/Fe are somewhat
different, in particular being anti-correlated with C/Fe. This occurs because
N is derived via the strength of the CN band, where higher N abundance can compensate for lower
C abundance, and vice versa. 

The marginalized error on each parameter is extracted as simply the standard deviation over
all realizations for a given $S/N$ (see Table~\ref{tab:ezerrmod}). Fitting these with a relation
of the form $\varepsilon_{\rm par} = a_{\rm par} (S/N)^{-1} +  b_{\rm par}$ yields a model
which we then apply to all galaxies as an approximation to the parameter errors. 

\begin{figure*}
\vskip -3mm
\includegraphics [angle=0,width=150mm]{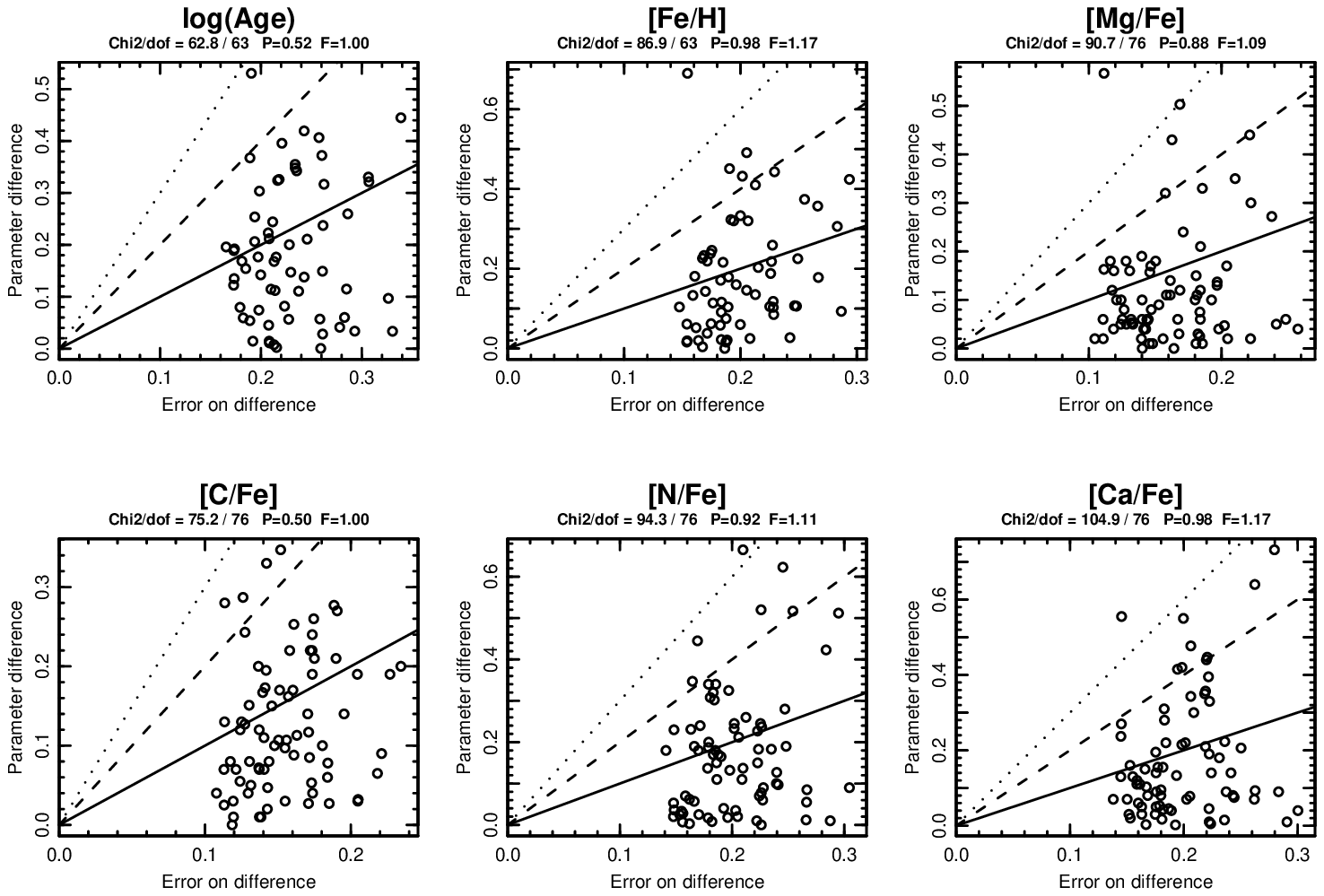}
\vskip -3mm
\caption{Reproducibility test on stellar population parameters derived from \eza, comparing parameters derived from the
two independent half-exposure spectra for each galaxy. The details are as in Figure~\ref{fig:divabix}.}
\label{fig:divabeza}
\end{figure*}

The level of repeatability between the half-exposure spectra can be used to assess the accuracy of the error model.
Figure~\ref{fig:divabeza} shows this test in a form analogous to Figure~\ref{fig:divabix}, but now comparing
stellar population parameters measured from the independent half-exposure spectra for each galaxy. 
As for the indices, we find that the measured parameters agree acceptably within the stated errors, and
the error model is confirmed to $\sim$10 per cent for most parameters. There is a hint that the errors on Ca/Fe
may be underestimated by $\sim$20 per cent. (For Fe/H, there is a similar underestimation at face value, but the 
comparison is clearly skewed by a single outlier.)

\section{Failed \eza\ fits}\label{sec:ezafails}

The following eight galaxies could not be fit with \eza\ to recover meaningful stellar population parameters: 

\begin{itemize}
\item
GMP2910, GMP3895, GMP4294 : 
These are the three objects in which the stellar H\,$\beta$ absorption is contaminated by nebular emission (see Figure~\ref{fig:emlspec}).  
\item
GMP3206 : 
This galaxy from the bright comparison sample has only a single extracted exposure of less 
than one hour integration. The spectrum is affected by bad pixels which are not adequately removed, leading to 
contamination of important indices. 
\item
GMP4200, GMP2805 : 
These galaxies are both from the bright comparison sample, and each has only a single exposure
of less than one hour total integration. They are the two galaxies lying slightly below the model grids in Figure~\ref{fig:hbfe52} (left). 
Their spectra do not show any sign of emission at H\,$\alpha$, so infilling of the stellar Hbeta index by low-level nebular
emission is not suspected. Nothing in the spectra suggest problems with bad pixels as in the previous case. The cause of the very low
measured Hbeta is unknown. 
\item
GMP3699 : 
Here, the inversion appears to be unstable, yielding unphysically large N/Fe. This faint galaxy 
has the lowest Fe5270 and Fe5335 indices in the sample, but does not have pathological index values; the reason for its failure is unknown. 
\item
GMP2615 : 
This bright-sample galaxy lies within the grid, but \eza\ generated no output. The reason for failure is unknown.
\end{itemize}

\section{Data tables}\label{sec:tables}

The data are presented in Table~\ref{tab:indexdat} (absorption line index data), Table~\ref{tab:comadata} (\eza\ and 
other derived parameters for the Coma sample), and Table~\ref{tab:shapleydata} (equivalent parameters for the Shapley
comparison sample). The Shapley index data can be found in Smith et al. (2007). 

\clearpage

\begin{table*}
\caption{Absorption line index data for the Coma sample. The galaxies are identified by their number in the 
Godwin et al. (1983, GMP) catalogue, and by the J2000 position. The following five columns are the 
Petrosian magnitudes in each of the SDSS bands (from Adelman-McCarthy et al. 2007). The remaining columns, 
on three subsequent lines, give the indices and errors, in the conventional units, i.e. magnitudes for CN1, CN2, Mg1 and Mg2;
angstroms for all others. The full version of this table will be provided in the electronic version of the journal. 
}\label{tab:indexdat}
\begin{tabular}{lcccccccccccc}
\hline
GMP  &  R.A.  &  Dec  &  $u$  &  $g$  &  $r$  &  $i$  &  $z$ \\ 
	 & HdA  & HdF  & CN1  & CN2  & Ca4227  & G4300  & HgA  &  \\ 
	 & HgF  & Fe4383  & Ca4455  & Fe4531  & Fe4668  & Hbeta  & Fe5015  &  \\ 
	 & Mgb5177  & Mg1  & Mg2  & Fe5270  & Fe5335  & Fe5406  &  \\ 
\hline \\ 
5361  &  12:56:36.0  &  +26:54:17  &  18.64  &  17.38  &  16.68  &  16.33  &  16.16  &  \\ 
	 & $+0.71\pm0.34$ & $+1.62\pm0.23$ & $-0.032\pm0.009$ & $+0.003\pm0.011$ & $+1.09\pm0.14$ & $+4.33\pm0.32$ & $-3.22\pm0.35$ &  \\ 
	 & $-0.10\pm0.19$ & $+3.35\pm0.39$ & $+1.03\pm0.16$ & $+3.12\pm0.29$ & $+3.18\pm0.48$ & $+2.60\pm0.19$ & $+4.48\pm0.35$ &  \\ 
	 & $+2.48\pm0.19$ & $+0.041\pm0.004$ & $+0.148\pm0.005$ & $+2.64\pm0.21$ & $+2.52\pm0.22$ & $+1.20\pm0.17$ &  \\ 
 \\ 
5296  &  12:56:40.9  &  +27:26:51  &  19.54  &  18.19  &  17.49  &  17.21  &  16.81  &  \\ 
	 & $+0.14\pm0.25$ & $+1.17\pm0.18$ & $-0.021\pm0.007$ & $+0.007\pm0.008$ & $+1.04\pm0.10$ & $+4.52\pm0.23$ & $-3.23\pm0.25$ &  \\ 
	 & $-0.09\pm0.14$ & $+3.66\pm0.28$ & $+1.19\pm0.12$ & $+2.81\pm0.22$ & $+3.38\pm0.35$ & $+2.05\pm0.14$ & $+4.40\pm0.25$ &  \\ 
	 & $+2.82\pm0.13$ & $+0.052\pm0.003$ & $+0.156\pm0.004$ & $+2.38\pm0.14$ & $+2.16\pm0.16$ & $+1.39\pm0.12$ &  \\ 
 \\ 
5254  &  12:56:47.4  &  +27:17:32  &  18.86  &  16.96  &  16.26  &  15.95  &  15.79  &  \\ 
	 & $+1.25\pm0.26$ & $+1.71\pm0.19$ & $-0.050\pm0.007$ & $-0.012\pm0.009$ & $+0.77\pm0.11$ & $+3.86\pm0.25$ & $-1.81\pm0.27$ &  \\ 
	 & $+0.80\pm0.15$ & $+3.21\pm0.31$ & $+0.85\pm0.13$ & $+2.56\pm0.24$ & $+2.62\pm0.37$ & $+2.82\pm0.15$ & $+4.25\pm0.26$ &  \\ 
	 & $+2.47\pm0.15$ & $+0.039\pm0.003$ & $+0.147\pm0.004$ & $+2.47\pm0.15$ & $+2.11\pm0.17$ & $+1.15\pm0.13$ &  \\ 
 \\ 
5217  &  12:56:50.8  &  +27:37:40  &  19.07  &  17.78  &  17.13  &  16.82  &  16.72  &  \\ 
	 & $+1.80\pm0.31$ & $+1.78\pm0.23$ & $-0.040\pm0.009$ & $-0.003\pm0.011$ & $+0.82\pm0.13$ & $+3.55\pm0.32$ & $-1.20\pm0.33$ &  \\ 
	 & $+0.93\pm0.18$ & $+2.71\pm0.39$ & $+0.82\pm0.17$ & $+2.49\pm0.30$ & $+3.66\pm0.48$ & $+2.66\pm0.18$ & $+3.60\pm0.34$ &  \\ 
	 & $+2.03\pm0.19$ & $+0.033\pm0.004$ & $+0.112\pm0.005$ & $+2.43\pm0.32$ & $+1.92\pm0.21$ & ------ &  \\ 
 \\ 
5178  &  12:56:55.9  &  +27:27:44  &  19.27  &  17.91  &  17.24  &  16.87  &  16.69  &  \\ 
	 & $+0.87\pm0.32$ & $+1.36\pm0.23$ & $-0.041\pm0.009$ & $+0.001\pm0.011$ & $+1.09\pm0.14$ & $+4.43\pm0.30$ & $-3.61\pm0.34$ &  \\ 
	 & $-0.13\pm0.18$ & $+4.13\pm0.37$ & $+0.67\pm0.18$ & $+2.38\pm0.29$ & $+3.13\pm0.45$ & $+2.08\pm0.18$ & $+4.08\pm0.33$ &  \\ 
	 & $+2.64\pm0.17$ & $+0.050\pm0.004$ & $+0.163\pm0.005$ & $+2.42\pm0.19$ & $+1.67\pm0.21$ & $+1.11\pm0.16$ &  \\ 
 \\ 
5076  &  12:57:07.7  &  +27:20:25  &  19.47  &  18.19  &  17.45  &  17.06  &  16.76  &  \\ 
	 & $-0.60\pm0.29$ & $+0.50\pm0.20$ & $+0.001\pm0.008$ & $+0.032\pm0.010$ & $+1.23\pm0.11$ & $+4.84\pm0.26$ & $-3.99\pm0.30$ &  \\ 
	 & $-0.71\pm0.16$ & $+3.85\pm0.32$ & $+1.08\pm0.13$ & $+2.57\pm0.24$ & $+4.18\pm0.38$ & $+1.98\pm0.15$ & $+4.23\pm0.28$ &  \\ 
	 & $+2.89\pm0.15$ & $+0.055\pm0.003$ & $+0.176\pm0.004$ & $+2.50\pm0.16$ & $+1.90\pm0.18$ & $+1.16\pm0.13$ &  \\ 
 \\ 
4980  &  12:57:18.6  &  +26:58:46  &  19.64  &  18.33  &  17.65  &  17.32  &  17.29  &  \\ 
	 & $-0.09\pm0.38$ & $+1.22\pm0.26$ & $-0.028\pm0.011$ & $-0.002\pm0.013$ & $+0.93\pm0.14$ & $+4.45\pm0.34$ & $-2.48\pm0.38$ &  \\ 
	 & $+0.34\pm0.21$ & $+3.76\pm0.43$ & $+0.67\pm0.18$ & $+2.44\pm0.33$ & $+3.71\pm0.51$ & $+2.88\pm0.20$ & $+3.98\pm0.36$ &  \\ 
	 & $+2.04\pm0.20$ & $+0.038\pm0.004$ & $+0.127\pm0.005$ & $+2.67\pm0.21$ & $+2.22\pm0.23$ & $+1.13\pm0.18$ &  \\ 
 \\ 
4967  &  12:57:19.6  &  +27:36:49  &  19.02  &  17.52  &  16.92  &  16.67  &  16.39  &  \\ 
	 & $+4.06\pm0.19$ & $+3.17\pm0.14$ & $-0.084\pm0.006$ & $-0.044\pm0.007$ & $+0.78\pm0.08$ & $+2.49\pm0.20$ & $+1.48\pm0.20$ &  \\ 
	 & $+2.47\pm0.11$ & $+2.39\pm0.25$ & $+0.69\pm0.11$ & $+2.98\pm0.19$ & $+2.40\pm0.33$ & $+3.54\pm0.13$ & $+3.90\pm0.24$ &  \\ 
	 & $+1.87\pm0.13$ & $+0.023\pm0.003$ & $+0.109\pm0.004$ & $+1.95\pm0.15$ & $+1.58\pm0.16$ & $+0.80\pm0.12$ &  \\ 
 \\ 
4937  &  12:57:23.6  &  +27:32:59  &  19.64  &  17.71  &  17.05  &  16.72  &  16.52  &  \\ 
	 & $+0.66\pm0.33$ & $+1.29\pm0.24$ & $-0.022\pm0.009$ & $+0.005\pm0.012$ & $+0.93\pm0.13$ & $+3.87\pm0.34$ & $-1.63\pm0.36$ &  \\ 
	 & $+0.90\pm0.20$ & $+3.54\pm0.39$ & $+0.81\pm0.17$ & $+2.97\pm0.41$ & $+3.20\pm0.49$ & $+2.83\pm0.19$ & $+4.26\pm0.35$ &  \\ 
	 & $+2.47\pm0.19$ & $+0.038\pm0.004$ & $+0.141\pm0.005$ & $+2.03\pm0.20$ & $+2.32\pm0.22$ & $+1.59\pm0.16$ &  \\ 
 \\ 
4888  &  12:57:27.5  &  +27:38:10  &  19.04  &  17.58  &  16.90  &  16.58  &  16.42  &  \\ 
	 & $+2.11\pm0.21$ & $+1.92\pm0.15$ & $-0.053\pm0.006$ & $-0.018\pm0.007$ & $+0.66\pm0.09$ & $+3.86\pm0.21$ & $-1.25\pm0.22$ &  \\ 
	 & $+0.95\pm0.12$ & $+3.10\pm0.27$ & $+0.90\pm0.11$ & $+2.70\pm0.20$ & $+3.16\pm0.32$ & $+2.59\pm0.13$ & $+3.96\pm0.24$ &  \\ 
	 & $+2.24\pm0.13$ & $+0.040\pm0.003$ & $+0.128\pm0.003$ & $+2.04\pm0.14$ & $+1.82\pm0.15$ & ------ &  \\ 
 \\ 
4768  &  12:57:38.7  &  +27:27:18  &  19.71  &  18.43  &  17.81  &  17.45  &  17.44  &  \\ 
	 & $+1.18\pm0.36$ & $+1.31\pm0.26$ & $-0.052\pm0.010$ & $-0.037\pm0.012$ & $+0.69\pm0.15$ & $+3.81\pm0.35$ & $-1.00\pm0.37$ &  \\ 
	 & $+0.86\pm0.21$ & $+2.30\pm0.44$ & $+0.50\pm0.18$ & $+1.93\pm0.45$ & $+3.02\pm0.53$ & $+2.04\pm0.22$ & $+2.82\pm0.38$ &  \\ 
	 & $+1.90\pm0.21$ & $+0.023\pm0.005$ & $+0.108\pm0.005$ & $+2.11\pm0.22$ & $+1.64\pm0.24$ & $+1.27\pm0.18$ &  \\ 
 \\ 

\hline
\end{tabular}
\end{table*}

\begin{table*}
\caption{Derived parameters for the Coma dwarf galaxy sample. Galaxies are identified by their number in the 
Godwin et al. (1983, GMP) catalogue. The luminosities given by the second column 
are based on the SDSS Petrosian magnitude and include the corrections described in Section~\ref{sec:sigphotom}. The velocity 
dispersion $\sigma$ is from the literature compilation. 
The signal-to-noise ratio $S/N$ is per angstrom, averaged over 4400--5400\,\AA\ in the rest frame, $cz_\odot$ is the heliocentric velocity. 
The \eza\ parameters follow; $t_{\rm SSP}$ is the SSP-equivalent age estimated from Hbeta and metal lines as in Section~\ref{sec:poppars}.
Superscripts on the first column indicate: $^{\rm a}$ galaxy with emission lines; 
$^{\rm b}$ other galaxy with failed \eza\ fits (see details in Appendix~\ref{sec:ezafails}). 
}\label{tab:comadata}
\begin{tabular}{lrrrrrrrrrrrr}
\hline
\multicolumn{1}{l}{GMP} &  
\multicolumn{1}{c}{$\log\frac{L_r}{L_\odot}$} & 
\multicolumn{1}{c}{$\log\sigma$}  &
\multicolumn{1}{c}{$S/N$}  &
\multicolumn{1}{c}{$cz_\odot$} & 
\multicolumn{1}{c}{$t_{\rm SSP}$} & 
\multicolumn{1}{c}{$\left[{\rm Fe/H}\right]$} &
\multicolumn{1}{c}{$\left[{\rm Mg/Fe}\right]$} & 
\multicolumn{1}{c}{$\left[{\rm C/Fe}\right]$} &  
\multicolumn{1}{c}{$\left[{\rm N/Fe}\right]$} & 
\multicolumn{1}{c}{$\left[{\rm Ca/Fe}\right]$} \\
\hline
5361$^{\rm }$ \phantom{\huge 1}  &  9.23  &  1.716  &   36  &   7886  &  $ 2.5^{+0.9}_{-0.7}$  &  $+0.06\pm0.12$  &  $-0.09\pm0.09$  &  $-0.28\pm0.09$  &  $+0.10\pm0.12$  &  $+0.02\pm0.12$  \\ 
5296$^{\rm }$ \phantom{\huge 1}  &  8.90  &  1.758  &   50  &   7323  &  $ 6.5^{+1.8}_{-1.4}$  &  $-0.32\pm0.10$  &  $+0.07\pm0.07$  &  $-0.12\pm0.07$  &  $+0.00\pm0.09$  &  $+0.06\pm0.09$  \\ 
5254$^{\rm }$ \phantom{\huge 1}  &  9.40  &  ---  &   48  &   7836  &  $ 2.2^{+0.6}_{-0.5}$  &  $-0.04\pm0.10$  &  $+0.00\pm0.07$  &  $-0.27\pm0.07$  &  $+0.07\pm0.09$  &  $-0.10\pm0.09$  \\ 
5217$^{\rm }$ \phantom{\huge 1}  &  9.05  &  ---  &   37  &   8084  &  $ 2.6^{+0.9}_{-0.7}$  &  $-0.24\pm0.12$  &  $-0.05\pm0.09$  &  $-0.03\pm0.09$  &  $-0.02\pm0.12$  &  $+0.07\pm0.11$  \\ 
5178$^{\rm }$ \phantom{\huge 1}  &  9.00  &  ---  &   39  &   7612  &  $ 6.7^{+2.3}_{-1.7}$  &  $-0.51\pm0.12$  &  $+0.15\pm0.08$  &  $-0.08\pm0.09$  &  $+0.03\pm0.11$  &  $+0.21\pm0.11$  \\ 
5076$^{\rm }$ \phantom{\huge 1}  &  8.92  &  ---  &   46  &   7314  &  $ 7.3^{+2.2}_{-1.7}$  &  $-0.39\pm0.10$  &  $+0.15\pm0.07$  &  $+0.03\pm0.08$  &  $+0.09\pm0.10$  &  $+0.28\pm0.10$  \\ 
4980$^{\rm }$ \phantom{\huge 1}  &  8.84  &  1.656  &   34  &   7298  &  $ 1.9^{+0.8}_{-0.5}$  &  $+0.07\pm0.13$  &  $-0.12\pm0.10$  &  $-0.11\pm0.10$  &  $+0.06\pm0.13$  &  $+0.03\pm0.12$  \\ 
4967$^{\rm }$ \phantom{\huge 1}  &  9.13  &  ---  &   54  &   7765  &  $ 1.5^{+0.4}_{-0.3}$  &  $-0.36\pm0.09$  &  $+0.04\pm0.06$  &  $-0.08\pm0.07$  &  $-0.01\pm0.09$  &  $+0.00\pm0.08$  \\ 
4937$^{\rm }$ \phantom{\huge 1}  &  9.08  &  ---  &   36  &   6050  &  $ 2.2^{+0.8}_{-0.6}$  &  $-0.19\pm0.12$  &  $+0.06\pm0.09$  &  $-0.12\pm0.09$  &  $+0.15\pm0.12$  &  $+0.13\pm0.12$  \\ 
4888$^{\rm }$ \phantom{\huge 1}  &  9.14  &  ---  &   53  &   8037  &  $ 2.9^{+0.8}_{-0.6}$  &  $-0.40\pm0.09$  &  $+0.09\pm0.06$  &  $-0.04\pm0.07$  &  $-0.11\pm0.09$  &  $+0.00\pm0.08$  \\ 
4768$^{\rm }$ \phantom{\huge 1}  &  8.78  &  ---  &   34  &   7602  &  $ 7.8^{+3.2}_{-2.2}$  &  $-0.66\pm0.13$  &  $+0.01\pm0.10$  &  $-0.01\pm0.10$  &  $-0.29\pm0.13$  &  $+0.00\pm0.12$  \\ 
4602$^{\rm }$ \phantom{\huge 1}  &  9.23  &  ---  &   47  &   6457  &  $ 5.6^{+1.7}_{-1.3}$  &  $-0.16\pm0.10$  &  $+0.04\pm0.07$  &  $-0.06\pm0.07$  &  $+0.08\pm0.10$  &  $+0.09\pm0.09$  \\ 
4591$^{\rm }$ \phantom{\huge 1}  &  8.94  &  1.447  &   42  &   6476  &  $ 3.3^{+1.1}_{-0.8}$  &  $-0.28\pm0.11$  &  $-0.10\pm0.08$  &  $-0.16\pm0.08$  &  $-0.08\pm0.10$  &  $-0.01\pm0.10$  \\ 
4565$^{\rm }$ \phantom{\huge 1}  &  8.62  &  1.644  &   53  &   8663  &  $ 3.5^{+0.9}_{-0.7}$  &  $-0.65\pm0.09$  &  $+0.20\pm0.06$  &  $-0.01\pm0.07$  &  $-0.14\pm0.09$  &  $+0.38\pm0.09$  \\ 
4557$^{\rm }$ \phantom{\huge 1}  &  8.59  &  ---  &   50  &   5346  &  $ 6.4^{+1.8}_{-1.4}$  &  $-0.62\pm0.10$  &  $+0.10\pm0.07$  &  $-0.20\pm0.07$  &  $+0.36\pm0.09$  &  $+0.26\pm0.09$  \\ 
4539$^{\rm }$ \phantom{\huge 1}  &  8.75  &  ---  &   39  &   7127  &  $ 5.2^{+1.8}_{-1.4}$  &  $-0.55\pm0.12$  &  $+0.10\pm0.09$  &  $-0.07\pm0.09$  &  $+0.13\pm0.11$  &  $+0.20\pm0.11$  \\ 
4469$^{\rm }$ \phantom{\huge 1}  &  9.37  &  ---  &   48  &   7492  &  $ 5.5^{+1.6}_{-1.2}$  &  $-0.17\pm0.10$  &  $+0.00\pm0.07$  &  $-0.01\pm0.07$  &  $-0.10\pm0.09$  &  $+0.06\pm0.09$  \\ 
4453$^{\rm }$ \phantom{\huge 1}  &  8.66  &  ---  &   37  &   6775  &  $ 3.4^{+1.3}_{-0.9}$  &  $-0.27\pm0.12$  &  $-0.04\pm0.09$  &  $-0.05\pm0.09$  &  $-0.44\pm0.12$  &  $-0.12\pm0.11$  \\ 
4418$^{\rm }$ \phantom{\huge 1}  &  8.60  &  1.447  &   30  &   6729  &  $ 3.7^{+1.6}_{-1.1}$  &  $-0.34\pm0.14$  &  $+0.12\pm0.11$  &  $-0.07\pm0.11$  &  $-0.33\pm0.14$  &  $-0.04\pm0.14$  \\ 
4381$^{\rm }$ \phantom{\huge 1}  &  8.74  &  1.467  &   51  &   7642  &  $ 8.3^{+2.3}_{-1.8}$  &  $-0.30\pm0.09$  &  $+0.08\pm0.06$  &  $-0.05\pm0.07$  &  $-0.11\pm0.09$  &  $+0.05\pm0.09$  \\ 
4383$^{\rm }$ \phantom{\huge 1}  &  9.29  &  ---  &   46  &   7722  &  $ 3.4^{+1.0}_{-0.8}$  &  $-0.31\pm0.10$  &  $+0.02\pm0.07$  &  $-0.11\pm0.07$  &  $-0.18\pm0.10$  &  $-0.10\pm0.10$  \\ 
4366$^{\rm }$ \phantom{\huge 1}  &  9.03  &  ---  &   51  &   5685  &  $ 8.0^{+2.2}_{-1.7}$  &  $-0.22\pm0.09$  &  $+0.00\pm0.06$  &  $-0.06\pm0.07$  &  $-0.11\pm0.09$  &  $-0.03\pm0.09$  \\ 
4294$^{\rm a}$ \phantom{\huge 1}  &  9.32  &  ---  &   44  &   8039  &  ---  &  ---  &  ---  &  ---  &  ---  &  ---  \\ 
4268$^{\rm }$ \phantom{\huge 1}  &  8.79  &  ---  &   38  &   6996  &  $ 5.4^{+1.9}_{-1.4}$  &  $-0.64\pm0.12$  &  $+0.20\pm0.09$  &  $+0.08\pm0.09$  &  $-0.10\pm0.11$  &  $+0.26\pm0.11$  \\ 
4200$^{\rm b}$ \phantom{\huge 1}  &  9.64  &  2.059  &   33  &   5698  &  ---  &  ---  &  ---  &  ---  &  ---  &  ---  \\ 
4175$^{\rm }$ \phantom{\huge 1}  &  8.98  &  1.669  &   62  &   4480  &  $ 3.8^{+0.9}_{-0.7}$  &  $-0.19\pm0.08$  &  $-0.04\pm0.05$  &  $-0.13\pm0.06$  &  $+0.02\pm0.08$  &  $+0.06\pm0.08$  \\ 
4129$^{\rm }$ \phantom{\huge 1}  &  9.04  &  ---  &   31  &   6163  &  $ 9.8^{+4.3}_{-3.0}$  &  $-0.42\pm0.14$  &  $+0.38\pm0.11$  &  $+0.14\pm0.10$  &  $+0.08\pm0.14$  &  $+0.03\pm0.13$  \\ 
4042$^{\rm }$ \phantom{\huge 1}  &  8.90  &  1.523  &   35  &   8843  &  $12.4^{+4.9}_{-3.5}$  &  $-0.81\pm0.13$  &  $+0.30\pm0.10$  &  $-0.06\pm0.10$  &  $-0.07\pm0.12$  &  $+0.46\pm0.12$  \\ 
4035$^{\rm }$ \phantom{\huge 1}  &  9.07  &  1.540  &   44  &   6634  &  $ 7.2^{+2.2}_{-1.7}$  &  $-0.40\pm0.11$  &  $+0.11\pm0.07$  &  $-0.07\pm0.08$  &  $-0.14\pm0.10$  &  $+0.02\pm0.10$  \\ 
4029$^{\rm }$ \phantom{\huge 1}  &  8.88  &  ---  &   28  &   8763  &  $ 9.9^{+4.9}_{-3.3}$  &  $-0.70\pm0.15$  &  $+0.13\pm0.12$  &  $-0.11\pm0.12$  &  $+0.27\pm0.15$  &  $+0.63\pm0.15$  \\ 
4003$^{\rm }$ \phantom{\huge 1}  &  8.97  &  ---  &   38  &   7095  &  $ 2.4^{+0.9}_{-0.6}$  &  $-0.19\pm0.12$  &  $+0.20\pm0.09$  &  $+0.08\pm0.09$  &  $-0.12\pm0.12$  &  $+0.16\pm0.11$  \\ 
3973$^{\rm }$ \phantom{\huge 1}  &  8.91  &  ---  &   48  &   6685  &  $ 3.5^{+1.0}_{-0.8}$  &  $-0.11\pm0.10$  &  $-0.04\pm0.07$  &  $-0.02\pm0.07$  &  $-0.23\pm0.10$  &  $+0.11\pm0.09$  \\ 
3969$^{\rm }$ \phantom{\huge 1}  &  8.84  &  1.447  &   46  &   7433  &  $ 7.2^{+2.1}_{-1.7}$  &  $-0.49\pm0.10$  &  $+0.03\pm0.07$  &  $+0.02\pm0.07$  &  $-0.14\pm0.10$  &  $+0.27\pm0.09$  \\ 
3895$^{\rm a}$ \phantom{\huge 1}  &  9.23  &  ---  &   56  &   8701  &  ---  &  ---  &  ---  &  ---  &  ---  &  ---  \\ 
3855$^{\rm }$ \phantom{\huge 1}  &  9.14  &  1.711  &   40  &   5736  &  $ 6.5^{+2.3}_{-1.7}$  &  $-0.40\pm0.11$  &  $+0.30\pm0.08$  &  $-0.06\pm0.08$  &  $+0.09\pm0.11$  &  $+0.24\pm0.11$  \\ 
3780$^{\rm }$ \phantom{\huge 1}  &  9.25  &  1.756  &   54  &   8040  &  $ 7.5^{+2.0}_{-1.6}$  &  $-0.27\pm0.09$  &  $+0.07\pm0.06$  &  $+0.05\pm0.07$  &  $-0.22\pm0.09$  &  $+0.08\pm0.08$  \\ 
3719$^{\rm }$ \phantom{\huge 1}  &  9.06  &  ---  &   60  &   7826  &  $ 9.6^{+2.3}_{-1.9}$  &  $-0.36\pm0.08$  &  $+0.28\pm0.05$  &  $+0.08\pm0.06$  &  $+0.12\pm0.08$  &  $+0.12\pm0.08$  \\ 
3699$^{\rm b}$ \phantom{\huge 1}  &  8.90  &  1.572  &   38  &   8583  &  ---  &  ---  &  ---  &  ---  &  ---  &  ---  \\ 
3616$^{\rm }$ \phantom{\huge 1}  &  8.46  &  ---  &   38  &   6320  &  $13.1^{+4.8}_{-3.5}$  &  $-0.15\pm0.12$  &  $+0.08\pm0.09$  &  $-0.08\pm0.09$  &  $-0.10\pm0.12$  &  $-0.02\pm0.11$  \\ 

\hline
\end{tabular}
\end{table*}
\begin{table*}
\contcaption{}
\begin{tabular}{lrrrrrrrrrrrr}
\hline
\multicolumn{1}{l}{GMP} &  
\multicolumn{1}{c}{$\log\frac{L_r}{L_\odot}$} & 
\multicolumn{1}{c}{$\log\sigma$}  &
\multicolumn{1}{c}{$S/N$}  &
\multicolumn{1}{c}{$cz_\odot$} & 
\multicolumn{1}{c}{$t_{\rm SSP}$} & 
\multicolumn{1}{c}{$\left[{\rm Fe/H}\right]$} &
\multicolumn{1}{c}{$\left[{\rm Mg/Fe}\right]$} & 
\multicolumn{1}{c}{$\left[{\rm C/Fe}\right]$} &  
\multicolumn{1}{c}{$\left[{\rm N/Fe}\right]$} & 
\multicolumn{1}{c}{$\left[{\rm Ca/Fe}\right]$} \\
\hline
3565$^{\rm }$ \phantom{\huge 1}  &  8.96  &  1.647  &   41  &   7231  &  $ 9.5^{+3.2}_{-2.4}$  &  $-0.43\pm0.11$  &  $+0.05\pm0.08$  &  $-0.01\pm0.08$  &  $-0.07\pm0.11$  &  $+0.19\pm0.11$  \\ 
3489$^{\rm }$ \phantom{\huge 1}  &  9.03  &  ---  &   38  &   5478  &  $ 8.3^{+3.0}_{-2.2}$  &  $-0.24\pm0.12$  &  $+0.22\pm0.09$  &  $+0.08\pm0.09$  &  $+0.01\pm0.12$  &  $+0.21\pm0.11$  \\ 
3473$^{\rm }$ \phantom{\huge 1}  &  8.74  &  1.644  &   40  &   4959  &  $10.4^{+3.6}_{-2.7}$  &  $-0.66\pm0.11$  &  $+0.12\pm0.08$  &  $-0.15\pm0.08$  &  $+0.30\pm0.11$  &  $+0.21\pm0.11$  \\ 
3463$^{\rm }$ \phantom{\huge 1}  &  9.18  &  ---  &   54  &   6556  &  $ 4.7^{+1.2}_{-1.0}$  &  $-0.12\pm0.09$  &  $+0.06\pm0.06$  &  $-0.13\pm0.07$  &  $-0.01\pm0.09$  &  $+0.03\pm0.08$  \\ 
3438$^{\rm }$ \phantom{\huge 1}  &  8.62  &  1.505  &   32  &   5971  &  $12.5^{+5.3}_{-3.7}$  &  $-0.59\pm0.13$  &  $-0.12\pm0.10$  &  $-0.12\pm0.10$  &  $-0.12\pm0.13$  &  $+0.02\pm0.13$  \\ 
3439$^{\rm }$ \phantom{\huge 1}  &  9.55  &  1.800  &   59  &   3679  &  $ 1.4^{+0.3}_{-0.3}$  &  $+0.07\pm0.09$  &  $+0.06\pm0.06$  &  $-0.01\pm0.06$  &  $+0.03\pm0.08$  &  $+0.12\pm0.08$  \\ 
3406$^{\rm }$ \phantom{\huge 1}  &  8.75  &  ---  &   42  &   7182  &  $ 5.8^{+1.9}_{-1.5}$  &  $-0.29\pm0.11$  &  $+0.00\pm0.08$  &  $-0.24\pm0.08$  &  $-0.05\pm0.11$  &  $-0.06\pm0.10$  \\ 
3387$^{\rm }$ \phantom{\huge 1}  &  8.92  &  1.644  &   42  &   7403  &  $ 8.8^{+2.9}_{-2.2}$  &  $-0.86\pm0.11$  &  $+0.20\pm0.08$  &  $+0.01\pm0.08$  &  $-0.10\pm0.11$  &  $+0.40\pm0.10$  \\ 
3383$^{\rm }$ \phantom{\huge 1}  &  9.00  &  ---  &   60  &   4651  &  $ 8.2^{+2.0}_{-1.6}$  &  $-0.44\pm0.09$  &  $+0.11\pm0.05$  &  $-0.12\pm0.06$  &  $+0.02\pm0.08$  &  $+0.11\pm0.08$  \\ 
3376$^{\rm }$ \phantom{\huge 1}  &  8.74  &  1.572  &   37  &   7069  &  $ 7.7^{+2.9}_{-2.1}$  &  $-0.50\pm0.12$  &  $+0.05\pm0.09$  &  $-0.19\pm0.09$  &  $+0.03\pm0.12$  &  $-0.10\pm0.12$  \\ 
3312$^{\rm }$ \phantom{\huge 1}  &  8.76  &  1.663  &   58  &   7214  &  $10.4^{+2.6}_{-2.1}$  &  $-0.58\pm0.09$  &  $+0.24\pm0.06$  &  $+0.00\pm0.06$  &  $+0.10\pm0.08$  &  $+0.10\pm0.08$  \\ 
3311$^{\rm }$ \phantom{\huge 1}  &  8.67  &  ---  &   33  &   6337  &  $ 3.2^{+1.3}_{-0.9}$  &  $-0.11\pm0.13$  &  $+0.01\pm0.10$  &  $-0.22\pm0.10$  &  $+0.07\pm0.13$  &  $+0.15\pm0.13$  \\ 
3298$^{\rm }$ \phantom{\huge 1}  &  9.51  &  1.710  &   60  &   6780  &  $ 4.8^{+1.2}_{-0.9}$  &  $-0.26\pm0.08$  &  $+0.19\pm0.05$  &  $+0.05\pm0.06$  &  $-0.09\pm0.08$  &  $+0.20\pm0.08$  \\ 
3292$^{\rm }$ \phantom{\huge 1}  &  9.19  &  1.778  &   63  &   4958  &  $ 5.3^{+1.2}_{-1.0}$  &  $-0.24\pm0.08$  &  $+0.10\pm0.05$  &  $-0.04\pm0.06$  &  $+0.03\pm0.08$  &  $+0.06\pm0.07$  \\ 
3269$^{\rm }$ \phantom{\huge 1}  &  9.66  &  1.994  &   30  &   8032  &  $ 9.0^{+4.0}_{-2.8}$  &  $-0.23\pm0.14$  &  $+0.22\pm0.11$  &  $+0.08\pm0.11$  &  $-0.14\pm0.14$  &  $+0.07\pm0.14$  \\ 
3262$^{\rm }$ \phantom{\huge 1}  &  9.56  &  1.885  &   58  &   3743  &  $ 8.0^{+2.0}_{-1.6}$  &  $+0.06\pm0.09$  &  $+0.13\pm0.06$  &  $+0.10\pm0.06$  &  $+0.14\pm0.08$  &  $+0.06\pm0.08$  \\ 
3248$^{\rm }$ \phantom{\huge 1}  &  8.89  &  ---  &   46  &   7656  &  $ 7.4^{+2.3}_{-1.7}$  &  $-0.55\pm0.10$  &  $+0.10\pm0.07$  &  $-0.10\pm0.08$  &  $+0.23\pm0.10$  &  $+0.30\pm0.10$  \\ 
3238$^{\rm }$ \phantom{\huge 1}  &  9.75  &  1.829  &   40  &   6730  &  $ 5.1^{+1.8}_{-1.3}$  &  $+0.02\pm0.11$  &  $+0.11\pm0.08$  &  $+0.04\pm0.08$  &  $+0.10\pm0.11$  &  $+0.10\pm0.11$  \\ 
3206$^{\rm b}$ \phantom{\huge 1}  &  9.69  &  1.946  &   26  &   6898  &  ---  &  ---  &  ---  &  ---  &  ---  &  ---  \\ 
3196$^{\rm }$ \phantom{\huge 1}  &  8.97  &  1.740  &   56  &   6797  &  $ 8.1^{+2.1}_{-1.7}$  &  $-0.29\pm0.09$  &  $+0.03\pm0.06$  &  $-0.18\pm0.06$  &  $-0.07\pm0.08$  &  $-0.08\pm0.08$  \\ 
3166$^{\rm }$ \phantom{\huge 1}  &  8.84  &  1.716  &   40  &   8389  &  $11.7^{+4.0}_{-3.0}$  &  $-0.91\pm0.11$  &  $+0.39\pm0.08$  &  $+0.00\pm0.08$  &  $+0.23\pm0.11$  &  $+0.41\pm0.11$  \\ 
3131$^{\rm }$ \phantom{\huge 1}  &  8.90  &  ---  &   35  &   7248  &  $12.7^{+5.0}_{-3.6}$  &  $-0.59\pm0.13$  &  $+0.05\pm0.10$  &  $-0.11\pm0.09$  &  $+0.06\pm0.12$  &  $+0.00\pm0.12$  \\ 
3121$^{\rm }$ \phantom{\huge 1}  &  9.38  &  1.690  &   56  &   7468  &  $ 7.7^{+2.0}_{-1.6}$  &  $-0.13\pm0.09$  &  $+0.06\pm0.06$  &  $+0.10\pm0.06$  &  $-0.03\pm0.08$  &  $+0.07\pm0.08$  \\ 
3098$^{\rm }$ \phantom{\huge 1}  &  8.71  &  ---  &   35  &   6781  &  $ 5.9^{+2.3}_{-1.6}$  &  $-0.37\pm0.13$  &  $+0.06\pm0.10$  &  $-0.06\pm0.09$  &  $+0.05\pm0.12$  &  $+0.00\pm0.12$  \\ 
3080$^{\rm }$ \phantom{\huge 1}  &  8.56  &  1.329  &   29  &   6660  &  $ 4.9^{+2.3}_{-1.6}$  &  $-0.27\pm0.15$  &  $+0.11\pm0.12$  &  $-0.10\pm0.11$  &  $+0.05\pm0.15$  &  $-0.04\pm0.14$  \\ 
3058$^{\rm }$ \phantom{\huge 1}  &  9.29  &  1.607  &   47  &   5815  &  $ 6.6^{+2.0}_{-1.5}$  &  $-0.65\pm0.10$  &  $+0.22\pm0.07$  &  $-0.02\pm0.07$  &  $+0.12\pm0.10$  &  $+0.37\pm0.09$  \\ 
2942$^{\rm }$ \phantom{\huge 1}  &  9.71  &  2.170  &   32  &   7529  &  $10.1^{+4.3}_{-3.0}$  &  $-0.47\pm0.14$  &  $+0.43\pm0.10$  &  $+0.25\pm0.10$  &  $+0.56\pm0.13$  &  $+0.45\pm0.13$  \\ 
2931$^{\rm }$ \phantom{\huge 1}  &  8.71  &  ---  &   57  &   7779  &  $ 5.5^{+1.4}_{-1.1}$  &  $-0.45\pm0.09$  &  $+0.13\pm0.06$  &  $-0.00\pm0.06$  &  $-0.16\pm0.08$  &  $+0.09\pm0.08$  \\ 
2910$^{\rm a}$ \phantom{\huge 1}  &  9.73  &  ---  &   27  &   5300  &  ---  &  ---  &  ---  &  ---  &  ---  &  ---  \\ 
2879$^{\rm }$ \phantom{\huge 1}  &  9.14  &  1.711  &   52  &   7334  &  $11.5^{+3.1}_{-2.5}$  &  $-0.60\pm0.09$  &  $+0.20\pm0.06$  &  $-0.05\pm0.07$  &  $+0.05\pm0.09$  &  $+0.16\pm0.09$  \\ 
2852$^{\rm }$ \phantom{\huge 1}  &  9.30  &  1.613  &   54  &   7387  &  $ 6.9^{+1.8}_{-1.5}$  &  $-0.02\pm0.09$  &  $-0.01\pm0.06$  &  $+0.01\pm0.07$  &  $-0.06\pm0.09$  &  $+0.05\pm0.08$  \\ 
2805$^{\rm b}$ \phantom{\huge 1}  &  9.75  &  2.096  &   31  &   6113  &  ---  &  ---  &  ---  &  ---  &  ---  &  ---  \\ 
2801$^{\rm }$ \phantom{\huge 1}  &  8.65  &  1.467  &   28  &   7094  &  $ 4.8^{+2.3}_{-1.6}$  &  $-0.65\pm0.15$  &  $+0.12\pm0.12$  &  $-0.05\pm0.11$  &  $+0.02\pm0.15$  &  $+0.40\pm0.15$  \\ 
2800$^{\rm }$ \phantom{\huge 1}  &  9.05  &  ---  &   28  &   6944  &  $ 2.6^{+1.3}_{-0.8}$  &  $-0.10\pm0.15$  &  $+0.01\pm0.12$  &  $-0.05\pm0.11$  &  $+0.01\pm0.15$  &  $+0.32\pm0.15$  \\ 
2799$^{\rm }$ \phantom{\huge 1}  &  8.71  &  1.643  &   46  &   5992  &  $10.8^{+3.3}_{-2.5}$  &  $-0.61\pm0.10$  &  $+0.12\pm0.07$  &  $-0.07\pm0.07$  &  $-0.09\pm0.10$  &  $+0.18\pm0.10$  \\ 
2784$^{\rm }$ \phantom{\huge 1}  &  9.08  &  1.799  &   42  &   7800  &  $ 7.1^{+2.3}_{-1.8}$  &  $-0.35\pm0.11$  &  $+0.14\pm0.08$  &  $-0.03\pm0.08$  &  $-0.16\pm0.11$  &  $+0.04\pm0.10$  \\ 
2764$^{\rm }$ \phantom{\huge 1}  &  8.76  &  ---  &   47  &   6762  &  $ 4.2^{+1.2}_{-1.0}$  &  $-0.37\pm0.10$  &  $+0.12\pm0.07$  &  $-0.02\pm0.07$  &  $+0.06\pm0.10$  &  $+0.12\pm0.09$  \\ 
2753$^{\rm }$ \phantom{\huge 1}  &  9.21  &  1.693  &   35  &   7847  &  $ 4.1^{+1.6}_{-1.1}$  &  $-0.38\pm0.13$  &  $+0.08\pm0.09$  &  $-0.10\pm0.09$  &  $-0.11\pm0.12$  &  $-0.06\pm0.12$  \\ 
2728$^{\rm }$ \phantom{\huge 1}  &  9.17  &  ---  &   57  &   7500  &  $ 8.5^{+2.1}_{-1.7}$  &  $-0.35\pm0.09$  &  $+0.05\pm0.06$  &  $+0.00\pm0.06$  &  $-0.07\pm0.08$  &  $+0.19\pm0.08$  \\ 
2692$^{\rm }$ \phantom{\huge 1}  &  9.13  &  1.691  &   46  &   7935  &  $ 5.0^{+1.5}_{-1.2}$  &  $-0.29\pm0.10$  &  $-0.01\pm0.07$  &  $-0.14\pm0.07$  &  $-0.04\pm0.10$  &  $-0.07\pm0.10$  \\ 
2676$^{\rm }$ \phantom{\huge 1}  &  8.68  &  ---  &   31  &   5532  &  $ 8.9^{+3.9}_{-2.7}$  &  $-0.94\pm0.14$  &  $+0.42\pm0.11$  &  $-0.07\pm0.10$  &  $-0.07\pm0.14$  &  $+0.01\pm0.13$  \\ 
2626$^{\rm }$ \phantom{\huge 1}  &  8.90  &  ---  &   47  &   5172  &  $ 5.3^{+1.6}_{-1.2}$  &  $-0.34\pm0.10$  &  $+0.16\pm0.07$  &  $-0.23\pm0.07$  &  $+0.21\pm0.10$  &  $+0.09\pm0.09$  \\ 

\hline
\end{tabular}
\end{table*}
\begin{table*}
\contcaption{}
\begin{tabular}{lrrrrrrrrrrrr}
\hline
\multicolumn{1}{l}{GMP} &  
\multicolumn{1}{c}{$\log\frac{L_r}{L_\odot}$} & 
\multicolumn{1}{c}{$\log\sigma$}  &
\multicolumn{1}{c}{$S/N$}  &
\multicolumn{1}{c}{$cz_\odot$} & 
\multicolumn{1}{c}{$t_{\rm SSP}$} & 
\multicolumn{1}{c}{$\left[{\rm Fe/H}\right]$} &
\multicolumn{1}{c}{$\left[{\rm Mg/Fe}\right]$} & 
\multicolumn{1}{c}{$\left[{\rm C/Fe}\right]$} &  
\multicolumn{1}{c}{$\left[{\rm N/Fe}\right]$} & 
\multicolumn{1}{c}{$\left[{\rm Ca/Fe}\right]$} \\
\hline
2615$^{\rm b}$ \phantom{\huge 1}  &  9.65  &  1.904  &   53  &   6640  &  ---  &  ---  &  ---  &  ---  &  ---  &  ---  \\
2591$^{\rm }$ \phantom{\huge 1}  &  8.99  &  ---  &   37  &   8741  &  $ 8.1^{+3.0}_{-2.2}$  &  $-0.69\pm0.12$  &  $+0.32\pm0.09$  &  $-0.05\pm0.09$  &  $+0.05\pm0.12$  &  $+0.40\pm0.11$  \\ 
2585$^{\rm }$ \phantom{\huge 1}  &  9.00  &  1.462  &   43  &   6961  &  $ 3.9^{+1.3}_{-0.9}$  &  $-0.46\pm0.11$  &  $+0.13\pm0.08$  &  $-0.04\pm0.08$  &  $+0.00\pm0.10$  &  $+0.14\pm0.10$  \\ 
2529$^{\rm }$ \phantom{\huge 1}  &  9.00  &  1.623  &   62  &   8610  &  $ 9.9^{+2.3}_{-1.9}$  &  $-0.52\pm0.08$  &  $+0.21\pm0.05$  &  $-0.01\pm0.06$  &  $-0.03\pm0.08$  &  $+0.15\pm0.08$  \\ 
2478$^{\rm }$ \phantom{\huge 1}  &  9.16  &  1.719  &   59  &   8769  &  $ 6.2^{+1.5}_{-1.2}$  &  $-0.39\pm0.09$  &  $+0.12\pm0.06$  &  $-0.08\pm0.06$  &  $-0.10\pm0.08$  &  $+0.16\pm0.08$  \\ 
2420$^{\rm }$ \phantom{\huge 1}  &  8.72  &  ---  &   39  &   8019  &  $ 2.7^{+1.0}_{-0.7}$  &  $-0.21\pm0.12$  &  $-0.05\pm0.09$  &  $-0.24\pm0.09$  &  $+0.01\pm0.11$  &  $+0.01\pm0.11$  \\ 
2411$^{\rm }$ \phantom{\huge 1}  &  8.84  &  ---  &   37  &   6850  &  $10.2^{+3.8}_{-2.8}$  &  $-0.64\pm0.12$  &  $+0.12\pm0.09$  &  $-0.04\pm0.09$  &  $-0.22\pm0.12$  &  $+0.24\pm0.11$  \\ 
2376$^{\rm }$ \phantom{\huge 1}  &  9.10  &  1.616  &   37  &   6000  &  $ 3.3^{+1.2}_{-0.9}$  &  $+0.04\pm0.12$  &  $-0.01\pm0.09$  &  $-0.17\pm0.09$  &  $-0.03\pm0.12$  &  $-0.01\pm0.12$  \\ 

\hline
\end{tabular}
\end{table*}

\begin{table*}
\caption{Stellar population parameters from \eza\ for the Shapley sample, with their luminosity and velocity dispersion
included for convenience. Other relevant parameters for these galaxies, including the individual line index
measurements, were tabulated by Smith et al. (2007).}\label{tab:shapleydata}
\begin{tabular}{lrrrrrrrrrrrr}
\hline
Galaxy ID & 
\multicolumn{1}{c}{$\log\frac{L_r}{L_\odot}$} & 
\multicolumn{1}{c}{$\log\sigma$}  &
\multicolumn{1}{c}{$t_{\rm SSP}$}  &
\multicolumn{1}{c}{$\left[{\rm Fe}/{\rm H}\right]$}  & 
\multicolumn{1}{c}{$\left[{\rm Mg}/{\rm Fe}\right]$}  & 
\multicolumn{1}{c}{$\left[{\rm C}/{\rm Fe}\right]$}  & 
\multicolumn{1}{c}{$\left[{\rm N}/{\rm Fe}\right]$}  & 
\multicolumn{1}{c}{$\left[{\rm Ca}/{\rm Fe}\right]$}  \\
\hline
132328.9-314242$^{\rm }$ \phantom{\huge 1}  &  10.08  &  2.067  &  $ 5.1^{+1.1}_{-0.9}$  &  $+0.11\pm0.08$  &  $+0.10\pm0.06$  &  $+0.03\pm0.06$  &  $+0.12\pm0.08$  &  $+0.06\pm0.08$  \\ 
132350.3-313519$^{\rm }$ \phantom{\huge 1}  &  9.60  &  ---  &  $ 3.3^{+1.2}_{-0.9}$  &  $-0.23\pm0.11$  &  $+0.07\pm0.10$  &  $+0.05\pm0.10$  &  $+0.01\pm0.13$  &  $+0.25\pm0.12$  \\ 
132406.9-314449$^{\rm }$ \phantom{\huge 1}  &  9.97  &  2.041  &  $11.2^{+2.6}_{-2.1}$  &  $-0.41\pm0.08$  &  $+0.31\pm0.07$  &  $+0.04\pm0.06$  &  $+0.07\pm0.08$  &  $+0.21\pm0.08$  \\ 
132418.2-314229$^{\rm }$ \phantom{\huge 1}  &  10.41  &  2.297  &  $ 9.3^{+0.9}_{-0.8}$  &  $-0.01\pm0.05$  &  $+0.22\pm0.03$  &  $+0.21\pm0.02$  &  $+0.21\pm0.03$  &  $+0.04\pm0.04$  \\ 
132425.9-314117$^{\rm }$ \phantom{\huge 1}  &  10.46  &  2.156  &  $ 7.5^{+1.0}_{-0.9}$  &  $+0.05\pm0.06$  &  $+0.06\pm0.04$  &  $+0.04\pm0.03$  &  $+0.14\pm0.05$  &  $-0.01\pm0.05$  \\ 
132431.0-315625$^{\rm }$ \phantom{\huge 1}  &  9.66  &  ---  &  $ 7.1^{+2.6}_{-1.9}$  &  $-0.29\pm0.11$  &  $-0.03\pm0.10$  &  $-0.45\pm0.10$  &  $+0.29\pm0.13$  &  $-0.19\pm0.12$  \\ 
132507.5-314625$^{\rm }$ \phantom{\huge 1}  &  9.44  &  ---  &  $ 6.4^{+2.3}_{-1.7}$  &  $-0.53\pm0.11$  &  $+0.16\pm0.10$  &  $-0.04\pm0.10$  &  $-0.09\pm0.13$  &  $+0.06\pm0.12$  \\ 
132642.2-312858$^{\rm }$ \phantom{\huge 1}  &  10.81  &  2.255  &  $10.6^{+1.5}_{-1.3}$  &  $+0.02\pm0.06$  &  $+0.08\pm0.04$  &  $+0.06\pm0.04$  &  $+0.23\pm0.05$  &  $+0.01\pm0.05$  \\ 
132656.0-312528$^{\rm }$ \phantom{\huge 1}  &  10.30  &  2.031  &  $ 6.6^{+0.7}_{-0.7}$  &  $-0.01\pm0.05$  &  $+0.15\pm0.04$  &  $+0.13\pm0.03$  &  $+0.04\pm0.04$  &  $+0.05\pm0.05$  \\ 
132658.3-313030$^{\rm }$ \phantom{\huge 1}  &  9.76  &  1.921  &  $ 8.6^{+1.5}_{-1.3}$  &  $-0.17\pm0.07$  &  $+0.19\pm0.05$  &  $+0.12\pm0.05$  &  $+0.13\pm0.06$  &  $+0.15\pm0.07$  \\ 
132721.8-312639$^{\rm }$ \phantom{\huge 1}  &  9.74  &  1.588  &  $ 4.0^{+0.8}_{-0.7}$  &  $+0.00\pm0.08$  &  $+0.03\pm0.06$  &  $-0.15\pm0.05$  &  $-0.03\pm0.07$  &  $-0.09\pm0.07$  \\ 
132722.0-314024$^{\rm }$ \phantom{\huge 1}  &  10.11  &  1.998  &  $ 6.6^{+0.8}_{-0.7}$  &  $-0.12\pm0.06$  &  $+0.18\pm0.04$  &  $+0.17\pm0.03$  &  $+0.13\pm0.04$  &  $+0.18\pm0.05$  \\ 
132723.5-312238$^{\rm }$ \phantom{\huge 1}  &  9.79  &  1.751  &  $ 7.6^{+3.4}_{-2.4}$  &  $-0.22\pm0.13$  &  $+0.07\pm0.11$  &  $-0.09\pm0.12$  &  $+0.06\pm0.15$  &  $-0.01\pm0.15$  \\ 
132724.2-313701$^{\rm }$ \phantom{\huge 1}  &  9.55  &  1.845  &  $11.6^{+4.9}_{-3.4}$  &  $-0.56\pm0.13$  &  $+0.34\pm0.11$  &  $-0.04\pm0.12$  &  $+0.11\pm0.14$  &  $+0.22\pm0.14$  \\ 
132731.3-313001$^{\rm }$ \phantom{\huge 1}  &  9.96  &  1.834  &  $ 5.6^{+1.2}_{-1.0}$  &  $-0.09\pm0.08$  &  $+0.22\pm0.06$  &  $+0.04\pm0.06$  &  $+0.01\pm0.08$  &  $+0.17\pm0.08$  \\ 
132732.8-311623$^{\rm }$ \phantom{\huge 1}  &  9.93  &  1.893  &  $ 6.1^{+1.2}_{-1.0}$  &  $-0.07\pm0.07$  &  $+0.11\pm0.06$  &  $+0.06\pm0.05$  &  $+0.03\pm0.07$  &  $+0.17\pm0.07$  \\ 
132735.1-312660$^{\rm }$ \phantom{\huge 1}  &  9.44  &  ---  &  $ 2.9^{+0.7}_{-0.6}$  &  $+0.10\pm0.09$  &  $+0.02\pm0.07$  &  $-0.24\pm0.07$  &  $+0.07\pm0.09$  &  $-0.08\pm0.09$  \\ 
132737.6-312749$^{\rm }$ \phantom{\huge 1}  &  10.08  &  2.083  &  $10.2^{+2.6}_{-2.1}$  &  $+0.00\pm0.09$  &  $+0.12\pm0.07$  &  $+0.03\pm0.07$  &  $+0.03\pm0.09$  &  $-0.05\pm0.09$  \\ 
132743.1-314841$^{\rm }$ \phantom{\huge 1}  &  9.39  &  1.708  &  $ 6.7^{+2.1}_{-1.6}$  &  $-0.28\pm0.10$  &  $+0.38\pm0.08$  &  $+0.24\pm0.09$  &  $-0.18\pm0.11$  &  $+0.29\pm0.11$  \\ 
132747.6-313829$^{\rm }$ \phantom{\huge 1}  &  9.90  &  2.144  &  $13.7^{+3.0}_{-2.5}$  &  $-0.18\pm0.08$  &  $+0.24\pm0.06$  &  $-0.08\pm0.06$  &  $+0.25\pm0.08$  &  $+0.02\pm0.08$  \\ 
132747.8-312451$^{\rm }$ \phantom{\huge 1}  &  10.47  &  2.357  &  $13.8^{+1.3}_{-1.2}$  &  $-0.04\pm0.05$  &  $+0.24\pm0.03$  &  $+0.26\pm0.02$  &  $+0.24\pm0.03$  &  $+0.10\pm0.04$  \\ 
132751.9-314128$^{\rm }$ \phantom{\huge 1}  &  9.44  &  1.719  &  $ 3.6^{+1.4}_{-1.0}$  &  $-0.18\pm0.12$  &  $+0.24\pm0.10$  &  $-0.10\pm0.11$  &  $+0.07\pm0.13$  &  $+0.29\pm0.13$  \\ 
132753.1-312646$^{\rm }$ \phantom{\huge 1}  &  9.56  &  1.744  &  $ 9.6^{+1.8}_{-1.5}$  &  $-0.27\pm0.07$  &  $+0.15\pm0.06$  &  $-0.12\pm0.05$  &  $+0.04\pm0.07$  &  $+0.22\pm0.07$  \\ 
132753.6-313246$^{\rm }$ \phantom{\huge 1}  &  9.64  &  ---  &  $10.2^{+2.4}_{-1.9}$  &  $-0.51\pm0.08$  &  $+0.33\pm0.07$  &  $+0.14\pm0.06$  &  $+0.02\pm0.08$  &  $+0.33\pm0.08$  \\ 
132755.3-313609$^{\rm }$ \phantom{\huge 1}  &  9.70  &  ---  &  $ 3.7^{+0.6}_{-0.5}$  &  $-0.34\pm0.07$  &  $+0.11\pm0.05$  &  $-0.13\pm0.04$  &  $+0.00\pm0.06$  &  $+0.08\pm0.06$  \\ 
132755.6-312857$^{\rm }$ \phantom{\huge 1}  &  10.36  &  2.178  &  $10.1^{+1.0}_{-0.9}$  &  $-0.07\pm0.05$  &  $+0.16\pm0.03$  &  $+0.08\pm0.02$  &  $+0.12\pm0.04$  &  $+0.02\pm0.04$  \\ 
132758.1-314116$^{\rm }$ \phantom{\huge 1}  &  10.06  &  1.886  &  $ 7.5^{+0.9}_{-0.8}$  &  $-0.20\pm0.06$  &  $+0.13\pm0.04$  &  $+0.10\pm0.03$  &  $-0.01\pm0.05$  &  $+0.11\pm0.05$  \\

\hline
\end{tabular}
\end{table*}
\begin{table*}
\contcaption{}
\begin{tabular}{lrrrrrrrrrrrr}
\hline
Galaxy ID & 
\multicolumn{1}{c}{$\log\frac{L_r}{L_\odot}$} & 
\multicolumn{1}{c}{$\log\sigma$}  &
\multicolumn{1}{c}{$t_{\rm SSP}$} & 
\multicolumn{1}{c}{$\left[{\rm Fe}/{\rm H}\right]$}  & 
\multicolumn{1}{c}{$\left[{\rm Mg}/{\rm Fe}\right]$}  & 
\multicolumn{1}{c}{$\left[{\rm C}/{\rm Fe}\right]$}  & 
\multicolumn{1}{c}{$\left[{\rm N}/{\rm Fe}\right]$}  & 
\multicolumn{1}{c}{$\left[{\rm Ca}/{\rm Fe}\right]$}  \\
\hline
132810.5-312310$^{\rm }$ \phantom{\huge 1}  &  10.55  &  2.267  &  $ 8.2^{+0.7}_{-0.6}$  &  $+0.10\pm0.05$  &  $+0.19\pm0.03$  &  $+0.26\pm0.02$  &  $+0.34\pm0.03$  &  $+0.14\pm0.04$  \\ 
132817.1-313339$^{\rm }$ \phantom{\huge 1}  &  10.56  &  2.323  &  $ 9.8^{+0.7}_{-0.7}$  &  $+0.03\pm0.04$  &  $+0.25\pm0.03$  &  $+0.31\pm0.01$  &  $+0.37\pm0.02$  &  $+0.15\pm0.03$  \\ 
132824.3-311439$^{\rm }$ \phantom{\huge 1}  &  10.33  &  2.213  &  $ 7.9^{+0.6}_{-0.6}$  &  $-0.16\pm0.05$  &  $+0.16\pm0.03$  &  $+0.19\pm0.02$  &  $+0.11\pm0.03$  &  $+0.11\pm0.04$  \\ 
132826.4-312800$^{\rm }$ \phantom{\huge 1}  &  10.23  &  2.242  &  $ 9.8^{+1.0}_{-0.9}$  &  $-0.07\pm0.05$  &  $+0.29\pm0.03$  &  $+0.27\pm0.02$  &  $+0.20\pm0.04$  &  $+0.13\pm0.04$  \\ 
132826.5-312939$^{\rm }$ \phantom{\huge 1}  &  9.35  &  1.501  &  $11.4^{+4.3}_{-3.1}$  &  $-0.09\pm0.12$  &  $+0.03\pm0.10$  &  $-0.36\pm0.11$  &  $-0.04\pm0.13$  &  $-0.18\pm0.13$  \\ 
132829.5-313605$^{\rm }$ \phantom{\huge 1}  &  9.47  &  1.707  &  $ 7.7^{+3.5}_{-2.4}$  &  $-0.09\pm0.13$  &  $+0.06\pm0.11$  &  $-0.07\pm0.12$  &  $-0.17\pm0.15$  &  $+0.26\pm0.15$  \\ 
132830.8-313231$^{\rm }$ \phantom{\huge 1}  &  10.14  &  2.171  &  $ 9.7^{+1.0}_{-0.9}$  &  $-0.04\pm0.05$  &  $+0.21\pm0.04$  &  $+0.15\pm0.03$  &  $+0.20\pm0.04$  &  $+0.06\pm0.04$  \\ 
132837.1-313358$^{\rm }$ \phantom{\huge 1}  &  10.06  &  1.824  &  $ 2.4^{+0.6}_{-0.4}$  &  $+0.06\pm0.08$  &  $+0.01\pm0.07$  &  $-0.16\pm0.06$  &  $+0.16\pm0.08$  &  $-0.03\pm0.09$  \\ 
132844.3-311414$^{\rm }$ \phantom{\huge 1}  &  10.47  &  2.312  &  $ 9.9^{+0.8}_{-0.8}$  &  $-0.11\pm0.05$  &  $+0.25\pm0.03$  &  $+0.24\pm0.02$  &  $+0.28\pm0.03$  &  $+0.16\pm0.04$  \\ 
132847.3-314135$^{\rm }$ \phantom{\huge 1}  &  10.40  &  2.047  &  $ 6.0^{+0.8}_{-0.7}$  &  $+0.11\pm0.06$  &  $+0.06\pm0.04$  &  $+0.03\pm0.04$  &  $+0.14\pm0.05$  &  $-0.01\pm0.06$  \\ 
132848.3-313947$^{\rm }$ \phantom{\huge 1}  &  10.23  &  2.273  &  $ 8.6^{+1.1}_{-0.9}$  &  $-0.06\pm0.06$  &  $+0.29\pm0.04$  &  $+0.18\pm0.03$  &  $+0.17\pm0.04$  &  $+0.07\pm0.05$  \\ 
132850.1-312934$^{\rm }$ \phantom{\huge 1}  &  9.53  &  1.784  &  $11.1^{+4.2}_{-3.0}$  &  $-0.75\pm0.12$  &  $+0.25\pm0.10$  &  $+0.10\pm0.10$  &  $+0.05\pm0.13$  &  $+0.38\pm0.13$  \\ 
132851.6-313525$^{\rm }$ \phantom{\huge 1}  &  9.63  &  1.880  &  $11.9^{+2.1}_{-1.8}$  &  $-0.54\pm0.07$  &  $+0.36\pm0.05$  &  $-0.02\pm0.05$  &  $+0.19\pm0.06$  &  $+0.29\pm0.07$  \\ 
132857.1-311908$^{\rm }$ \phantom{\huge 1}  &  9.62  &  1.899  &  $ 6.7^{+1.5}_{-1.2}$  &  $-0.11\pm0.08$  &  $+0.16\pm0.07$  &  $+0.02\pm0.06$  &  $+0.16\pm0.08$  &  $+0.19\pm0.08$  \\ 
132857.2-314202$^{\rm }$ \phantom{\huge 1}  &  9.98  &  2.127  &  $ 7.8^{+1.4}_{-1.2}$  &  $-0.06\pm0.07$  &  $+0.18\pm0.06$  &  $+0.29\pm0.05$  &  $+0.14\pm0.07$  &  $+0.14\pm0.07$  \\ 
132901.0-313322$^{\rm }$ \phantom{\huge 1}  &  9.77  &  1.888  &  $15.4^{+3.4}_{-2.8}$  &  $-0.66\pm0.08$  &  $+0.36\pm0.06$  &  $+0.18\pm0.06$  &  $+0.04\pm0.08$  &  $+0.27\pm0.08$  \\ 
132906.2-313208$^{\rm }$ \phantom{\huge 1}  &  10.11  &  1.990  &  $ 7.7^{+1.1}_{-1.0}$  &  $-0.05\pm0.06$  &  $+0.13\pm0.05$  &  $+0.02\pm0.04$  &  $+0.19\pm0.05$  &  $+0.13\pm0.06$  \\ 
132906.4-311712$^{\rm }$ \phantom{\huge 1}  &  10.12  &  2.055  &  $ 6.6^{+0.9}_{-0.8}$  &  $+0.11\pm0.06$  &  $+0.10\pm0.04$  &  $+0.09\pm0.03$  &  $+0.13\pm0.05$  &  $+0.04\pm0.05$  \\ 
132909.1-314303$^{\rm }$ \phantom{\huge 1}  &  9.59  &  1.702  &  $ 9.6^{+3.2}_{-2.4}$  &  $-0.22\pm0.11$  &  $+0.04\pm0.09$  &  $-0.14\pm0.09$  &  $+0.20\pm0.11$  &  $+0.16\pm0.11$  \\ 
132914.7-314934$^{\rm }$ \phantom{\huge 1}  &  10.06  &  1.942  &  $ 8.0^{+1.2}_{-1.1}$  &  $-0.10\pm0.07$  &  $+0.09\pm0.05$  &  $+0.10\pm0.04$  &  $+0.05\pm0.05$  &  $+0.05\pm0.06$  \\ 
132921.5-312514$^{\rm }$ \phantom{\huge 1}  &  9.83  &  ---  &  $ 3.3^{+0.6}_{-0.5}$  &  $-0.18\pm0.07$  &  $+0.07\pm0.05$  &  $-0.07\pm0.05$  &  $-0.16\pm0.06$  &  $-0.02\pm0.07$  \\ 
132930.7-314915$^{\rm }$ \phantom{\huge 1}  &  10.09  &  1.804  &  $ 3.4^{+0.5}_{-0.4}$  &  $+0.03\pm0.06$  &  $+0.02\pm0.05$  &  $-0.03\pm0.04$  &  $-0.01\pm0.05$  &  $-0.05\pm0.06$  \\ 
132933.5-314738$^{\rm }$ \phantom{\huge 1}  &  9.68  &  ---  &  $ 3.2^{+0.9}_{-0.7}$  &  $-0.17\pm0.09$  &  $+0.10\pm0.07$  &  $-0.02\pm0.07$  &  $-0.06\pm0.09$  &  $+0.24\pm0.10$  \\ 
133218.7-315004$^{\rm }$ \phantom{\huge 1}  &  9.82  &  1.719  &  $ 8.6^{+5.0}_{-3.2}$  &  $-0.24\pm0.16$  &  $+0.08\pm0.14$  &  $+0.00\pm0.16$  &  $-0.10\pm0.19$  &  $-0.07\pm0.18$  \\ 
133222.5-314832$^{\rm }$ \phantom{\huge 1}  &  9.72  &  2.112  &  $14.3^{+3.1}_{-2.6}$  &  $-0.22\pm0.08$  &  $+0.26\pm0.06$  &  $+0.03\pm0.06$  &  $+0.15\pm0.08$  &  $-0.06\pm0.08$  \\ 
133239.3-314953$^{\rm }$ \phantom{\huge 1}  &  10.43  &  2.234  &  $ 6.4^{+0.8}_{-0.7}$  &  $+0.14\pm0.06$  &  $+0.08\pm0.04$  &  $+0.16\pm0.03$  &  $+0.20\pm0.04$  &  $+0.04\pm0.05$  \\ 
133245.7-314911$^{\rm }$ \phantom{\huge 1}  &  10.15  &  2.188  &  $10.1^{+1.3}_{-1.1}$  &  $-0.12\pm0.06$  &  $+0.18\pm0.04$  &  $+0.09\pm0.03$  &  $+0.15\pm0.05$  &  $+0.03\pm0.05$  \\ 
133248.4-315540$^{\rm }$ \phantom{\huge 1}  &  9.78  &  1.685  &  $ 8.2^{+2.9}_{-2.1}$  &  $-0.34\pm0.11$  &  $+0.18\pm0.09$  &  $+0.07\pm0.10$  &  $+0.01\pm0.12$  &  $+0.34\pm0.12$  \\ 
133256.1-314218$^{\rm }$ \phantom{\huge 1}  &  9.74  &  1.693  &  $11.0^{+3.6}_{-2.7}$  &  $-0.29\pm0.11$  &  $+0.11\pm0.09$  &  $+0.05\pm0.09$  &  $-0.14\pm0.11$  &  $-0.02\pm0.11$  \\ 
133312.2-314214$^{\rm }$ \phantom{\huge 1}  &  10.04  &  ---  &  $ 2.9^{+0.7}_{-0.5}$  &  $+0.10\pm0.08$  &  $+0.00\pm0.06$  &  $+0.06\pm0.06$  &  $-0.04\pm0.08$  &  $-0.01\pm0.08$  \\ 
133313.3-313413$^{\rm }$ \phantom{\huge 1}  &  10.16  &  2.014  &  $ 5.2^{+0.9}_{-0.7}$  &  $+0.10\pm0.07$  &  $+0.08\pm0.05$  &  $+0.01\pm0.04$  &  $-0.04\pm0.06$  &  $-0.05\pm0.06$  \\ 
133333.8-313710$^{\rm }$ \phantom{\huge 1}  &  10.07  &  2.121  &  $ 6.8^{+1.0}_{-0.9}$  &  $-0.15\pm0.06$  &  $+0.27\pm0.05$  &  $+0.15\pm0.04$  &  $+0.03\pm0.05$  &  $+0.08\pm0.06$  \\ 
133339.6-314847$^{\rm }$ \phantom{\huge 1}  &  9.53  &  1.808  &  $ 9.6^{+3.7}_{-2.7}$  &  $-0.29\pm0.12$  &  $+0.21\pm0.10$  &  $-0.17\pm0.11$  &  $+0.10\pm0.13$  &  $+0.11\pm0.13$  \\ 
133347.9-313322$^{\rm }$ \phantom{\huge 1}  &  10.43  &  2.203  &  $ 9.0^{+1.0}_{-0.9}$  &  $+0.03\pm0.06$  &  $+0.19\pm0.04$  &  $+0.16\pm0.03$  &  $+0.15\pm0.04$  &  $+0.04\pm0.05$  \\ 
133409.6-314239$^{\rm }$ \phantom{\huge 1}  &  9.92  &  1.856  &  $ 4.8^{+1.4}_{-1.1}$  &  $-0.14\pm0.10$  &  $+0.15\pm0.08$  &  $+0.04\pm0.08$  &  $-0.04\pm0.10$  &  $+0.13\pm0.10$  \\ 
132802.4-314340$^{\rm }$ \phantom{\huge 1}  &  10.20  &  2.096  &  $ 8.2^{+1.4}_{-1.2}$  &  $-0.18\pm0.07$  &  $+0.25\pm0.05$  &  $+0.11\pm0.05$  &  $+0.15\pm0.06$  &  $+0.21\pm0.07$  \\
132802.6-314521$^{\rm }$ \phantom{\huge 1}  &  10.84  &  2.455  &  $13.6^{+0.9}_{-0.8}$  &  $+0.01\pm0.04$  &  $+0.31\pm0.03$  &  $+0.38\pm0.01$  &  $+0.36\pm0.02$  &  $+0.04\pm0.03$  \\
132804.0-313836$^{\rm }$ \phantom{\huge 1}  &  9.79  &  1.890  &  $11.5^{+2.9}_{-2.3}$  &  $-0.27\pm0.09$  &  $+0.13\pm0.07$  &  $-0.14\pm0.07$  &  $+0.22\pm0.09$  &  $+0.15\pm0.09$  \\
132806.6-314146$^{\rm }$ \phantom{\huge 1}  &  9.81  &  1.588  &  $ 6.8^{+1.6}_{-1.3}$  &  $-0.11\pm0.08$  &  $+0.09\pm0.07$  &  $+0.04\pm0.06$  &  $+0.19\pm0.08$  &  $+0.11\pm0.08$  \\

\hline
\end{tabular}
\end{table*}

\end{document}